\documentclass[aps,prb,twocolumn,superscriptaddress]{revtex4-2}

\usepackage{amsmath,amssymb}
\usepackage{graphicx}
\usepackage{comment}

\usepackage{xcolor} 
\usepackage{soul}
\definecolor{myyellow}{RGB}{255,255,0}
\sethlcolor{myyellow}

\usepackage{comment}
\usepackage{amsmath}
\usepackage{amssymb}
\usepackage{graphicx}
\usepackage{float}
\usepackage{dsfont}
\usepackage{epstopdf}
\usepackage{soul}
\usepackage{color}
\usepackage{braket}
\usepackage{subfigure}
\usepackage{setspace}
\usepackage{enumitem}
\usepackage[title]{appendix}
\usepackage[normalem]{ulem}

\emergencystretch=\maxdimen
\usepackage{titlesec}
\titlespacing\subsection{0pt}{12pt plus 4pt minus 2pt}{12pt plus 2pt minus 2pt}

\usepackage{verbatim}

\usepackage[colorlinks=true,linkcolor=magenta,urlcolor=blue,citecolor=blue]{hyperref}

\begin{document}

\title{Airy Resonances in Photonic Crystal Superpotentials}

\author{Zeyu Zhang*}
\affiliation{Department of Physics, The Pennsylvania State University, University Park, PA, USA}
\thanks{These authors contributed equally}

\author{Brian Gould*}
\affiliation{Department of Physics, The Pennsylvania State University, University Park, PA, USA}
\thanks{These authors contributed equally}

\author{Maria Barsukova}
\affiliation{Department of Physics, The Pennsylvania State University, University Park, PA, USA}

\author{Mikael C. Rechtsman}
\affiliation{Department of Physics, The Pennsylvania State University, University Park, PA, USA}

\date{\today}

\begin{abstract}
Airy wavefunctions are associated with one of the simplest scenarios in wave mechanics: a quantum bouncing ball.  In other words, they are the eigenstates of the time-independent Schr\"odinger equation with a linear potential.  In the domain of optics, laser beams that are spatially shaped as Airy functions (`Airy beams') have been shown to exhibit a prominent lobe that follows a curved path, rather than propagating in a straight line, and which has self-healing properties in the presence of obstacles.  Here, we observe the presence of Airy {\it resonances} in two-dimensional photonic crystals composed of a lattice of holes in a silicon slab.  Analogously to electrons in a linear potential, these Airy resonances arise due to a linear spatial variation in the lattice constant of the holes.  We map the electromagnetic description of the photonic crystal onto a 2D non-Hermitian Schr\"odinger equation with a linear potential, which we call a `superpotential'.  The non-Hermiticity appears in the form of a complex effective mass due to out-of-plane radiation and fundamentally alters the collective optical response of the Airy resonances.      
\end{abstract}

\maketitle

% Introduction
Airy functions are the eigenstates of the time-independent Schr\"odinger equation with a linear potential \cite{vallee2010airy}. In the domain of optics, laser beams that are spatially shaped as Airy functions are known as Airy beams \cite{balazs1979nonspreading, siviloglou2007accelerating, siviloglou2007observation}. Due to their non-diffracting, accelerating, and self-healing nature \cite{broky2008self}, Airy beams have been widely used for abrupt auto-focusing \cite{chremmos2011pre}, as well as in optical micromanipulation of small particles \cite{baumgartl2008optically}, among other applications \cite{ellenbogen2009nonlinear, polynkin2009curved, dolev2012surface, rose2013airy, jia2014isotropic, vettenburg2014light}.  In the context of lattices, Airy beams have been theoretically predicted to be realizable in photonic crystals \cite{kaminer2013self}, and predicted and realized in waveguide arrays \cite{Ramy2011, qi2014observation, makris2014accelerating}. Airy resonances, on the other hand, refer to confined resonant states whose wavefunction can be described by the Airy function, typically emerging in systems with a linear potential \cite{nesvizhevsky2002quantum, delgoffe2019tilted}.  While Airy beams have been extensively studied, Airy resonances remain rarely explored, especially in photonics. Airy resonances have been realized in photonic cavities \cite{delgoffe2019tilted}. However, the system size was limited to cavity modes, which are only several unit cells in only one dimension. 

Here, we directly realize Airy resonances in two-dimensional photonic crystal slabs which are global modes of the whole system. We theoretically demonstrate that if a slow variation is imposed on a periodic photonic crystal, the states near a quadratic band edge are effectively governed by a Schr\"odinger equation with an additional spatially varying potential term, which is referred to here as ``superpotential”. In particular, by slowly varying the lattice constant as a function of position, we experimentally introduce a linear superpotential and directly observe the Airy resonances that result.  Importantly, we examine the effect of the non-Hermiticity that arises due to the out-of-plane radiation of the photonic crystal slab.  Because the band on which we focus has a bound state in the continuum (BIC) at the $\Gamma$ ($\mathbf{k}=\mathbf{0}$) point, the effective mass associated with the photon dispersion is complex, meaning that the loss scales with the square of the in-plane lattice momentum.  The superpotential formalism discussed here allows us to incorporate the non-Hermitian effects  \cite{makris2008beam, ruter2010observation, el2018non}, allowing for the calculation of both resonance frequencies and linewidths for the Airy modes. Additionally, we observe that the light leaking out-of-plane from the Airy resonances propagates in free space as an accelerating Airy beam. 

% End introduction

% Figure 1 discussion

\begin{figure*}[ht]
  \begin{center}
    \includegraphics[width=16 cm]{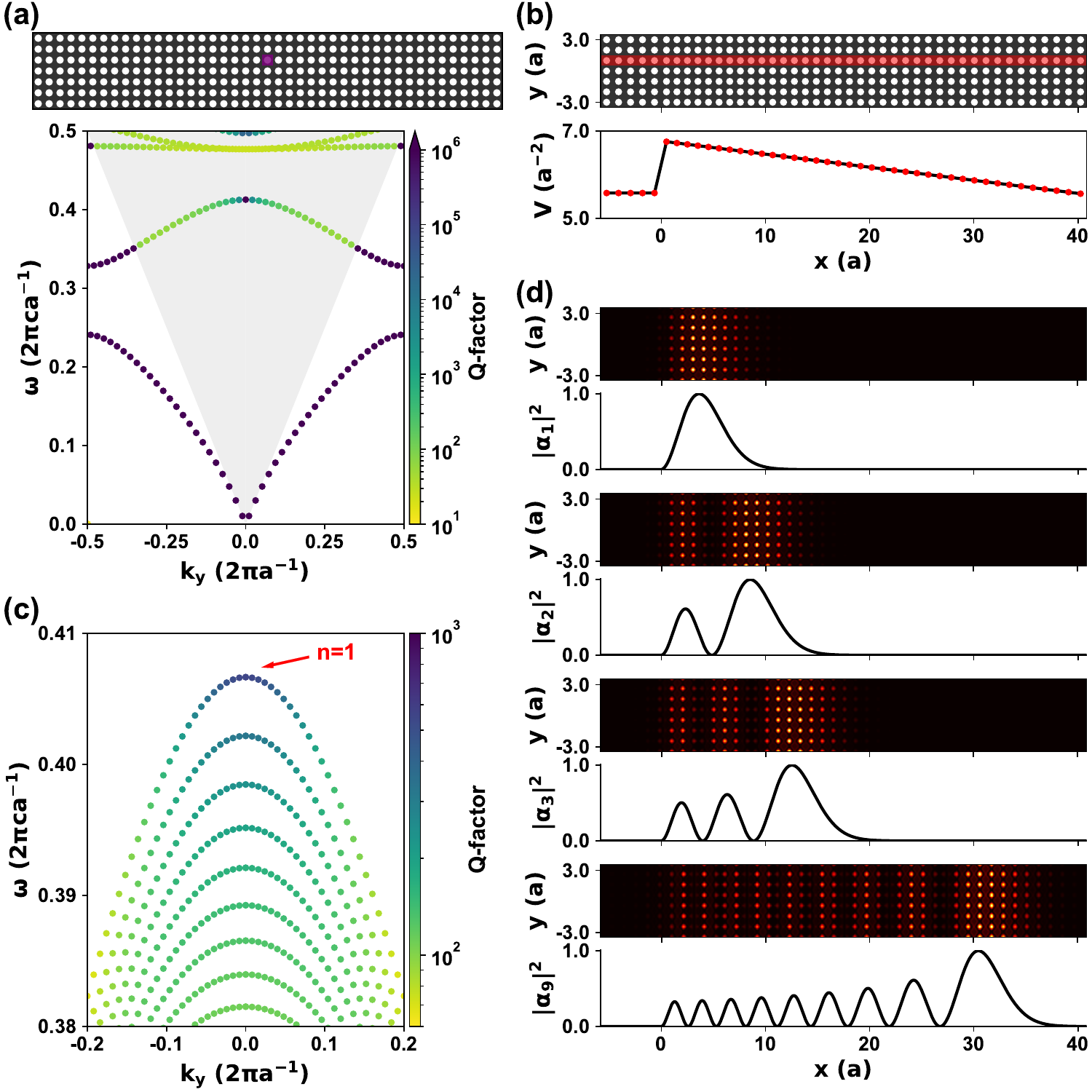}
    \caption{Photonic crystal slab superpotential. (a) The in-plane geometry of a periodic photonic crystal slab (upper panel) and its simulated band structure (lower panel). (b) The photonic crystal structure (upper panel) after a linear superpotential ($\kappa=0.311a^{-1}$) has been introduced. The lattice constant in the $x$ direction $a_x$ is slowly varied as a function of $x$. The corresponding superpotential is shown in the lower panel. (c) The simulated band structure of the structure in (b). The first $n=1$ Airy state is marked with a red arrow. (d) The eigenstate profiles ($\left|H_z\right|^2$ at the center of the slab) of the structure in (b) at $k_y=0$. The $n=1$, $n=2$, $n=3$ and $n=9$ Airy states are shown from top to bottom. The corresponding theoretical Airy function solutions are also shown in the lower panels.
}
    \label{figure1}
  \end{center}
\end{figure*}

Our starting point is a periodic photonic crystal structure in a silicon slab ($\varepsilon=12.11$) on a silica substrate ($\varepsilon=2.25$) with a quadratic band. The in-plane geometry is composed of circular air holes in a square lattice. The radii of the holes are $r_0=0.34a$ and the thickness of the slab is $h=0.35a$ with the lattice constant $a=629nm$. The upper panel of Fig. \ref{figure1}a shows the two-dimensional photonic crystal slab structure used in the experiment, with the unit cell highlighted in purple. 
The eigenmodes in a photonic crystal can be described by the Maxwell’s equation \cite{Joannopoulos:08:Book}:
\begin{equation}
\mathbf{\nabla}\times \left[\varepsilon(\mathbf{r})^{-1}\mathbf{\nabla}\times\mathbf{H}\right] =E\mathbf{H},
\label{equation1}
\end{equation}
where $\varepsilon(\mathbf{r})$ is the dielectric function with $\mathbf{r}=(x,y,z)$ the three spatial coordinates, $\mathbf{H}$ is the magnetic field profile of the eigenmode, and $E=\left(\omega/c\right)^2$ with $\omega$ the frequency of the eigenmode and $c$ the speed of light. 

We numerically compute the photonic band structure by solving Eq. \eqref{equation1} in the transverse electric (TE)-like polarization using the guided mode expansion method, as implemented in the open-source software package \textsc{Legume} \cite{Legume}. An isolated quadratically dispersing band is centered at $\Gamma$, as shown in the lower panel of Fig. \ref{figure1}a.  For mathematical convention, below we characterize the band dispersion and the frequency of the band by the eigenvalue of the Maxwell’s equation in Eq. \eqref{equation1}, namely $E=(\omega/c)^2$, instead of the frequency $\omega$ itself. In analogy with how massive particles behave in quantum mechanics, we refer to $E$ as `energy' (although it has the unit of $(a^{-2})$). The band in the lower panel of Fig. \ref{figure1}a is isotropic in the vicinity of $\Gamma$, such that:
\begin{equation}
E_{\mathbf{k}}=E_0-\frac{1}{2m}\left(k_x^2+k_y^2\right),
\label{equation2}
\end{equation}
where $E_\mathbf{k}$ is the energy of the band at wavevector $\mathbf{k}=\left(k_x,k_y\right)$ and $m$ is the (complex) effective mass. In simulations, the band tip energy is $E_0=6.76a^{-2}$, corresponding to frequency $\omega_0=0.414\ (2\pi c a^{-1})$, and the effective mass is $\frac{1}{2m}=0.651+0.109i$. The effective mass is a complex number, because the band is quadratically dispersive in both the real and imaginary parts of the energy, so that the loss scales as $O(k^2)$ due to the out-of-plane radiation. This effect can be directly observed in Fig. \ref{figure1}a: the Q-factor of the band decreases as $k$ increases near the band tip at $\Gamma$. The $O(k^2)$-dependent loss arises due to the symmetry-protected BIC at $\Gamma$-point \cite{hsu2016bound, koshelev2018asymmetric}. The non-Hermiticity that arises from the imaginary part of the effective mass governs the linewidth of the Airy resonances, as we describe later.

Next, we add a superpotential $V(\mathbf{r})$ to the system. This is realized by slowly varying the lattice constant as a function of position. By computing the shift of the band tip energy $E_0$ as a function of the lattice constant in the periodic structure, we establish a quantitative measure of the local band tip energy. Effectively, the local tip energy $E_0$ is no longer a constant - which is the case in periodic structure - but made to depend on the position such that $E_0=E_0(\mathbf{r})$. The superpotential is then realized and defined as the local band tip energy $V(\mathbf{r})=E_0(\mathbf{r})$. We require the superpotential to only vary on a length scale which is large compared with the lattice constant, $a$, namely $V=V(\kappa \mathbf{r})$, where $\kappa$ determines the variation scale of the potential with $\kappa a\ll 1$. To theoretically determine the resonances of photonic crystals with superpotentials, we employ a two-scale expansion of Eq. \eqref{equation1} using $\kappa$ as the expansion parameter. The solution to Eq. \eqref{equation1} can then be written to the leading order in $\kappa$ as
\begin{equation}
\mathbf{H}=\alpha(\kappa\mathbf{r})\mathbf{H_\Gamma(r)}+O(\kappa^1)
,
\label{envelope}
\end{equation}
where $\mathbf{H_\Gamma(r)}$ is the magnetic field profile in the periodic structure at the tip of the quadratic band $\mathbf{k}=\Gamma$, which varies on the scale of the lattice constant $a$, and $\alpha(\kappa\mathbf{r})$ is a slowly-varying envelope function. We theoretically prove (more details can be found in Supplemental Information Section 1 \cite{SuppInfo}) that under the slowly varying superpotential $V(\kappa\mathbf{r})$, Eq. \eqref{equation1} can be effectively simplified as a Schr\"odinger equation to leading order:
\begin{equation}
\begin{aligned}
&\left[\frac{\kappa^2}{2m}\frac{\partial^2}{\partial (\kappa x)^2}
+\frac{\kappa^2}{2m}\frac{\partial^2}{\partial (\kappa y)^2}
+V(\kappa\mathbf{r})\right]
\alpha(\kappa\mathbf{r})\\
&=\left(E+ O(\kappa^3)\right)\alpha(\kappa\mathbf{r}),
\label{equation3-0}
\end{aligned}
\end{equation}

Then, this superpotential formalism is employed to generate a linear potential, for which the eigenstate is theoretically predicted to be an Airy function. Fig. \ref{figure1}b shows how this potential is realized: the lattice constant in the $x$ direction $a_x$ is varied slowly at different positions along $x$. The upper panel shows the dielectric function after the superpotential is introduced in the structure. In the $x>0$ region, the lattice constant $a_x$ gradually increases from $a_x=a$ at $x=0$ to $a_x=1.15a$ at the right edge of the sample. In the left region where $x<0$, the lattice constant is fixed at $a_x=1.15a$, which serves as a hard-wall barrier. The corresponding superpotential $V(x)$ is shown in the lower panel. In the region $x>0$, the superpotential has the form of $V(\kappa x)=E_0-\kappa^2 (\kappa x)$, with $\kappa=0.311a^{-1}$. The superpotential breaks periodicity in the $x$ direction, such that the unit cell in Fig. \ref{figure1}a becomes a large supercell in Fig. \ref{figure1}b as highlighted in red. Since the superpotential preserves the periodicity in the $y$ direction, the lattice momentum coordinate $k_y$ remains well-defined.
The solution to Eq. \eqref{equation3-0} can be analytically calculated:
\begin{equation}
\scalebox{0.88}{$\displaystyle
\begin{aligned}
&E_{n,k_y}
=E_0
-\left(\frac{1}{2m}\right)^{\frac{1}{3}}\left[\frac{3}{2}\pi\left(n-\frac{1}{4}\right)\right]^{\frac{2}{3}}\kappa^2
-\frac{k_y^2}{2m},\\
&\alpha_{n,k_y}
=\left\{
\begin{aligned}
e^{i k_y y}\cdot 
\mathrm{Ai}\left(\left(\frac{1}{2m}\right)^{-\frac{1}{3}}\kappa \left(x-x_n\right)\right) & , & x>0\\
0 & , & x<0
\end{aligned}
\right.
\label{equation4}
\end{aligned}
$}
\end{equation}
where $x_n=\left(\frac{1}{2m}\right)^{\frac{1}{3}}\left[\frac{3}{2}\pi\left(n-\frac{1}{4}\right)\right]^{\frac{2}{3}}\kappa^{-1}$ and $\mathrm{Ai}(\cdot)$ is the Airy function. A more detailed derivation can be found in the Supplemental Information Section 2 \cite{SuppInfo}. Figure \ref{figure1}c shows the simulated band structure of the sample in Fig. \ref{figure1}b. The quadratic band in Fig. \ref{figure1}a splits into discrete Airy levels as predicted in Eq. \eqref{equation4}. At fixed $k_y$, the Airy states can be labeled by their mode index $n=1,2,3,\cdots$, and the $n=1$ state is marked in the Fig. \ref{figure1}c by an arrow. In order to further confirm that the envelope of the eigenmode profile is indeed an Airy function, we compare the wavefunctions themselves (upper panels) with the Airy function solutions in Eq. \eqref{equation4} (lower panels) in Fig. \ref{figure1}d and find good quantitative agreement. Due to the cutoff at $x=0$ in Eq. \eqref{equation4}, the wavefunction $\alpha_n$ is a part of the Airy function that contains the first $n$ lobes, as shown in Fig. \ref{figure1}d. When $n\to\infty$, we approach the continuous limit in which the mode separation $E_{n+1}-E_n$ tends to zero and the eigenfunction $\alpha_n$ converges to a complete Airy function, as can be seen from the bottom two panels of Fig. \ref{figure1}d.

% End figure 1 discussion

% Measurements discription

\begin{figure*}[ht]
  \begin{center}
    \includegraphics[width=16 cm]{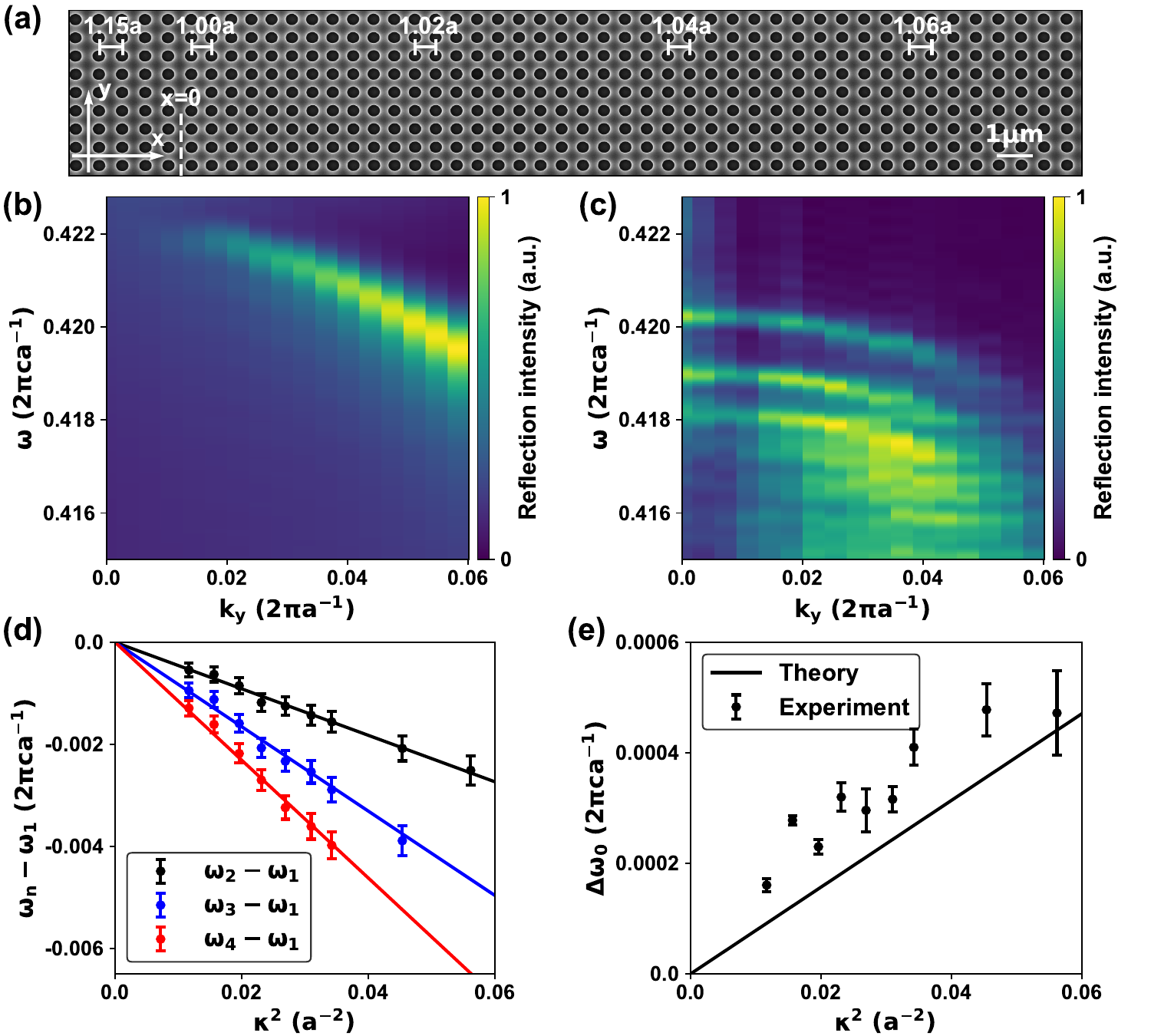}
    \caption{Observation of Airy resonances. (a) An SEM image of the sample with $\kappa=0.237a^{-1}$. (b) The isolated quadratic band in the control sample with no superpotential. (c) Several Airy bands observed in an Airy sample with $\kappa = 0.152a^{-1}$ (d) The frequency separation of Airy states vs. superpotential strength. The solid lines show the theoretical frequency separation between Airy states, and the dots with error bars show the measured values. (e) The linewidth of the first $n=1$ Airy state vs. superpotential strength. The solid line shows the linewidth predicted from superpotential formalism, and the dots with error bars show the measured values. The experimental linewidths are consistently $\sim 10^{-4}\ (2\pi ca^{-1})$ larger than the theoretical linewidths. This discrepancy is likely caused by some additional loss in our system not associated with the superpotential such as scattering from fabrication-induced roughness or absorption.}
    \label{figure2}
  \end{center}
\end{figure*}

For our experiments, we fabricate nine silicon-on-insulator photonic crystals with Airy superpotentials imposed by varying the lattice constant between unit cells (as shown in Fig. \ref{figure1}b), each with a different potential strength controlled by varying $\kappa$. The lattice constant in the $y$ direction is fixed at $a_y=a=629nm$. Figure \ref{figure2}a shows a part of the SEM image of the $\kappa=0.237a^{-1}$ sample. In the $x<0$ region, the lattice constant is $a_x=1.15a$. While in the $x>0$ region, the lattice constant $a_x$ gradually increases from $a_x=a$ at $x=0$ to $a_x=1.15a$ at the right edge of the sample (which is beyond the right edge of Fig. \ref{figure2}a). We also fabricate one control sample with no superpotential (a periodic square lattice with circular holes). More details on the fabrication methods can be found in Supplemental Information Section 3 \cite{SuppInfo}. In order to observe Airy resonances, we probe our samples with collimated laser light. We take real-space images of our sample as we vary both the frequency and angle (which determines the in-plane wavevector) of the probe beam. The images do not measure the electromagnetic near-field at the sample's surface, but only measure the light distribution that radiates away from the sample. Using this method, we can observe the spatially-varying photonic crystal resonances in our sample. More details of the experiential methods can be found in Supplemental Information Sections 4-7 \cite{SuppInfo}. %done - Brian 

% End measurements discription 

% Figure 2 discussion

Figure \ref{figure2}b shows the measured band structure of the periodic sample without superpotential. The total reflection intensity is obtained by averaging the intensity of all pixels in the real-space images. We can see that both the frequency and the linewidth of the band in the control sample scale with $k^2$, so the band energy is analogous to the energy of a free particle with a complex effective mass, as described by Eq. \eqref{equation2}. By fitting the measured band frequency and linewidth with a quadratic function of $k$, we experimentally measure the tip frequency and the effective mass of this band to be $\omega_0=0.422\ (2\pi c a^{-1})$ and $\frac{1}{2m} = 0.628 + 0.124i$, respectively. 

Figure \ref{figure2}c shows the measured band structure of the Airy sample with superpotential strength $\kappa=0.152a^{-1}$. The Airy sample exhibits several quadratically dispersive bands, which match predictions from theory (Eq. \eqref{equation4}) and simulations (Fig. \ref{figure1}c).   From the definition $E=\omega^2/c^2$, the Airy resonances' frequency and linewidth can be determined from the mode energies as calculated in the superpotential formalism in Eq. \eqref{equation4} to be $\omega_{n, k_y} = c\sqrt{\mathrm{Re}(E_{n,k_y})} $ and $\Delta \omega_{n, k_y} = - c \frac{\mathrm{Im}(E_{n,ky})}{\sqrt{\mathrm{Re}(E_{n,ky})}}$. Here, we have assumed $\mathrm{Im}(E_{n,ky})\ll\mathrm{Re}(E_{n,ky})$, which is true in our experiments. In Fig. \ref{figure2}d we directly compare the experimentally-measured separation in frequency of Airy states with the theoretical values predicted by the superpotential formalism in nine samples with different potential strengths ($\kappa$). We see excellent agreement between the theory -- which contains no free parameters -- and our measurements, confirming that the bands are indeed Airy resonances.

The Airy sample in Fig. \ref{figure2}c has three distinct bands present. However, the theory and simulations predict there should be many more bands below the third band. We attribute this discrepancy to the intrinsic non-Hermiticity of the photonic crystal, which causes the resonances to have finite linewidths; specifically, the superpotential formalism in Eq. \eqref{equation4} predicts that the linewidths of the Airy states scale with $\left(n-\frac{1}{4}\right)^{\frac{2}{3}}$. Below the frequency of the third band, the linewidths of the Airy states become sufficiently large that it is not possible to distinguish individual states by eye. However, the reflected intensity below the frequency of the third band is significantly larger than the reflected intensity above the first band, indicating the presence of additional resonances with large linewidths below the third band. 

To further quantify the non-Hermitian effects of the complex effective mass on the Airy resonances, we compare the linewidths of the first $n=1$ Airy state obtained from measurements and theory in Fig. \ref{figure2}e. The experimentally measured linewidths of the first Airy state are consistently slightly larger than the predicted values. However, the linewidth still scales with $\kappa^2$, so the excess linewidth is likely caused by some additional constant loss (for instance, scattering from fabrication-induced roughness or absorption) in our sample, which is not accounted for in the theory. We do not extract the linewidths of Airy states with $n>1$; these states have sufficiently large linewidths that multiple Airy resonances interfere with each other, making it difficult to determine the linewidths of individual states. 

% End figure 2 discussion

% Figure 3 discussion

\begin{figure*}[ht]
  \begin{center}
    \includegraphics[width=16 cm]{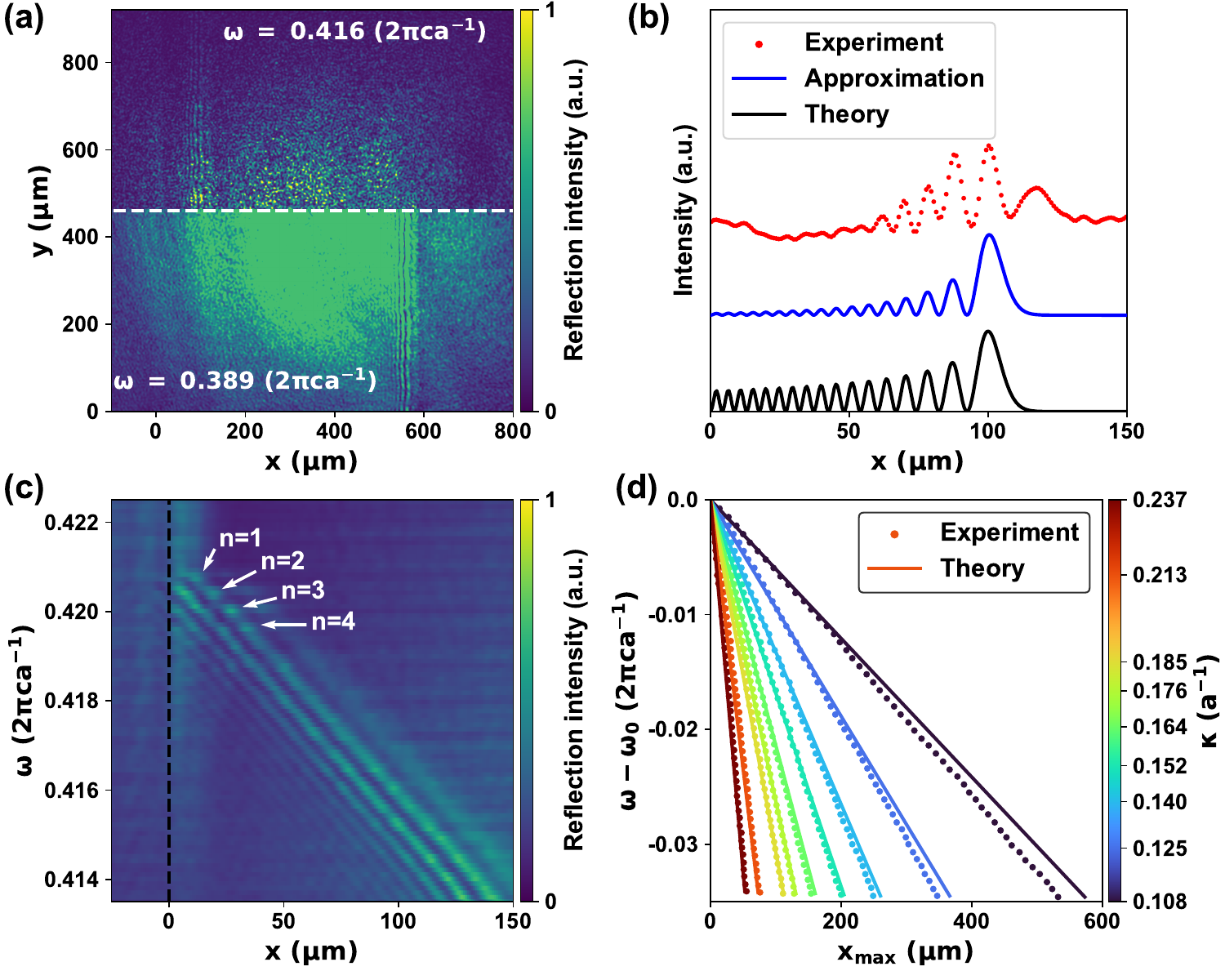}
    \caption{Spatial profiles of Airy resonances. (a) The intensity distribution of light at the surface of an Airy sample with $\kappa = 0.108a^{-1}$ at two different frequencies. The main feature of the resonance - several parallel ``lobes" - moves smoothly across the sample as the frequency of the probe beam is swept. (b) The spatial profile of an Airy resonance at $\omega = 0.416\ (2\pi ca^{-1})$. The red dots show the experimentally measured spatial profile. The black line shows $\left|\alpha_{17}(x)\right|^2$, which is predicted in the theory. The blue line shows the modified Airy envelope which is approximated by adding an exponential decay factor on top of the Airy function. (c) The intensity distribution of light at the surface of the Airy sample with $\kappa = 0.108a^{-1}$ averaged in the $y$ direction at different probe beam frequencies. (d) The position of the brightest lobe in the Airy resonance as a function of probe beam frequency for nine samples with different potential strengths. The solid lines and dots show the theoretical predictions and experimental measurements, respectively.} 
    \label{figure3}
  \end{center}
\end{figure*}

To further characterize the Airy resonances, we analyze the real-space images of light at the surface of the samples. Figure \ref{figure3}a shows the intensity of light at two different frequencies. The measured images capture light that was resonantly absorbed and leaked from the resonance as well as light that is specularly reflected off the sample, and the latter process is responsible for the slowly varying background intensity of light over the entire sample. A few wavelengths above the sample, light that leaks out of the Airy resonances loses the short-range spatial variation from $\mathbf{H_{\Gamma}}(\mathbf{r})$ in Eq. \eqref{envelope} (See Supplemental Information Section 6 for more details), so we can only observe the Airy resonance envelope function $\alpha_{n, k_y}(\kappa\mathbf{r})$. Over a wide range of frequencies, a resonant mode with a main bright lobe followed by several lobes of decaying intensity to the left of the main lobe was observed. Figure \ref{figure3}b shows that the experimentally-measured distance between the lobes qualitatively matches theoretical predictions  (red and black curves, respectively). However, for Airy states with $n>4$, the intensity of the lobes decays faster to the left of the main lobe than the theoretically-predicted Airy resonances, as shown in Fig. \ref{figure3}b and c. Again, we attribute this discrepancy to the non-Hermiticity in our samples. For $n>4$, the linewidths of the Airy states are sufficiently large that a probe beam with a single frequency can excite multiple Airy resonances that interfere with each other. As a result, the envelope function obtained in the experimental real-space image is no longer a single Airy function, but a superposition of multiple Airy functions. Moreover, we find that this superposition effect can be effectively approximated by adding an extra exponential decay on top of the Airy wavefunction:
\begin{equation}
    I\left(\omega,x\right)\propto
    \left|e^{\gamma \left(x-x_\omega\right)}\cdot
    \mathrm{Ai}\left(\frac{1}{x_l}(x-x_\omega )\right)\right|^2,
    \label{gamma_eq}
\end{equation}
where $I\left(\omega,x\right)$ is the intensity distribution of the envelope function, $\gamma$ is a parameter which describes how quickly the Airy function decays to the left of the main lobe, $x_l = (2m)^{-1/3}\kappa^{-1}$ is the length scale of the Airy resonance, and $x_\omega = 2c^{-2}\kappa^{-3}\omega_0(\omega_0-\omega)$ is the offset of the Airy function from the potential barrier at $k_y=0$. Supplemental Information Section 6 \cite{SuppInfo} shows how we obtain this expression from the superposition of Airy resonances. Then, the decaying parameter $\gamma$ is fitted by the experimental data and the approximated intensity profile is shown in the blue curve in Fig. \ref{figure3}b and matches the experimental curve better than the corresponding black curve from the Hermitian theory. 

The non-Hermitian effects from finite linewidths of Airy resonances can also be seen in Fig. \ref{figure3}c, which shows the measured spatial profiles of the envelope function $\alpha(\kappa \mathbf{x})$ at different frequencies. The four bright spots in the brightest lobe (marked by white arrows) indicate the first four Airy resonances. For resonances with $n\leq 4$, the spatial profile does not decay quickly to the left of the bright lobe. States with $n>4$ cannot be distinguished from each other due to their large linewidths, and the intensity of these states decay faster to the left of the bright lobe due to the interference between multiple Airy states as explained earlier. As the frequency of the probe beam decreases, the lobes move to the right of the sample away from the potential barrier at $x=0$. This motion corresponds to lower frequency light exciting Airy resonances with higher $n$, which extend further away from the potential barrier. The position of the main lobe as a function of frequency $\omega$ can be calculated from Eq. \eqref{equation4} as:
\begin{equation}
x_{\mathrm{max}}\left(\omega\right)
=-\frac{2\omega_0}{c^2\kappa^3}\omega+x_\mathrm{offset}
,
\end{equation}
where $\omega_0 = c\sqrt{\mathrm{Re}(E_0)}$ is the band tip frequency, $x_{\mathrm{max}}$ is the position of the main lobe, and $x_\mathrm{offset}$ is an offset in $x$ (more details justifying $x_\mathrm{offset}$ can be found in Supplemental Information Section 2 \cite{SuppInfo}). Figure \ref{figure3}d compares the position of the main lobe in experiment (dots) and theory (lines) for samples with nine different potential strengths ($\kappa$) indicated by the color. Figure \ref{figure3}d shows excellent agreement between experiment and theory for all samples.

% End figure 3 discussion

% Figure 4 discussion

Finally, we turn our attention to light propagation in free space as it leaks out-of-plane from Airy resonances. After light leaks out of the Airy resonances, it evolves in free space according to the paraxial approximation of Maxwell's equations \cite{PhysRevA.11.1365},

\begin{equation}
\frac{\partial^2 \mathbf{H_0}}{\partial x^2}
+\frac{\partial^2 \mathbf{H_0}}{\partial y^2}
+2i\frac{\omega}{c}\frac{\partial \mathbf{H_0}}{\partial z}=0
,
\label{c31-0}
\end{equation}
where $z$ is the out-of-plane direction, $\mathbf{H_0}(x, y, z) = \mathbf{H}(x, y, z)e^{-i\frac{\omega}{c}z}$ is the magnetic field (in $z$) which is assumed to vary slowly in the $z$ direction, with $\mathbf{H}(x, y, z)$ the magnetic profile in Eq. \eqref{equation1}. By solving the paraxial equation in Eq. \eqref{c31-0} with the initial condition at $z=0$ described in Eq. \eqref{gamma_eq}, we obtain an analytical solution to the light intensity distribution at each propagation distance $z$:
\begin{equation}
\scalebox{0.95}{$\displaystyle
\begin{aligned}
&\left|\mathbf{H_0}(x,y,z)\right|^2 \propto \\
&\left|e^{\gamma \left(x-x_\omega\right)}
\cdot
\mathrm{Ai}\left(\frac{1}{x_l}\left(x-x_\omega  + i\frac{c\gamma}{\omega}z - \frac{mc^2\kappa^3}{2\omega^2}z^2\right)\right)\right|^2
.
\label{paraxial_sol-2}
\end{aligned}
$}
\end{equation}

Note that the intensity distribution at each $z$ is still a decaying Airy function, but the argument of the Airy function has an additional offset whose real part depends on $z^2$. Therefore, the lobes of the light that leak out of the Airy resonances will propagate in free space along a parabolic trajectory given by:
%Note that at $z=0$, this equation matches Eq. \ref{gamma_eq}. Therefore light that leaks out of the Airy resonances will propagate in free space as an Airy beam. In this solution, each Airy lobe bends in free space as it travels along the parabolic trajectory given by 
\begin{equation}
    x(z) = \left.x(z)\right|_{z=0} + \frac 12 \left(\frac{mc^2}{\omega^2}\kappa^3\right)z^2
    ,
    \label{quadratic}
\end{equation}
where $x(z)$ is the lobe position at $z$, and $\left.x(z)\right|_{z=0}$ is the corresponding lobe position at $z=0$. The rate of beam bending, $\beta$, is defined as $\beta=\frac{\partial^2 x(z)}{\partial z^2}$, such that it can be determined via theory as $\beta = \frac{mc^2}{\omega^2}\kappa^3$. 

Recall from Eq. \eqref{envelope} that the spatial profiles  of Airy resonances are a long-length-scale envelope function determined by the superpotential ($\alpha_n(\kappa \mathbf{r})$) multiplied by a short-length-scale Bloch mode determined by the slab resonances of a periodic photonic crystal ($\mathbf{H_\Gamma}(\mathbf{r})$). According to the Bloch theorem, the short-length-scale component of the resonance $\mathbf{H_\Gamma}(\mathbf{r})$ has both a spatially uniform component that can radiate to free space (Fourier component corresponding to reciprocal lattice vector $\mathbf{G}=(0,0)$), and spatially varying components that decay exponentially outside the slab (Fourier components corresponding to  $\mathbf{G}\neq (0,0)$, which cannot radiate to free space because they are under the light line). Supplemental Information Section 2 \cite{SuppInfo} outlines why this occurs. Hence, despite the fact that the wavefunction of the Airy resonances has a fast-oscillating Bloch term, only the long-length-scale Airy envelope function (Fig. \ref{figure3}b and Eq. \eqref{paraxial_sol-2}) dictates the behavior in the radiation. Thus, the Airy resonances can be effectively treated as a light source for (finite-energy) Airy beam \cite{siviloglou2007observation}, and experience the same acceleration as when propagating in free space.

\begin{figure*}[ht]
  \begin{center}
    \includegraphics[width=16 cm]{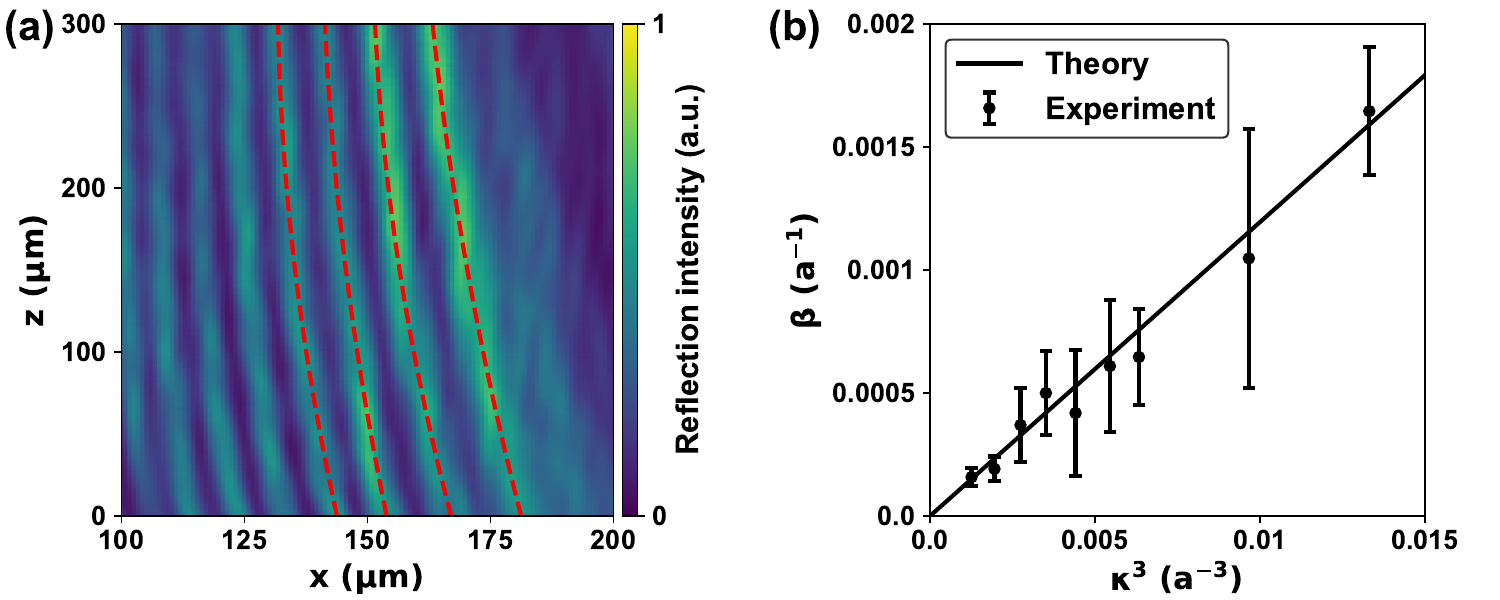}
    \caption{Accelerating Airy beam generated from Airy resonances. (a) The intensity of light averaged in the $y$-direction as the distance above the sample ($z$) is swept. The dashed red curves are the quadratic fits for the bending of the lobes. For this measurement, the Airy sample with $\kappa = 0.108a^{-1}$ was used and the frequency of the probe beam was $\omega = 0.411\ (2\pi c a^{-1})$. (b) The beam acceleration as a function of superpotential strength. The solid line and dots with error bars represent the theory and experimental values of the beam acceleration respectively.}
    \label{figure4}
  \end{center}
\end{figure*}

Figure \ref{figure4}a shows the bending of an Airy beam radiated from the Airy photonic crystal with $\kappa = 0.108a^{-1}$. The lobes follow parabolic trajectories (which are marked with red dashed lines) as expected in Eq. \eqref{quadratic}. The experimentally measured values for the bending of the Airy beams were determined by fitting the paths of the first several Airy lobes with a quadratic function for each of the nine Airy samples. Figure \ref{figure4}b shows that the experimentally observed Airy beam bending in the nine Airy samples matches the theoretical predicted accelerations, confirming that the light leaking out of the Airy resonances is indeed Airy beam.  %done - Brian 

% End figure 4 discussion

% Final summary/conclusion

In conclusion, we directly observe Airy resonances in two-dimensional silicon photonic crystal slabs using a linear superpotential. We experimentally demonstrate three signatures of Airy resonances: Airy mode frequency separation (and linewidth), Airy eigenstate spatial profile, and the bending of Airy mode radiation in free space. Due to the non-Hermiticity of photonic crystal slabs arising from the complex effective mass, the Airy resonances we observed had finite linewidths which were predicted using the superpotential formalism. The non-Hermiticity significantly altered the response of the photonic crystal due to interference of multiple Airy resonances; specifically, the light leaking out of the resonances becomes more localized around the brightest Airy lobe. The non-Hermiticity arising from complex effective mass provides a new platform to investigate the interplay between non-Hermiticity and external potentials in aperiodic photonic systems. Furthermore, the realization of Airy resonances in a photonic crystal slab could be used for the compact on-chip Airy beam generation.  Moreover, we theoretically developed the general formalism of superpotentials in photonic crystals using a two-scale expansion, which simplifies Maxwell’s equations in an isolated band to a non-Hermitian Schr\"odinger equation with a complex effective mass. The theoretical framework developed here may be widely applied to any slowly-varying aperiodic photonic structure; for example, it may be useful in photonic crystal surface emitting laser design and for resonant nonlinear optics of photonic crystal slabs.

% End conclusion

\noindent Acknowledgements. We thank K. Makris for input on the manuscript. The authors acknowledge the Nanofabrication Lab within the Materials Research Institute at Penn State. We thank Sachin Vaidya, Koorosh Sadri and Abhijit Chaudhari for fruitful discussions. We gratefully acknowledge funding support from the Air Force Office of Scientific Research under grant number FA9550-22-1-0339 and the Army Research Office under grant number W911NF-22-2-0103.  

\onecolumngrid
\clearpage

\clearpage

\section*{Supplementary Information for: \texorpdfstring{\\ Airy Resonances in Photonic Crystal Superpotentials}{Lg}}

\label{supplementary}

\renewcommand{\theequation}{S\arabic{equation}}
\renewcommand{\thefigure}{S\arabic{figure}}
\setcounter{equation}{0}
\setcounter{figure}{0}

\tableofcontents

\clearpage

\section*{Overview} %done
\addcontentsline{toc}{section}{Overview}

This Supplementary Information section acts as a companion to the main text, where we provide additional details on theoretical background, simulation results, and experimental methods. \\

In Section 1, we present the general theoretical framework for superpotentials in detail. We find that if a slow variation is imposed on a periodic photonic crystal, the states near a quadratic band edge are effectively governed by a Schr\"odinger equation with an additional spatially varying potential term, which is referred as ``superpotential”. This superpotential formalism can be applied to any aperiodic photonic system with slow spatial variations compared to the lattice constant.\\

In Section 2, we show how to realize Airy resonances through the superpotential framework. In particular, by slowly varying the lattice constant as a function of position, we introduce a linear superpotential and realize Airy resonances in a photonic crystal slab. We also theoretically introduce the three signatures of Airy resonances: the Airy mode frequency separation (and linewidth); the Airy spatial mode profile; and the bending of the Airy mode radiation in free space. The experimental data related to these three signatures of Airy resonances are explained in more detail in Section 5, Section 6 and Section 7.\\

In Section 3, we introduce the fabrication method that we used to create silicon-on-insulator photonic crystal slabs. A standard electron beam lithography process followed by an inductively coupled plasma etch is used to etch holes into a $220nm$ thick silicon layer with sub-$10nm$ precision.\\

In Section 4, we describe the experimental setup that characterizes the photonic modes of the sample by measuring its real-space image at various angles and frequencies. Then, we present the method for extracting the resonance frequency and linewidth of an isolated single band from the reflection intensity spectrum obtained in the experiment.\\

In Section 5, we explain the experimental data related to the first signature of the Airy resonances: the Airy mode frequency separation and linewidth. As predicted by the superpotential framework, an isolated quadratic band in a periodic photonic crystal will split into several discrete Airy modes after a linear superpotential is imposed. Moreover, each Airy mode has a finite linewidth due to the imaginary part of the effective mass of the band. We present a method to extract the frequency separation and the linewidth of Airy modes from the experimental data. The results show a good quantitative consistency with the theoretical and simulation results. \\

In Section 6, we explain the experimental data related to the second signature of the Airy resonances: the Airy spatial mode profile. The envelopes of Airy resonance eigenstates are predicted to be truncated Airy functions. However, we find in the experiment that the intensity of the lobes in the real-space image decays faster than the actual Airy function. We attribute this to the finite linewidths of Airy modes. Due to the finite linewidths, a plane wave with a single frequency can excite multiple Airy modes at the same time, so the optical response of the photonic crystal is a superposition of several Airy eigenfunctions with different intensity (and phase). We estimate this effect by summing the Airy eigenfunctions with a Gaussian weight function, and again observe good agreement between the experiment and the theory.\\

In Section 7, we explain the experimental data related to the third signature of the Airy resonances: the bending of the Airy mode radiation in free space. Due to its finite linewidth, the light in an Airy resonance will eventually leak out of the slab and propagate in free space with an Airy function as its initial spatial profile. According to the paraxial equation, the position of the Airy function lobes will follow a quadratic path as it propagates away from the slab. We present the experimental method to measure the bending. The bending is quantitatively consistent with the theoretical and simulation results.

\clearpage

\section*{Section 1: General theoretical framework for superpotentials}
\addcontentsline{toc}{section}{Section 1: General theoretical framework for superpotentials}

In this section, we use multi-scale analysis to derive a general formula describing photonic crystal resonances when we add a slow-varying potential to a two-dimensional photonic crystal slab. We start from the general dispersion relation of a band in a periodic photonic crystal, and we use the information from the periodic structure to evaluate the effective equation when a slow-varying aperiodic superpotential is imposed on the system.\\

In a 2D photonic crystal slab, the eigenmodes are described by the Maxwell's equations:
\begin{equation}
\mathbf{\nabla}\times \left[\varepsilon(\mathbf{r})^{-1}\mathbf{\nabla}\times\mathbf{H}(\mathbf{r})\right]=E\mathbf{H}(\mathbf{r}),
\label{a1}
\end{equation}
where $\varepsilon(\mathbf{r})$ is the dielectric function (with $\mathbf{r}=(x,y,z)$ the three spatial coordinates), $\mathbf{H(r)}$ is the magnetic field profile of the eigenmode, and $E=\left(\omega/c\right)^2$ is the energy of the eigenmode (with $\omega$ the eigen-frequency of the mode and $c$ the speed of light). Two-dimensional photonic crystal slabs have both guided (i.e., bound) and radiative (i.e., resonant) modes that are below and above the light line, respectively \cite{photonicbook}. In our work, we focus on radiative modes that are above the light line. The radiative loss associated with these modes may be treated as a negative imaginary part of the energy $\mathrm{Im}(E)$ in the 2D description. 

For simplicity, from this point onward the energy $E=\left(\omega/c\right)^2$ is used, which is the eigenvalue of the Maxwell's equation in Eq. \eqref{a1}, instead of the frequency $\omega$, to describe the photonic band. Assuming that the real part of the energy is much larger than the imaginary part of the energy ($\mathrm{Re}(E)\gg\mathrm{Im}(E)$), the relation between the frequency and the energy is:

\begin{equation}
\mathrm{Re}(\omega)=c\sqrt{\mathrm{Re}(E)},
\label{a1-1}
\end{equation}
\begin{equation}
\Delta\omega=2\left|\mathrm{Im}(\omega)\right|
=-c\frac{\mathrm{Im}(E)}{\sqrt{\mathrm{Re}(E)}},
\label{a1-2}
\end{equation}
where $\mathrm{Re}(\omega)$ is the (real) frequency and $\Delta\omega$ is the linewidth (in frequency) of a photonic mode. For simplicity, from this point onward, $\omega$ is used to represent $\mathrm{Re}(\omega)$ unless otherwise specified.

Without loss of generality, we assume that the photonic crystal slab has a periodicity in both the $x$ and the $y$ directions and has a finite thickness in the $z$ direction. We assume one solution to Eq. \eqref{a1} at in-plane wavevector $\mathbf{k}=(k_x,k_y)$ is
\begin{equation}
\mathbf{H}=\mathbf{H_k},\quad E=E_{\mathbf{k}}.
\label{a2}
\end{equation}

First, the energy ($E=\left(\omega/c\right)^2$) is estimated in the vicinity of $\mathbf{k}$ to evaluate the band energy dispersion. Assume $\mathbf{\Delta k}=\left(\Delta k_x, \Delta k_y\right)$ is a small deviation of the wavevector. According to the Bloch's theorem, the solution to Eq. \eqref{a1} at wavevector $\mathbf{k}+\mathbf{\Delta k}$ is:
\begin{equation}
\mathbf{H}=e^{i(\mathbf{k}+\mathbf{\Delta k})\cdot\mathbf{r}}\mathbf{\Psi(r)}=e^{i\mathbf{\Delta k\cdot r}}\left[e^{i\mathbf{k}\cdot\mathbf{r}}\mathbf{\Psi(r)}\right],
\label{a3}
\end{equation}
where $\mathbf{\Psi(r)}$ is a periodic function. Notice that the pre-factor $e^{i\mathbf{k}\cdot\mathbf{r}}$ in the last term of Eq. \eqref{a3} can be absorbed into $\mathbf{\Psi(r)}$. In the following text, $\mathbf{\Psi(r)}$ is used to represent $e^{i\mathbf{k}\cdot\mathbf{r}}\mathbf{\Psi(r)}$, and $\mathbf{\Psi(r)}$ is not required to be a periodic function. Then, Eq. \eqref{a3} becomes:
\begin{equation}
\mathbf{H}=e^{i\mathbf{\Delta k}\cdot\mathbf{r}}\mathbf{\Psi(r)}.
\label{a4}
\end{equation}

Using the form of solution in Eq. \eqref{a4}, Eq. \eqref{a1} becomes an eigenvalue problem of $\mathbf{\Psi(r)}$:
\begin{equation}
\mathbf{\nabla}\times \left[\varepsilon^{-1}\mathbf{\nabla}\times\mathbf{\Psi}\right]
+i\mathbf{\Delta k}\times\left[\varepsilon^{-1} (\mathbf{\nabla}\times\mathbf{\Psi})\right]
+\mathbf{\nabla}\times\left[\varepsilon^{-1}(i\mathbf{\Delta k}\times \mathbf{\Psi})\right]
+i\mathbf{\Delta k}\times\left[\varepsilon^{-1} (i\mathbf{\Delta k}\times\mathbf{\Psi})\right]
=E\mathbf{\Psi}.
\label{a5}
\end{equation}

Here, we use the facts that $\mathbf{\nabla}\times\left(e^{i\mathbf{\Delta k}\cdot\mathbf{r}}\mathbf{\Psi}\right)=e^{i\mathbf{\Delta k}\cdot\mathbf{r}}(\mathbf{\nabla}\times\mathbf{\Psi})+\mathbf{\nabla}\left(e^{i\mathbf{\Delta k}\cdot\mathbf{r}}\right)\times\mathbf{\Psi}$ and $\mathbf{\nabla}\left(e^{i\mathbf{\Delta k}\cdot\mathbf{r}}\right)=i\mathbf{\Delta k}e^{i\mathbf{\Delta k}\cdot\mathbf{r}}$. Notice that $\mathbf{\Delta k}=(\Delta k_x,\Delta k_y)$ is a two-dimensional vector, so $\mathbf{\Delta k}$ is always treated as $\mathbf{\Delta k}=(\Delta k_x, \Delta k_y,0)$ when performing cross-product or dot product in Eqs. \eqref{a3}, \eqref{a4}, and \eqref{a5}.

Since $\Delta k_x$ and $\Delta k_y$ are two independent small quantities, Eq. \eqref{a5} can be rewritten into different orders of $\Delta k_x$ and $\Delta k_y$ by expanding $\mathbf{\Delta k}=\Delta k_x\mathbf{\widehat{x}}+\Delta k_y\mathbf{\widehat{y}}$:
\begin{equation}
\left[L_0+\left(\Delta k_x\right) L_{1x}+\left(\Delta k_y\right) L_{1y}+\left(\Delta k_x\right)^2 L_{2xx}+\left(\Delta k_y\right)^2L_{2yy}+\left(\Delta k_x\right) \left(\Delta k_y\right) L_{2xy}\right]\mathbf{\Psi}=E\mathbf{\Psi},
\label{a6}
\end{equation}
where the operators $L_0$, $L_{1x}$, $L_{1y}$, $L_{2xx}$, $L_{2yy}$ and $L_{2xy}$ are defined as follows:
\begin{equation}
L_0\mathbf{G}=\mathbf{\nabla}\times \left[\varepsilon^{-1}\mathbf{\nabla}\times\mathbf{G}\right]=
\begingroup
\renewcommand*{\arraystretch}{1.5}
\left(
\begin{matrix}
    \frac{\partial}{\partial y}\left[\varepsilon^{-1}\left(\frac{\partial G_y}{\partial x}-\frac{\partial G_x}{\partial y}\right)\right]-\frac{\partial}{\partial z}\left[\varepsilon^{-1}\left(\frac{\partial G_x}{\partial z}-\frac{\partial G_z}{\partial x}\right)\right]\\
    \frac{\partial}{\partial z}\left[\varepsilon^{-1}\left(\frac{\partial G_z}{\partial y}-\frac{\partial G_y}{\partial z}\right)\right]-\frac{\partial}{\partial x}\left[\varepsilon^{-1}\left(\frac{\partial G_y}{\partial x}-\frac{\partial G_x}{\partial y}\right)\right]\\
    \frac{\partial}{\partial x}\left[\varepsilon^{-1}\left(\frac{\partial G_x}{\partial z}-\frac{\partial G_z}{\partial x}\right)\right]-\frac{\partial}{\partial y}\left[\varepsilon^{-1}\left(\frac{\partial G_z}{\partial y}-\frac{\partial G_y}{\partial z}\right)\right]
\end{matrix}
\right)\endgroup,
\label{a7}
\end{equation}

\begin{equation}
L_{1x}\mathbf{G}=i\mathbf{\widehat{x}}\times[\varepsilon^{-1}(\nabla\times \mathbf{G})]+i\nabla\times[\varepsilon^{-1}(\mathbf{\widehat{x}}\times\mathbf{G})]
=i
\begingroup
\renewcommand*{\arraystretch}{1.5}
\left(
\begin{matrix}
    \frac{\partial}{\partial y}\varepsilon^{-1} G_y
    +\frac{\partial}{\partial z}\varepsilon^{-1} G_z\\
    -\varepsilon^{-1} \frac{\partial}{\partial x} G_y
    +\varepsilon^{-1} \frac{\partial}{\partial y} G_x
    -\frac{\partial}{\partial x}\varepsilon^{-1} G_y\\
    \varepsilon^{-1}\frac{\partial}{\partial z} G_x
    -\varepsilon^{-1}\frac{\partial}{\partial x} G_z
    -\frac{\partial}{\partial x}\varepsilon^{-1} G_z
\end{matrix}
\right)\endgroup,
\label{a8}
\end{equation}

\begin{equation}
L_{1y}\mathbf{G}=i\mathbf{\widehat{y}}\times[\varepsilon^{-1}(\nabla\times \mathbf{G})]+i\nabla\times[\varepsilon^{-1}(\mathbf{\widehat{y}}\times\mathbf{G})]
=i
\begingroup
\renewcommand*{\arraystretch}{1.5}
\left(
\begin{matrix}
    \varepsilon^{-1}\frac{\partial}{\partial x} G_y
    -\varepsilon^{-1}\frac{\partial}{\partial y} G_x
    -\frac{\partial}{\partial y}\varepsilon^{-1} G_x\\
    \frac{\partial}{\partial z}\varepsilon^{-1} G_z
    +\frac{\partial}{\partial x}\varepsilon^{-1} G_x\\
    -\varepsilon^{-1}\frac{\partial}{\partial y} G_z
    +\varepsilon^{-1}\frac{\partial}{\partial z} G_y
    -\frac{\partial}{\partial y}\varepsilon^{-1} G_z
\end{matrix}
\right)\endgroup,
\label{a9}
\end{equation}

\begin{equation}
L_{2xx}\mathbf{G}=-\mathbf{\widehat{x}}\times[\varepsilon^{-1}(\mathbf{\widehat{x}}\times\mathbf{G})]
=-
\begingroup
\renewcommand*{\arraystretch}{1.5}
\left(
\begin{matrix}
    0\\
    -\varepsilon^{-1} G_y\\
    -\varepsilon^{-1} G_z
\end{matrix}
\right)\endgroup,
\label{a10}
\end{equation}

\begin{equation}
L_{2yy}\mathbf{G}=-\mathbf{\widehat{y}}\times[\varepsilon^{-1}(\mathbf{\widehat{y}}\times\mathbf{G})]
=-
\begingroup
\renewcommand*{\arraystretch}{1.5}
\left(
\begin{matrix}
    -\varepsilon^{-1} G_x\\
    0\\
    -\varepsilon^{-1} G_z
\end{matrix}
\right)\endgroup,
\label{a11}
\end{equation}

\begin{equation}
L_{2xy}\mathbf{G}=-\mathbf{\widehat{x}}\times[\varepsilon^{-1}(\mathbf{\widehat{y}}\times\mathbf{G})]-\mathbf{\widehat{y}}\times[\varepsilon^{-1}(\mathbf{\widehat{x}}\times\mathbf{G})]
=-
\begingroup
\renewcommand*{\arraystretch}{1.5}
\left(
\begin{matrix}
    \varepsilon^{-1} G_y\\
    \varepsilon^{-1} G_x\\
    0
\end{matrix}
\right)\endgroup.
\label{a12}
\end{equation}

Here, $\mathbf{G}$ is an arbitrary three-dimensional vector field $\mathbf{G(r)}:\mathbb{R}^3\to\mathbb{C}^3 $, and $\mathbf{\widehat{x}}=(1,0,0)$ and $\mathbf{\widehat{y}}=(0,1,0)$ are the unit vectors in the $x$ and $y$ directions, respectively.

The form of Eq. \eqref{a6} gives us the intuition to also expand the eigenfunction $\mathbf{\Psi}$ and the eigenvalue $E$ in series of $\Delta k_x$ and $\Delta k_y$:

\begin{align}
\mathbf{\Psi}
&=\mathbf{\Psi_0}
+\Delta k_x\mathbf{\Psi_{1x}}
+\Delta k_y\mathbf{\Psi_{1y}}
+\left(\Delta k_x\right)
^2\mathbf{\Psi_{2xx}}
+\left(\Delta k_y\right)^2\mathbf{\Psi_{2yy}}
+\left(\Delta k_x\right)\left(\Delta k_y\right)\mathbf{\Psi_{2xy}}
+o\left(\left|\mathbf{\Delta k}\right|^2\right),
\label{a13}\\
E
&=E_0
+\Delta k_xE_{1x}
+\Delta k_y E_{1y}
+\left(\Delta k_x\right)
^2 E_{2xx}
+\left(\Delta k_y\right)
^2 E_{2yy}
+\left(\Delta k_x\right)\left(\Delta k_y\right) E_{2xy}
+o\left(\left|\mathbf{\Delta k}\right|^2\right).
\label{a14}
\end{align}

Using the information in Eqs. \eqref{a13} and \eqref{a14}, Eq. \eqref{a6} can be separated to different orders.

The leading-order $O(1)$ equation is:
\begin{equation}
L_0\mathbf{\Psi_0}=E_0\mathbf{\Psi_0}.
\label{a15}
\end{equation}

Since $L_0$ is the Maxwell operator the same as in Eq. \eqref{a1}, we directly get the solution to Eq. \eqref{a15} from Eq. \eqref{a2}:
\begin{equation}
\mathbf{\Psi_0}=\mathbf{H_k},\quad E_0=E_k.
\label{a16}
\end{equation}

The $O(\Delta k_x^1)$ equation is:
\begin{equation}
L_0\mathbf{\Psi_{1x}}+L_{1x}\mathbf{\Psi_0}=E_0\mathbf{\Psi_{1x}}+E_{1x}\mathbf{\Psi_0}.
\label{a17}
\end{equation}

Using the information in Eq. \eqref{a16}, we get the solution to Eq. \eqref{a17}:
\begin{align}
E_{1x}&=\left<\mathbf{H_k}\left|L_{1x}\right|\mathbf{H_k}\right>,
\label{a18}\\
\mathbf{\Psi_{1x}}&=-\left(L_0-E_{\mathbf{k}}\right)^{-1}\left[\left(L_{1x}-E_{1x}\right)\mathbf{H_k}\right].
\label{a19}
\end{align}

The $O(\Delta k_y^1)$ equation is:
\begin{equation}
L_0\mathbf{\Psi_{1y}}+L_{1y}\mathbf{\Psi_0}=E_0\mathbf{\Psi_{1y}}+E_{1y}\mathbf{\Psi_0}.
\label{a20}
\end{equation}

Using the information in Eq. \eqref{a16}, we get the solution to Eq. \eqref{a20}:
\begin{align}
E_{1y}&=\left<\mathbf{H_k}\left|L_{1y}\right|\mathbf{H_k}\right>,
\label{a21}\\
\mathbf{\Psi_{1y}}&=-\left(L_0-E_k\right)^{-1}\left[\left(L_{1y}-E_{1y}\right)\mathbf{H_k}\right].
\label{a22}
\end{align}

The $O(\Delta k_x^2)$ equation is:
\begin{equation}
L_0\mathbf{\Psi_{2xx}}+L_{1x}\mathbf{\Psi_{1x}}+L_{2xx}\mathbf{\Psi_0}=E_0\mathbf{\Psi_{2xx}}+E_{1x}\mathbf{\Psi_{1x}}+E_{2xx}\mathbf{\Psi_0}.
\label{a23}
\end{equation}

Using the information in Eq. \eqref{a16}, we get the solution to Eq. \eqref{a23}:
\begin{equation}
E_{2xx}=\left<\mathbf{H_k}\left|L_{2xx}\right|\mathbf{H_k}\right>+\left<\mathbf{H_k}\left|(L_{1x}-E_{1x})\right|\mathbf{\Psi_{1x}}\right>.
\label{a24}
\end{equation}

Here, only an expression for the eigenvalue $E_{2xx}$ is calculated, but not for the eigenvector $\mathbf{\Psi_{2xx}}$, because the information in $E_{2xx}$ is enough for evaluating the dispersion (up to the second order).

The $O(\Delta k_y^2)$ equation is:
\begin{equation}
L_0\mathbf{\Psi_{2yy}}+L_{1y}\mathbf{\Psi_{1y}}+L_{2yy}\mathbf{\Psi_0}=E_0\mathbf{\Psi_{2yy}}+E_{1y}\mathbf{\Psi_{1y}}+E_{2yy}\mathbf{\Psi_0}.
\label{a25}
\end{equation}

Using the information in Eq. \eqref{a16}, we get the solution to Eq. \eqref{a25}:
\begin{equation}
E_{2yy}=\left<\mathbf{H_k}\left|L_{2yy}\right|\mathbf{H_k}\right>+\left<\mathbf{H_k}\left|(L_{1y}-E_{1y})\right|\mathbf{\Psi_{1y}}\right>.
\label{a26}
\end{equation}

The $O(\Delta k_x \Delta k_y)$ equation is:
\begin{equation}
L_0\mathbf{\Psi_{2xy}}+L_{1x}\mathbf{\Psi_{1y}}+L_{1y}\mathbf{\Psi_{1x}}+L_{2xy}\mathbf{\Psi_0}=E_0\mathbf{\Psi_{2xy}}+E_{1x}\mathbf{\Psi_{1y}}+E_{1y}\mathbf{\Psi_{1x}}+E_{2xy}\mathbf{\Psi_0}.
\label{a27}
\end{equation}

Using the information in Eq. \eqref{a16}, we get the solution to Eq. \eqref{a27}:
\begin{equation}
E_{2xy}=
\left<\mathbf{H_k}\left|L_{2xy}\right|\mathbf{H_k}\right>
+\left<\mathbf{H_k}\left|(L_{1x}-E_{1x})\right|\mathbf{\Psi_{1y}}\right>
+\left<\mathbf{H_k}\left|(L_{1y}-E_{1y})\right|\mathbf{\Psi_{1x}}\right>.
\label{a28}
\end{equation}

Then, according to Eq. \eqref{a14}, we obtain the energy dispersion of a band near an arbitrary wavevector $\mathbf{k}$:
\begin{equation}
E\left(\mathbf{k}+\mathbf{\Delta k}\right)=E_0+E_{1x}\Delta k_x+E_{1y}\Delta k_y+E_{2xx}\Delta k_x^2+E_{2yy}\Delta k_y^2+E_{2xy}\Delta k_x \Delta k_y
+o\left(\left|\mathbf{\Delta k}\right|^2\right),
\label{a29}
\end{equation}
where the coefficients $E_0$, $E_{1x}$, $E_{1y}$, $E_{2xx}$, $E_{2yy}$ and $E_{2xy}$ can be found in Eqs. \eqref{a16}, \eqref{a18}, \eqref{a21}, \eqref{a24}, \eqref{a26}, and \eqref{a28}.

In our experiment, we intentionally use a photonic band with a quadratically dispersive energy. Likewise in our theory, we require that the energy is quadratically dispersive (to the leading order) at a given wavevector $\mathbf{k}=\mathbf{k_0}$, which means:
\begin{equation}
E(\mathbf{k_0}+\mathbf{\Delta k})
=E_{\mathbf{k_0}}
-\frac{1}{2m_{xx}}\Delta k_x^2
-\frac{1}{2m_{yy}} \Delta k_y^2
-\frac{1}{m_{xy}} \Delta k_x \Delta k_y
+o\left(\left|\mathbf{\Delta k}\right|^2\right).
\label{a30}
\end{equation}

Comparing Eq. \eqref{a30} with the result in Eq. \eqref{a29}, we directly get:
\begin{align}
E_{1x}
&=\left<\mathbf{H_{k_0}}\left|L_{1x}\right|\mathbf{H_{k_0}}\right>
=0,
\label{a31}
\\
E_{1y}
&=\left<\mathbf{H_{k_0}}\left|L_{1y}\right|\mathbf{H_{k_0}}\right>
=0,
\label{a32}
\\
E_{2xx}
&=\left<\mathbf{H_{k_0}}\left|L_{2xx}\right|\mathbf{H_{k_0}}\right>+\left<\mathbf{H_{k_0}}\left|L_{1x}\right|\mathbf{\Psi_{1x}}\right>
=-\frac{1}{2m_{xx}},
\label{a33}
\\
E_{2yy}
&=\left<\mathbf{H_{k_0}}\left|L_{2yy}\right|\mathbf{H_{k_0}}\right>+\left<\mathbf{H_{k_0}}\left|L_{1y}\right|\mathbf{\Psi_{1y}}\right>
=-\frac{1}{2m_{yy}},
\label{a34}
\\
E_{2xy}
&=\left<\mathbf{H_{k_0}}\left|L_{2xy}\right|\mathbf{H_{k_0}}\right>
+\left<\mathbf{H_{k_0}}\left|L_{1x}\right|\mathbf{\Psi_{1y}}\right>
+\left<\mathbf{H_{k_0}}\left|L_{1y}\right|\mathbf{\Psi_{1x}}\right>
=-\frac{1}{m_{xy}},
\label{a35}
\\
\mathbf{\Psi_{1x}}&=-\left(L_0-E_{\mathbf{k_0}}\right)^{-1}L_{1x}\mathbf{H_{k_0}},
\label{a36}
\\
\mathbf{\Psi_{1y}}&=-\left(L_0-E_{\mathbf{k_0}}\right)^{-1}L_{1y}\mathbf{H_{k_0}}.
\label{a37}
\end{align}\\

Next, we induce an extra potential to a quadratic band in a periodic photonic crystal slab. The potential is induced by slowly varying the unit cell at different positions, $\mathbf{r}=(x,y,z)$, in the photonic crystal. The slow variation of the unit cell could be realized by varying the size of the holes, the shape of the holes, or the position of the holes. 

To define and quantify the value of the potential, first a reference position is selected for the potential. For example, at $\mathbf{r}_{\mathrm{ref}}=(0,0,0)$. The value of the potential is set to $U=0$ at $\mathbf{r}=\mathbf{r}_{\mathrm{ref}}$. We assume that the unit cell at the reference position extends periodically with respect to its primitive vectors, and define the corresponding periodic dielectric function as the reference dielectric function $\varepsilon_{\mathrm{ref}}(\mathbf{r})$. The energy of the quadratic band tip in this periodic reference dielectric function is $E_{\mathrm{ref}}$. At another arbitrary position $\mathbf{r}=(x,y,z)$, we assume that the energy of the band tip is $V(\mathbf{r})$ if the unit cell at this position extends periodically with respect to its local primitive vectors. Then the potential at $\mathbf{r}$ is defined as: 
\begin{equation}
U(\mathbf{r})=V(\mathbf{r})-E_{\mathrm{ref}}.
\label{b0}
\end{equation}

When a potential $U(\mathbf{r})$ is induced in the system, the Maxwell's equations in Eq. \eqref{a1} can be modified as:
\begin{equation}
\mathbf{\nabla}\times \left[\varepsilon_{\mathrm{ref}}(\mathbf{r})^{-1}\mathbf{\nabla}\times\mathbf{H(r)}\right]+U(\mathbf{r})\mathbf{H(r)}=E\mathbf{H(r)},
\label{b1}
\end{equation}
where $\varepsilon_{\mathrm{ref}}(\mathbf{r})$ is the reference dielectric function as mentioned earlier, satisfying: 
\begin{equation}
\mathbf{\nabla}\times \left[\varepsilon_{\mathrm{ref}}(\mathbf{r})^{-1}\mathbf{\nabla}\times\mathbf{H_{k_0}(r)}\right]=E_{\mathrm{ref}}\mathbf{H_{k_0}(r)},
\label{b1-1}
\end{equation}
where $\mathbf{k_0}$ is the wavevector of the quadratic band tip, and $\mathbf{H_{k_0}(r)}$ is the corresponding magnetic profile at $\mathbf{k}=\mathbf{k_0}$.

In our theory and experiment, we require the potential to take the form of:
\begin{equation}
U(\mathbf{r})=\kappa^2 \cdot f(\kappa x, \kappa y),
\label{b2}
\end{equation}
where $\kappa$ is a small parameter satisfying $\kappa a\ll 1$ with $a$ denoted as the lattice constant, and $f$ is an arbitrary function. From the form of potential in Eq. \eqref{b2}, we can see that the strength of the potential is small and has an order of $\kappa^2$. The potential is independent of $z$, which means it is uniform in the $z$ direction. The potential only depends on $\kappa x$ and $\kappa y$, which means that it only varies on a length scale which is large compared to the lattice constant.

Assuming that the potential has an order of $\kappa^2$ as in Eq. \eqref{b2}, Eq. \eqref{b1} becomes:
\begin{equation}
\mathbf{\nabla}\times \left[\varepsilon_{\mathrm{ref}}(\mathbf{r})^{-1}\mathbf{\nabla}\times\mathbf{H(r)}\right]+\kappa^2 f(\kappa x, \kappa y)\mathbf{H(r)}=E\mathbf{H(r)}.
\label{b3}
\end{equation}

We notice that, other than the spatial scale $\mathbf{r}=(x,y,z)$, there is another slow-varying scale emerging from the left-hand-side of Eq. \eqref{b3}, which is:
\begin{equation}
\mathbf{R}=(X,Y,0)\equiv (\kappa x,\kappa y, 0).
\label{b4}
\end{equation}

This suggests that the solution to $\mathbf{H}$ in Eq. \eqref{b3} needs to incorporate both spatial scales: $\mathbf{r}$ and $\mathbf{R}$. Since $\kappa a\ll 1$, the eigenvalues and the eigenvectors in Eq. \eqref{b3} can be expanded in series of $\kappa$. Hence, we choose the ansatz of $\mathbf{H}$ and $E$ to be in the form of:
\begin{align}
\mathbf{H}
&=\mathbf{H_0}(\mathbf{R}, \mathbf{r})
+\kappa\mathbf{H_1}(\mathbf{R}, \mathbf{r})
+\kappa^2\mathbf{H_2}(\mathbf{R}, \mathbf{r})
+O(\kappa^3),
\label{b5}
\\
E
&=E_0+\kappa E_1+\kappa^2 E_2+O(\kappa^3).
\label{b6}
\end{align}

Notice that
\begin{equation}
\mathbf{\nabla}\times \mathbf{H}(\kappa \mathbf{r}, \mathbf{r})
=(\mathbf{\nabla_r}+\kappa\mathbf{\nabla_R})\times \mathbf{H}(\mathbf{R}, \mathbf{r}),
\label{b7}
\end{equation}
where $\mathbf{\nabla}=\left(\frac{d}{dx}, \frac{d}{dy}, \frac{d}{dz}\right)$, $\mathbf{\nabla_r}=\left(\frac{\partial}{\partial x},\frac{\partial}{\partial y},\frac{\partial}{\partial z}\right)$ and $\mathbf{\nabla_R}=(\frac{\partial}{\partial X},\frac{\partial}{\partial Y},0)$. This means that the operator $\mathbf{\nabla}$ in Eq. \eqref{b1} can be replaced with $\mathbf{\nabla_r}+\kappa \mathbf{\nabla_R}$, where $\mathbf{r}$ and $\mathbf{R}$ can be treated as independent variables.

Notice that
\begin{equation}
\begin{aligned}
&(\mathbf{\nabla_r}+\kappa \mathbf{\nabla_R})\times \left[\varepsilon_{\mathrm{ref}}^{-1}(\mathbf{\nabla_r}+\kappa \mathbf{\nabla_R})\times\mathbf{H}\right]
\\=&
\left(L_0-i\kappa L_{1x}\frac{\partial}{\partial X}
-i\kappa L_{1y}\frac{\partial}{\partial Y}
-\kappa^2 L_{2xx}\frac{\partial^2}{\partial X^2}
-\kappa^2 L_{2yy}\frac{\partial^2}{\partial Y^2}
-\kappa^2 L_{2xy}\frac{\partial^2}{\partial X \partial Y}\right)
\mathbf{H},
\end{aligned}
\label{b9}
\end{equation}
where the operators $L_0$, $L_{1x}$, $L_{1y}$, $L_{2xx}$, $L_{2yy}$ and $L_{2xy}$ are the same as we defined earlier in Eqs. \eqref{a7}, \eqref{a8}, \eqref{a9}, \eqref{a10}, \eqref{a11} and \eqref{a12}. Then Eq. \eqref{b3} can be modified as:
\begin{equation}
\left(L_0-i\kappa L_{1x}\frac{\partial}{\partial X}
-i\kappa L_{1y}\frac{\partial}{\partial Y}
-\kappa^2 L_{2xx}\frac{\partial^2}{\partial X^2}
-\kappa^2 L_{2yy}\frac{\partial^2}{\partial Y^2}
-\kappa^2 L_{2xy}\frac{\partial^2}{\partial X \partial Y}+\kappa^2 \cdot f(X,Y)\right)\mathbf{H}=E\mathbf{H}.
\label{b10}
\end{equation}

Next, we use the information in Eqs. \eqref{b5} and \eqref{b6} to expand Eq. \eqref{b10} in series of $\kappa$.

The $O(\kappa^0)$ equation is:
\begin{equation}
L_0\mathbf{H_0(R,r)}=E_0\mathbf{H_0(R,r)}.
\label{b11}
\end{equation}

According to Eq. \eqref{b1-1}, we directly get the solution to Eq. \eqref{b11}:
\begin{equation}
\mathbf{H_0(R,r)}=\alpha_0(\mathbf{R})\mathbf{H_{k_0}(r)},\quad E_0=E_\mathrm{ref},
\label{b12}
\end{equation}
where $\alpha_0(\mathbf{R})$ is an arbitrary function (which will be determined later) that only depends on the slow variable $\mathbf{R}=(X, Y, 0)$.

The $O(\kappa^1)$ equation is:
\begin{equation}
L_0\mathbf{H_1}
-iL_{1x}\frac{\partial}{\partial X}\mathbf{H_0}
-iL_{1y}\frac{\partial}{\partial Y}\mathbf{H_0}
=E_0\mathbf{H_1}+E_1\mathbf{H_0}.
\label{b13}
\end{equation}

Using the information in Eq. \eqref{b12} and multiplying both sides of Eq. \eqref{b13} by $\left<\mathbf{H_{k_0}}\right|$, we get:
\begin{equation}
-i\left(\frac{\partial}{\partial X}\alpha_0\right)\left<\mathbf{H_{k_0}}\left|L_{1x}\right|\mathbf{H_{k_0}}\right>
-i\left(\frac{\partial}{\partial Y}\alpha_0\right)\left<\mathbf{H_{k_0}}\left|L_{1y}\right|\mathbf{H_{k_0}}\right>
=
E_1\alpha_0.
\label{b14}
\end{equation}

From Eqs. \eqref{a31} and \eqref{a32}, we know that:
\begin{equation}
\left<\mathbf{H_{k_0}}\left|L_{1x}\right|\mathbf{H_{k_0}}\right>
=\left<\mathbf{H_{k_0}}\left|L_{1y}\right|\mathbf{H_{k_0}}\right>
=0.
\label{b15}
\end{equation}

This directly gives us:
\begin{equation}
E_1=0.
\label{b16}
\end{equation}

Then, equation \eqref{b13} becomes:
\begin{equation}
(L_0-E_{\mathrm{ref}})\mathbf{H_1}
=i\left(\frac{\partial}{\partial X}\alpha_0\right)\left(L_{1x}\mathbf{H_{k_0}}\right)
+i\left(\frac{\partial}{\partial Y}\alpha_0\right)\left(L_{1y}\mathbf{H_{k_0}}\right),
\label{b17}
\end{equation}
which means:
\begin{equation}
\mathbf{H_1}
=i\left(\frac{\partial}{\partial X}\alpha_0\right)\left[\left(L_0-E_{\mathrm{ref}}\right)^{-1} L_{1x}\mathbf{H_{k_0}}\right]
+i\left(\frac{\partial}{\partial Y}\alpha_0\right)\left[\left(L_0-E_{\mathrm{ref}}\right)^{-1} L_{1y}\mathbf{H_{k_0}}\right].
\label{b18}
\end{equation}

Recalling Eqs. \eqref{a36} and \eqref{a37}, we have:
\begin{equation}
\mathbf{H_1}
=\left(-i\frac{\partial}{\partial X}\alpha_0\right)\mathbf{\Psi_{1x}}
+\left(-i\frac{\partial}{\partial Y}\alpha_0\right)\mathbf{\Psi_{1y}}.
\label{b19}
\end{equation}

The $O(\kappa^2)$ equation is:
\begin{equation}
\begin{aligned}
&L_0\mathbf{H_2}
-iL_{1x}\frac{\partial}{\partial X}\mathbf{H_1}
-iL_{1y}\frac{\partial}{\partial Y}\mathbf{H_1}
-L_{2xx}\frac{\partial^2}{\partial X^2}\mathbf{H_0}
-L_{2yy}\frac{\partial^2}{\partial Y^2}\mathbf{H_0}
-L_{2xy}\frac{\partial^2}{\partial X\partial Y}\mathbf{H_0}+f\mathbf{H_0}
\\=&E_0\mathbf{H_2}+E_1\mathbf{H_1}+E_2\mathbf{H_0}.
\label{b20}
\end{aligned}
\end{equation}

Using the information in Eqs. \eqref{b12}, \eqref{b16} and \eqref{b19}, and multiplying both sides of Eq. \eqref{b13} by $\left<\mathbf{H_{k_0}}\right|$ , we get:
\begin{equation}
\scalebox{0.85}{$\displaystyle
\begin{aligned}
&-\left(\frac{\partial^2}{\partial X^2}\alpha_0\right)\left<\mathbf{H_{k_0}}\left|L_{1x}\right|\mathbf{\Psi_{1x}}\right>
-\left(\frac{\partial^2}{\partial X\partial Y}\alpha_0\right)\left<\mathbf{H_{k_0}}\left|L_{1x}\right|\mathbf{\Psi_{1y}}\right>
-\left(\frac{\partial^2}{\partial X \partial Y}\alpha_0\right)\left<\mathbf{H_{k_0}}\left|L_{1y}\right|\mathbf{\Psi_{1x}}\right>
-\left(\frac{\partial^2}{\partial Y^2}\alpha_0\right)\left<\mathbf{H_{k_0}}\left|L_{1y}\right|\mathbf{\Psi_{1y}}\right>
\\&-\left(\frac{\partial^2}{\partial X^2}\alpha_0\right)\left<\mathbf{H_{k_0}}\left|L_{2xx}\right|\mathbf{H_{k_0}}\right>
-\left(\frac{\partial^2}{\partial Y^2}\alpha_0\right)\left<\mathbf{H_{k_0}}\left|L_{2yy}\right|\mathbf{H_{k_0}}\right>
-\left(\frac{\partial^2}{\partial X\partial Y}\alpha_0\right)\left<\mathbf{H_{k_0}}\left|L_{2xy}\right|\mathbf{H_{k_0}}\right>
+f\alpha_0
\\&=
E_2 \alpha_0.
\label{b21}
\end{aligned}
$}
\end{equation}

Using the information in Eqs. \eqref{a33}, \eqref{a34} and \eqref{a35}, Eq. \eqref{b21} can be simplified as:
\begin{equation}
\left[\frac{1}{2m_{xx}}\frac{\partial^2}{\partial X^2}
+\frac{1}{2m_{yy}}\frac{\partial^2}{\partial Y^2}
+\frac{1}{m_{xy}}\frac{\partial^2}{\partial X\partial Y}
+f(X,Y)\right]
\alpha_0(X,Y)
=E_2\alpha_0(X,Y).
\label{b22}
\end{equation}

Equation \eqref{b22} is a Schr\"odinger-like equation that effectively describes the behavior of states near a quadratic band tip when a slow-varying superpotential is added to the photonic crystal.

In our experiment, our quadratic band is isotropic in the vicinity of $\mathbf{k}=\Gamma$, such that $\frac{1}{2m_{xx}}=\frac{1}{2m_{yy}}=\frac{1}{2m}$ and $\frac{1}{m_{xy}}=0$. Using the information in Eqs. \eqref{b4}, \eqref{b0}, \eqref{b2}, \eqref{b6}, \eqref{b12} and \eqref{b16}, Eq. \eqref{b22} can be rewritten as:
\begin{equation}
\begin{aligned}
&\left[\frac{\kappa^2}{2m}\frac{\partial^2}{\partial (\kappa x)^2}
+\frac{\kappa^2}{2m}\frac{\partial^2}{\partial (\kappa y)^2}
+V(\kappa\mathbf{r})\right]
\alpha_0(\kappa\mathbf{r})
=\left(E+ O(\kappa^3)\right)\alpha_0(\kappa\mathbf{r}),
\label{equation3}
\end{aligned}
\end{equation}
where $V(\kappa \mathbf{r})$ is the local band tip energy as described earlier and $E=\left(\omega/c\right)^2$ is the energy of the band. Equation \eqref{equation3} is the form presented in Eq. (4) of the main text.

Also notice from Eqs. \eqref{b5} and \eqref{b12} that:
\begin{equation}
\mathbf{H}=\alpha_0(\kappa \mathbf{r})\mathbf{H_{k_0}(r)}+O(\kappa^1),
\label{b23}
\end{equation}
which means the $\alpha_0(\kappa \mathbf{r})$ obtained from the eigenvalue problem in Eq. \eqref{b22} is the envelope function of the eigenstate in the leading order. Since only the leading order of the envelope function $\alpha_0(\kappa \mathbf{r})$ is considered in this paper, the subscript in $\alpha_0$ is omitted for simplicity from this point onward, as well as in the main text. $\alpha(\kappa \mathbf{r})$ is used to represent the envelope function of the eigenstate in the leading order. Equation \eqref{b23} is the same as Eq. (3) in the main text.

As we can see from Eqs. \eqref{equation3} and \eqref{b23}, the superpotential framework can be applied to any form of the external slow-varying potential $V(\kappa \mathbf{r})$ in aperiodic photonic crystals, which is a powerful tool to engineer the mode frequency separation and the eigenmode profile in an interpretable and efficient way.

\clearpage

\section*{Section 2: Linear superpotential and Airy resonances}
\addcontentsline{toc}{section}{Section 2: Linear superpotential and Airy resonances}

In this section, we show how to apply the theoretical framework of superpotential from the previous section to realize Airy resonances. Specifically, we slowly vary the lattice constant in the $x$ direction, $a_x$, as a function of position $x$, such that the superpotential $V(\kappa \mathbf{r})$ is linear. At the end of this section, we also propose three signatures of Airy resonances that can be measured and verified in the experiment.\\

\begin{figure}[H]
    \centering
    \subfigure[]{\includegraphics[width=0.23\textwidth]{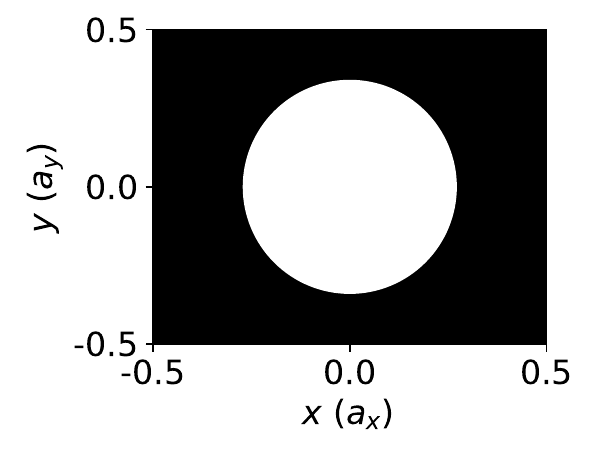}}
    \subfigure[]{\includegraphics[width=0.23\textwidth]{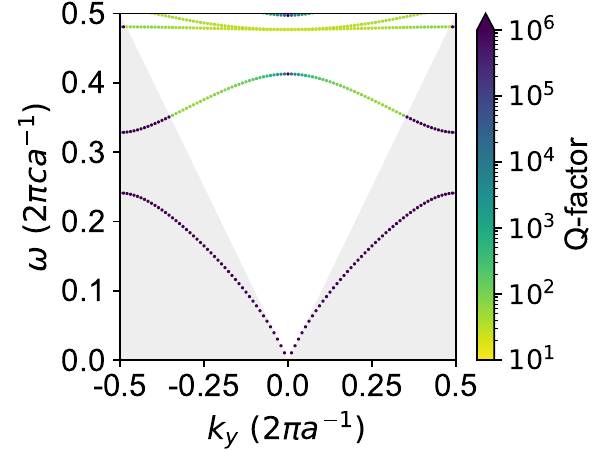}}
    \subfigure[]{\includegraphics[width=0.23\textwidth]{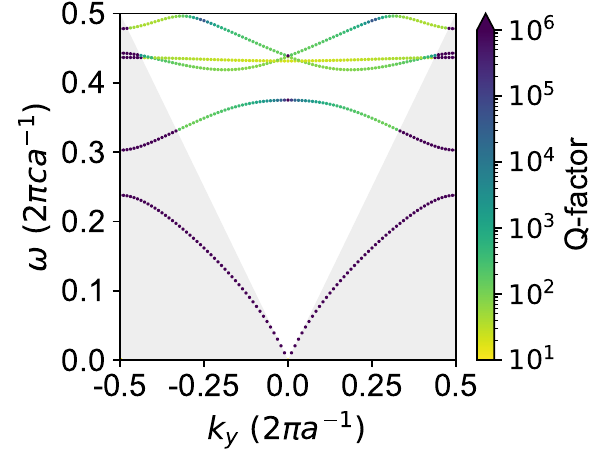}}
    \subfigure[]{\includegraphics[width=0.23\textwidth]{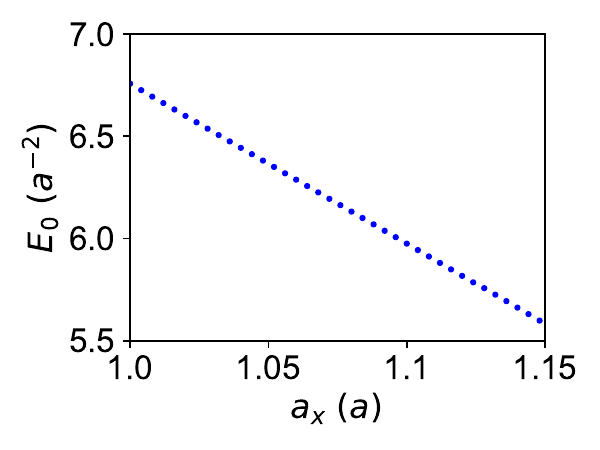}}

\caption{The periodic photonic structure and the definition of the local potential. (a) The unit cell (in the $x-y$ plane, and $z$ is the out-of-plane direction) of our photonic crystal slab. The structure contains a circular air hole ($\varepsilon=1.0$) with $r_0=0.34a$ in a rectangular lattice, where $a_x$ and $a_y$ are the lattice constant in the $x$ and $y$ directions. The slab is made of silicon ($\varepsilon=12.11$) with a thickness of $h=0.35a$ and sits on top of a silica substrate ($\varepsilon=2.25$). (b) The simulated band structure of $a_x=a_y=a$ along the $k_x=0$ line. Only the transverse electric (TE)-like modes are shown. (c) The simulated band structure of $a_x=1.15a$ and $a_y=a$ along the $k_x=0$ line. (d) The tip energy ($E=\left(\omega/c\right)^2$) of the quadratic band as a function of $a_x$ when the lattice constant in the $y$ direction is fixed at $a_y=a$.}
\label{figs1}
\end{figure}

As shown in Eq. \eqref{a30}, the superpotential formalism requires an isolated quadratic band in the periodic structure. Figure \ref{figs1}(a) shows the unit cell used in the experiment. The structure contains circular air holes arranged to form a rectangular lattice in the $x-y$ plane in a silicon slab. The slab has a finite thickness $h=0.35a$ in the $z$ direction. The structure hosts an isolated quadratic band in the vicinity of $\mathbf{k}=\Gamma$. We find that the band tip energy changes as the lattice constant in the $x$ direction, $a_x$, is varied while the lattice constant in the $y$ direction is fixed at $a_y=a$. In Fig. \ref{figs1}(b), the lattice constant in the $x$ direction is $a_x=a$. The band tip energy is $E_0=6.76a^{-2}$. In Fig. \ref{figs1}(c), as $a_x$ increases to $a_x=1.15a$, the band tip energy decreases to $E_0=5.58a^{-2}$. More generally, we show in Fig. \ref{figs1}(d) that the tip energy $E_0$ decreases as the lattice constant $a_x$ increases. The simulation results are obtained from an open-source software package \textsc{Legume} \cite{Legume}, which implements the guided mode expansion method to solve for the photonic modes in a slab geometry.

Next, we use the square lattice (corresponding to $a_x=a_y=a$) as our reference structure at $\mathbf{r}=(0,0,0)$. There is an isolated quadratic band in the periodic structure, with the reference tip energy $E_\mathrm{ref}=6.76a^{-2}$ as shown in Fig. \ref{figs1}(b). The effective mass of this quadratic band is $\frac{1}{2m}=0.651+0.109i$, such that in the vicinity of $\mathbf{k}=\Gamma$:
\begin{equation}
E_{\mathbf{k}}=E_{\mathrm{ref}}-\frac{1}{2m}|\mathbf{k}|^2.
\label{c1}
\end{equation}

Here, we emphasize that the effective mass is a complex number. This is because the band is quadratically dispersive in both the real part of the energy and the imaginary part of the energy. The imaginary part of the energy corresponds to the out-of-plane radiation loss in the slab system. The radiation loss scales as $O\left(\left|\mathbf{k}\right|^2\right)$, which arises due to the symmetry-protected bound state
in the continuum (BIC) at $\mathbf{k}=\Gamma$ \cite{hsu2016bound, koshelev2018asymmetric}. The non-Hermiticity that arises from the imaginary part of the effective mass governs the linewidth of the Airy resonances and also has an impact on the spatial mode profile of Airy resonances, as we describe later in Section 5 and Section 6.

\begin{figure}[H]
    \centering    \subfigure{\includegraphics[width=0.96\textwidth]{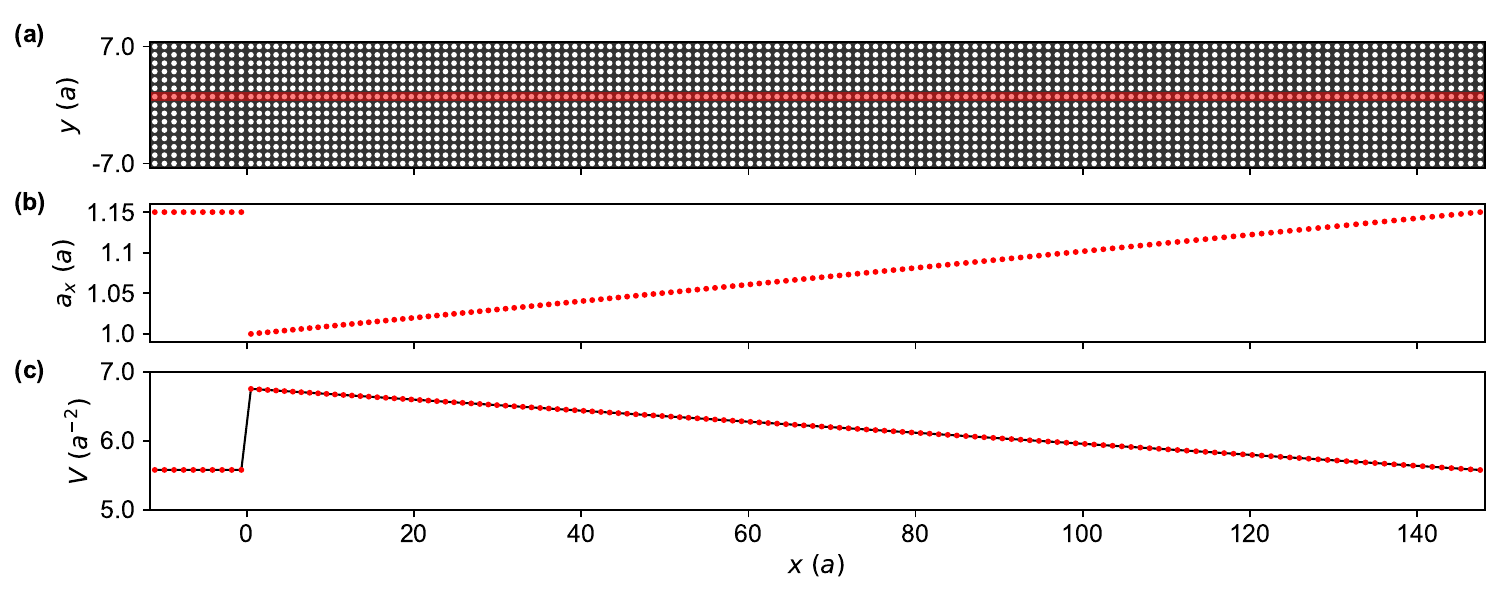}}

\caption{Adding superpotential into a photonic crystal slab. (a) The dielectric function after we add superpotential. White represents air with $\varepsilon=1$, and black represents silicon with $\varepsilon=12.11$. The strength of the superpotential is $\kappa=0.2a^{-1}$. (b) The local lattice constant $a_x$ as a function of the position $x$ of the structure in (a). (c) The value of the superpotential $V$ as a function of position $x$ of the structure in (a)}
\label{figs2}
\end{figure}

Then, a superpotential is added to the system by slowly varying the lattice constant $a_x$ as a function of the position $x$ in the dielectric function, as shown in Fig. \ref{figs2}(a). The deformation breaks the periodicity in the $x$ direction, so the unit cell becomes a supercell as highlighted in red. The local lattice constant $a_x$ is defined as the separation between adjacent holes in the $x$ direction. In Fig. \ref{figs2}(b), the lattice constant $a_x$ is no longer a constant but a slow-varying function of the position $x$. As shown in Fig. \ref{figs1}(d), the local band tip energy is shifted when the lattice constant is changed. As a result of the deformation in Fig. \ref{figs1}(a), the local tip energy is again a function of the position $x$ as shown in Fig. \ref{figs2}(c), which acts as the superpotential $V(\mathbf{r})$ derived in the earlier text.

In particular, the deformation shown in Fig. \ref{figs2}(a) results in a linear superpotential shown in \ref{figs2}(c), which is a linear function with a barrier at $x=0$:
\begin{equation}
V(\mathbf{r})=
\left\{
\begin{aligned}
E_{\mathrm{b}} & , & x<0\\
E_{\mathrm{ref}}-\kappa^3x & , & x>0
\end{aligned}
\right.
\label{c2}
\end{equation}
where $E_\mathrm{ref}=6.76a^{-2}$ is the reference energy, $E_\mathrm{b}=5.58a^{-2}$ is the energy beyond the barrier for $x<0$, and $\kappa$ is the length scale of the slow-varying superpotential. In the structure in Fig. \ref{figs2}(a), $\kappa=0.2a^{-1}$.

Substituting Eq. \eqref{c2} into Eq. \eqref{equation3}, the Schr\"odinger-like equation for $\alpha(\mathbf{r})$ becomes:
\begin{equation}
\left\{
\begin{aligned}
\left[
\frac{1}{2m}\left(\frac{\partial^2}{\partial x^2}+\frac{\partial^2}{\partial y^2}\right)
+E_\mathrm{b}
\right]
\alpha(\mathbf{r})
=E\cdot \alpha(\mathbf{r}) & , & x<0\\
\left[
\frac{1}{2m}\left(\frac{\partial^2}{\partial x^2}+\frac{\partial^2}{\partial y^2}\right)
+E_\mathrm{ref}-\kappa^3x
\right]
\alpha(\mathbf{r})
=E\cdot \alpha(\mathbf{r}) & , & x>0
\end{aligned}
\right.
\label{c3}
\end{equation}

Here, for simplicity, we omit the $O(\kappa^3)$ term in the energy. Notice that the superpotential in Eq. \eqref{c2} only breaks the periodicity in the $x$ direction but still preserves the translation symmetry in the $y$ direction. The wavevector in the $y$ direction, $k_y$, is still a conserved quantity, which means:
\begin{equation}
\alpha\left(\mathbf{r}\right)=e^{ik_y\cdot y}\alpha(x).
\label{c4}
\end{equation}

Equation \eqref{c3} then becomes:
\begin{equation}
\left\{
\begin{aligned}
\left[
\frac{1}{2m}\frac{\partial^2}{\partial x^2}
-\frac{k_y^2}{2m}
+E_\mathrm{b}
\right]
\alpha(x)
=E\cdot \alpha(x) & , & x<0\\
\left[
\frac{1}{2m}\frac{\partial^2}{\partial x^2}
-\frac{k_y^2}{2m}
+E_\mathrm{ref}-\kappa^3x
\right]
\alpha(x)
=E\cdot \alpha(x) & , & x>0
\end{aligned}
\right.
\label{c5}
\end{equation}

When $x<0$, the solution to Eq. \eqref{c5} is 
\begin{equation}
\alpha(x)
=C_E\cdot 
e^{\left[2m\left(E-E_\mathrm{b}\right)+k_y^2\right]^{\frac{1}{2}}\cdot x},\quad x<0
\label{c6}
\end{equation}
where $C_E$ is a normalization constant for the solution at energy $E$. We already consider the fact that $\alpha(x)\to 0$ when $x\to-\infty$.

When $x>0$, the solution to Eq. \eqref{c5} is 
\begin{equation}
\alpha(x)
=D_E\cdot 
\mathrm{Ai}\left[
\left(\frac{1}{2m}\right)^{-\frac{1}{3}}\kappa \left(x-x_E\right)
\right],\quad x>0
\label{c7}
\end{equation}
where $D_E$ is a normalization constant, $\mathrm{Ai}\left[\cdot\right]$ is the Airy function, and $x_E=\frac{1}{\kappa^3}\left(E_\mathrm{ref}-E-\frac{k_y^2}{2m}\right)$ is a position offset. Here, we use the fact that $f(x)=\mathrm{Ai}(x)$ is the solution to the following differential equation \cite{vallee2010airy}:
\begin{equation}
\frac{d^2f(x)}{dx^2}-x\cdot f(x)=0.
\label{c8}
\end{equation}

At the boundary $x=0$, the wavefunction and its first-order derivative should be continuous. Hence, we have:
\begin{equation}
\alpha(x=0^-)=\alpha(x=0^+),
\label{c9}
\end{equation}
\begin{equation}
\alpha^{\prime}(x=0^-)=\alpha^{\prime}(x=0^+),
\label{c10}
\end{equation}
which gives the connection condition:
\begin{equation}
C_E=\mathrm{Ai}\left[-\left(\frac{1}{2m}\right)^{-\frac{1}{3}}\kappa \cdot x_E\right]
\cdot D_E,
\label{c11}
\end{equation}
\begin{equation}
\left[2m\left(E-E_\mathrm{b}\right)+k_y^2\right]^{\frac{1}{2}}
\cdot C_E=
\left(\frac{1}{2m}\right)^{-\frac{1}{3}}\kappa
\cdot\mathrm{Ai}^{\prime}\left[-\left(\frac{1}{2m}\right)^{-\frac{1}{3}}\kappa \cdot x_E\right]
\cdot D_E.
\label{c12}
\end{equation}

As shown in Eq. \eqref{b2}, the superpotential is very weak in strength ($V(\mathbf{r})-E_\mathrm{ref}$ is in the order of $O(\kappa^2)$), which means the energy $E$ of the Airy mode is very close to the reference energy $E_\mathrm{ref}$. Moreover, the superpotential framework requires $\kappa$ to be a small parameter that $\kappa a\ll1$. Hence, within our experimental parameters, the following relation always holds:
\begin{equation}
\left(E-E_\mathrm{b}\right)^{\frac{1}{2}}
\approx
\left(E_\mathrm{ref}-E_\mathrm{b}\right)^{\frac{1}{2}}
=
1.09a^{-1}
\gg
\kappa.
\label{c13}
\end{equation}

Combining Eqs. \eqref{c12} and \eqref{c13}, we get:
\begin{equation}
1
\gg
\frac{C_E}{D_E}
\approx
0.
\label{c14}
\end{equation}

Since $C_E$ is very small, we treat $C_E=0$ in the later text and only focus on the wavefunction $\alpha(x)$ when $x>0$. This is equivalent to setting $E_\mathrm{b}=-\infty$, which is an ideal hard-wall barrier.

Eq. \eqref{c11} then becomes:
\begin{equation}
\mathrm{Ai}\left[-\left(\frac{1}{2m}\right)^{-\frac{1}{3}}\kappa \cdot x_E\right]
=\frac{C_E}{D_E}
= 0.
\label{c15}
\end{equation}

Equation \eqref{c15} indicates that the energy of Airy modes cannot take arbitrary values, but needs to be discrete to satisfy Eq. \eqref{c15}. We use the notation $E_n$ to represent the $n^{\mathrm{th}}$ energy level with $n=1,2,3,\cdots$ the mode index.

Notice that $x_n=\frac{1}{\kappa^3}\left(E_\mathrm{ref}-E_n-\frac{k_y^2}{2m}\right)>0$, so the argument of the Airy function in Eq. \eqref{c15} is always smaller than zero. According to the asymptotic behavior of the Airy function $\mathrm{Ai}(z)$ when $z<0$ \cite{vallee2010airy}:
\begin{equation}
\mathrm{Ai}(z)\sim
\frac{1}{\sqrt{\pi}\left(-z\right)^{\frac{1}{4}}}\mathrm{sin}\left(\frac{2}{3}\left(-z\right)^{\frac{3}{2}}+\frac{\pi}{4}\right),
\quad z<0
\label{c16}
\end{equation}
the $n^\mathrm{th}$ zero of the Airy function $\mathrm{Ai}(z)$ is 
\begin{equation}
z_n=-\left[\frac{3}{2}\pi\left(n-\frac{1}{4}\right)\right]^{\frac{2}{3}},
\quad n=1,2,3,\cdots
\label{c17}
\end{equation}

According to Eq. \eqref{c15}, we have:
\begin{equation}
-\left(\frac{1}{2m}\right)^{-\frac{1}{3}}\kappa \cdot x_n
=z_n=-\left[\frac{3}{2}\pi\left(n-\frac{1}{4}\right)\right]^{\frac{2}{3}},
\label{c18}
\end{equation}
which means
\begin{equation}
x_n
=\left(\frac{1}{2m}\right)^{\frac{1}{3}}\left[\frac{3}{2}\pi\left(n-\frac{1}{4}\right)\right]^{\frac{2}{3}}\kappa^{-1},
\label{c19}
\end{equation}
and 
\begin{equation}
E_n
=E_\mathrm{ref}-\kappa^3 x_n-\frac{k_y^2}{2m}
=E_\mathrm{ref}
-\left(\frac{1}{2m}\right)^{\frac{1}{3}}\left[\frac{3}{2}\pi\left(n-\frac{1}{4}\right)\right]^{\frac{2}{3}}\kappa^{2}
-\frac{k_y^2}{2m}.
\label{c20}
\end{equation}

Since $E_n$ also depends on $k_y$, we use both $n$ and $k_y$ as subscripts of $E$, as well as $\alpha$, to show the discrete Airy bands:
\begin{equation}
E_{n,k_y}
=E_\mathrm{ref}
-\left(\frac{1}{2m}\right)^{\frac{1}{3}}\left[\frac{3}{2}\pi\left(n-\frac{1}{4}\right)\right]^{\frac{2}{3}}\kappa^{2}
-\frac{k_y^2}{2m}.
\label{c21}
\end{equation}
\begin{equation}
\alpha_{n,k_y}\left(\mathbf{r}\right)
=\left\{
\begin{aligned}
D_n\cdot e^{ik_y\cdot y}
\cdot \mathrm{Ai}\left[
\left(\frac{1}{2m}\right)^{-\frac{1}{3}}\kappa \left(x-x_n\right)
\right] & , & x>0\\
0 & , & x<0
\end{aligned}
\right.
\label{c22}
\end{equation}

Equations \eqref{c21} and \eqref{c22} are the same as Eq. (5) in the main text. For simplicity, we omit the normalization factor $D_n$ from Eq. \eqref{c22} in the main text.

\begin{figure}[H]
    \centering
    \subfigure[]{\includegraphics[width=0.23\textwidth]{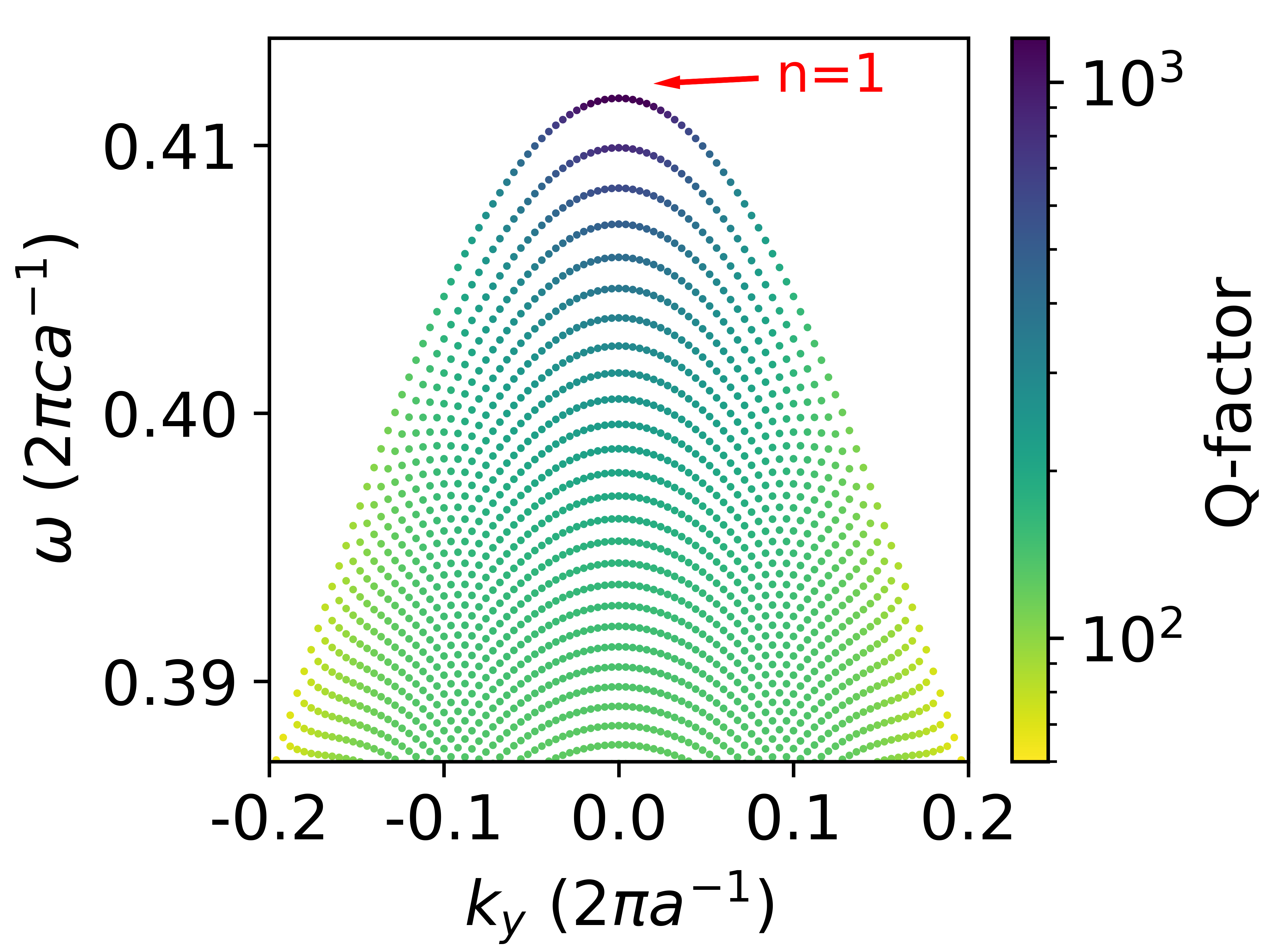}}
    \subfigure[]{\includegraphics[width=0.23\textwidth]{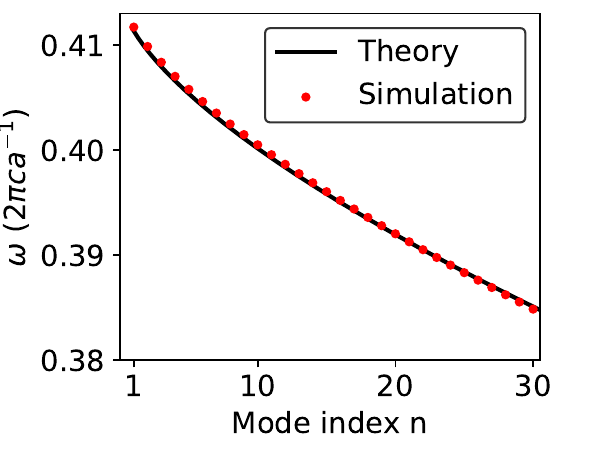}}
    \subfigure[]{\includegraphics[width=0.23\textwidth]{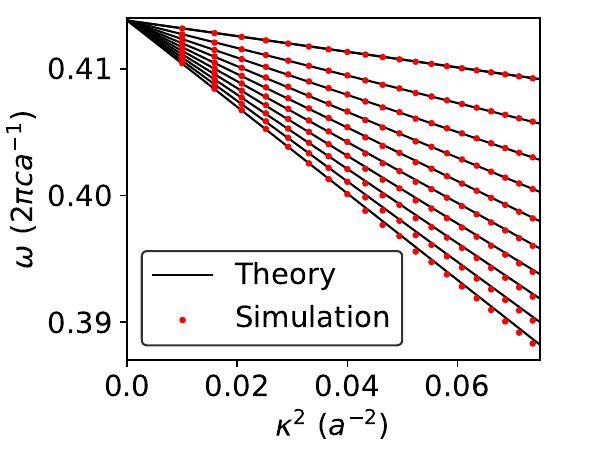}}
    \subfigure[]{\includegraphics[width=0.23\textwidth]{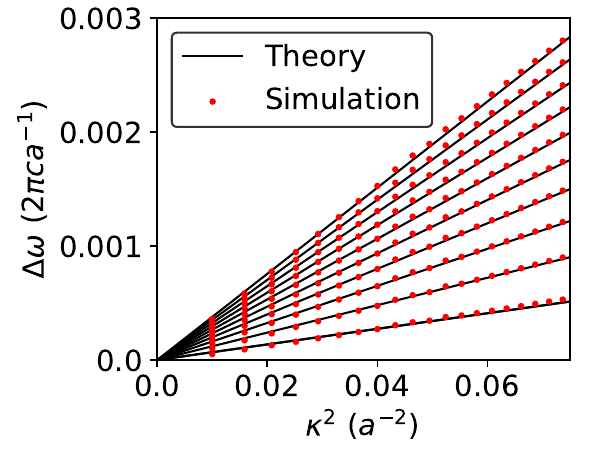}}

\caption{The first signature of Airy resonances: The Airy state frequency separation and linewidth. (a) The (TE-like) band structure of the Airy sample in Fig. \ref{figs2}(a). (b) The Airy state eigen-frequency $\omega_n$ at $k_y=0$ in (a). Only the first $30$ Airy states are shown. (c) The Airy state eigen-frequency $\omega_n$ at different potential strength $\kappa$. Only the first $10$ Airy states are shown. (d) The Airy state linewidth $\Delta\omega_n$ at different potential strength $\kappa$. Only the first $10$ Airy states are shown.}
\label{figs3}
\end{figure}

Equation \eqref{c21} is the first signature of Airy resonances. It describes the frequency separation as well as the linewidth of Airy modes. Next, the results in Eq. \eqref{c21} are verified in simulations.

According to Eqs. \eqref{a1-1} and \eqref{a1-2}, Eq. \eqref{c21} can be expressed in the frequency domain:
\begin{equation}
\omega_{n,k_y}
=c\cdot\sqrt{\mathrm{Re}\left(E_{n,k_y}\right)}
=c\cdot\sqrt{E_\mathrm{ref}}-\frac{c}{2\sqrt{E_\mathrm{ref}}}
\left[\mathrm{Re}\left[\left(\frac{1}{2m}\right)^{\frac{1}{3}}\right]\left[\frac{3}{2}\pi\left(n-\frac{1}{4}\right)\right]^{\frac{2}{3}}\kappa^{2}
+\mathrm{Re}\left(\frac{1}{2m}\right)\cdot k_y^2\right]+O\left(\kappa^3\right).
\label{c23}
\end{equation}
\begin{equation}
\Delta\omega_{n,k_y}
=-c\frac{\mathrm{Im}\left(E_{n,k_y}\right)}{\sqrt{\mathrm{Re}\left(E_{n,k_y}\right)}}
=\frac{c}{\sqrt{E_\mathrm{ref}}}
\left[\mathrm{Im}\left[\left(\frac{1}{2m}\right)^{\frac{1}{3}}\right]\left[\frac{3}{2}\pi\left(n-\frac{1}{4}\right)\right]^{\frac{2}{3}}\kappa^{2}
+\mathrm{Im}\left(\frac{1}{2m}\right)\cdot k_y^2\right]+O\left(\kappa^3\right).
\label{c24}
\end{equation}

Figure \ref{figs3}(a) shows the band structure of the Airy sample in Fig. \ref{figs2}(a) with potential strength $\kappa=0.2a^{-1}$. Compared with the periodic band structure in Fig. \ref{figs1}(b), a single quadratic band splits into several discrete energy levels in Fig. \ref{figs3}(a), as predicted in Eq. \eqref{c23} by mode index $n$. The first ($n=1$) Airy mode is marked with a red arrow in the figure. Each energy level is quadratically dispersive in $k_y$, which is consistent with the $k_y^2$ term in Eq. \eqref{c23}.

We then focus on the Airy state frequency separation and linewidth at $k_y=0$. At $k_y=0$, Eqs. \eqref{c23} and \eqref{c24} become:
\begin{equation}
\omega_{n}
=\omega_\mathrm{ref}-\frac{\left(\frac{3}{2}\pi\right)^{\frac{2}{3}}c^2\cdot \mathrm{Re}\left[\left(\frac{1}{2m}\right)^{\frac{1}{3}}\right]}{2\omega_\mathrm{ref}}
\left(n-\frac{1}{4}\right)^{\frac{2}{3}}\kappa^{2}
+O\left(\kappa^3\right),
\label{c25}
\end{equation}
\begin{equation}
\Delta\omega_{n}
=\frac{\left(\frac{3}{2}\pi\right)^{\frac{2}{3}}c^2\cdot \mathrm{Im}\left[\left(\frac{1}{2m}\right)^{\frac{1}{3}}\right]}{\omega_\mathrm{ref}}
\left(n-\frac{1}{4}\right)^{\frac{2}{3}}\kappa^{2}+O\left(\kappa^3\right),
\label{c26}
\end{equation}
where $\omega_\mathrm{ref}=c\sqrt{E_\mathrm{ref}}$ is the reference frequency corresponding to the reference tip energy $E_\mathrm{ref}$ as defined in Eq. \eqref{c1}.

Figure \ref{figs3}(b) shows the eigen-frequencies $\omega_n$ of the first $30$ Airy states at a fixed value of potential strength $\kappa=0.2a^{-1}$. The simulation is obtained by directly solving the Maxwell's equation in \eqref{a1} by \textsc{Legume} \cite{Legume}, while the theory is calculated from Eq. \eqref{c25} with the band parameters: $\omega_\mathrm{ref}=0.414\ (2\pi ca^{-1})$ and $\frac{1}{2m}=0.651+0.109i$. We find good quantitative agreement between the theory and the simulation. When the potential strength $\kappa$ is fixed, the Airy state frequency separation $\omega_n-\omega_\mathrm{ref}$ is proportional to $\left(n-\frac{1}{4}\right)^{\frac{2}{3}}$.

Figure \ref{figs3}(c) shows the eigen-frequencies $\omega_n$ of the first $10$ Airy states at different values of $\kappa$. As predicted in Eq. \eqref{c25}, the Airy state frequency separation $\omega_n-\omega_\mathrm{ref}$ is proportional to $\kappa^2$, which is consistent with the simulation results. The theoretical lines with different slopes in Fig. \ref{figs3}(c) represent different mode index $n$. From top to bottom, the lines represent mode indices $n=1,2,3,\cdots$. The $n=1$ state has the highest frequency.

Figure \ref{figs3}(a) shows that all of the Airy bands have a finite Q-factor, and higher $n$ modes have lower Q-factor. This can be explained in Eq. \eqref{c26}. Due to the BIC at $\mathbf{k}=\Gamma$ in the reference structure, the effective mass is no longer a real number but a complex number. This gives a finite imaginary part to the energy $E_n$ and causes a finite linewidth in frequency $\Delta \omega_n$. From Eq. \eqref{c26}, the linewidth is proportional to both $\left(n-\frac{1}{4}\right)^{\frac{2}{3}}$ and $\kappa^2$, which indicates that linewidths are larger for larger $n$ modes and in higher $\kappa$ structures. Equation \eqref{c26} is then verified in Fig. \ref{figs3}(d), which shows a perfect match between the simulation and the theory.\\

Equation \eqref{c22} is the second signature of Airy resonances. It predicts that the wavefunctions of Airy resonances are Airy functions. Next, Eq. \eqref{c22} is verified in simulations.

\begin{figure}[H]
    \centering
    \subfigure{\includegraphics[width=0.96\textwidth]{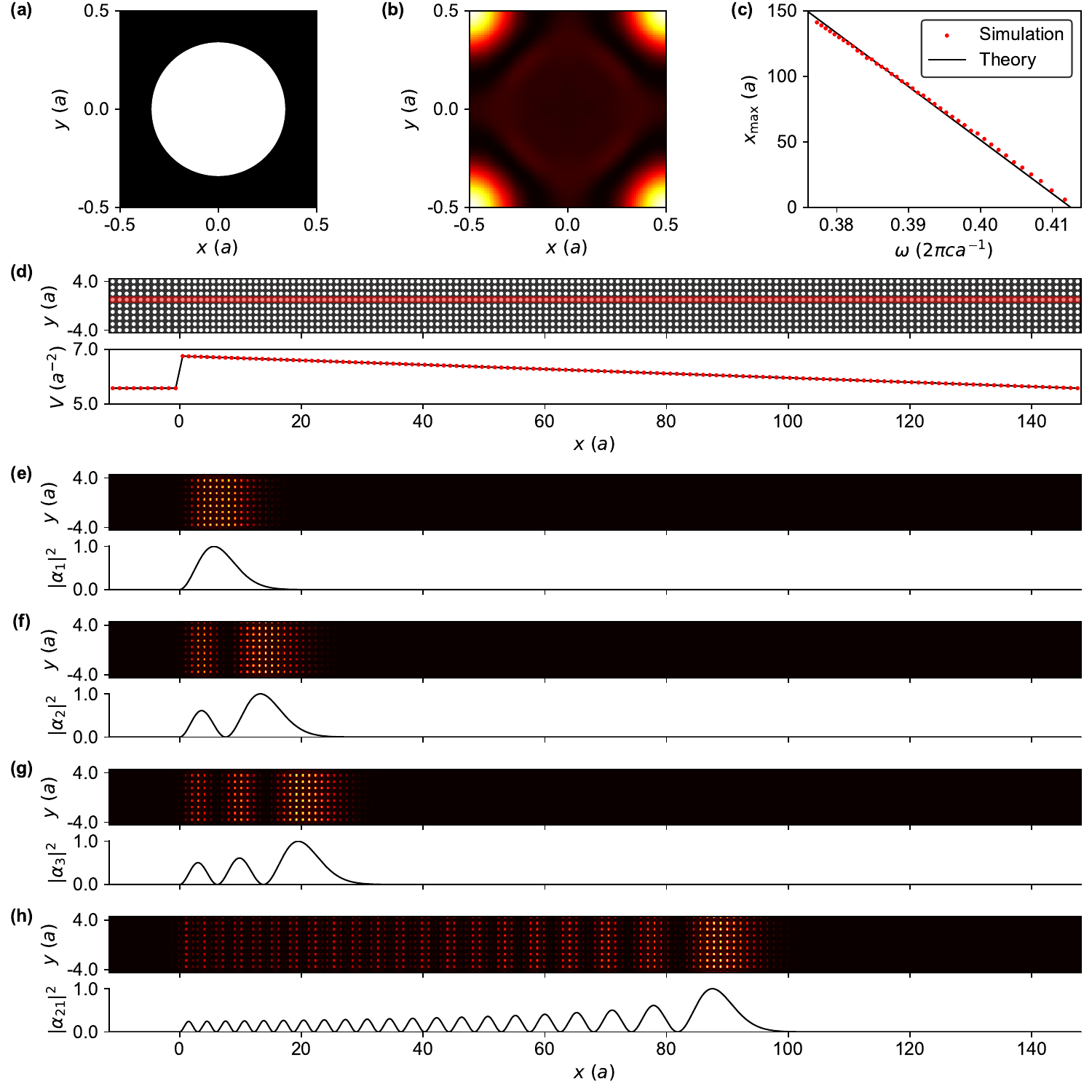}}

\caption{The second signature of Airy resonances: the Airy mode eigenfunction profile. (a) The unit cell of the reference structure. White represents air with $\varepsilon=1$, and black represents silicon with $\varepsilon=12.11$. (b) The simulated eigenfunction profile ($\left|H_z\right|^2$ at the center of the slab) at $\mathbf{k}=\Gamma$ of the structure in (a). (c) The position of the main lobe $x_{\mathrm{max}}(n)$ as a function of eigen-frequency $\omega_n$. (d) The dielectric function (upper panel) and the corresponding superpotential (lower panel) of an Airy sample with $\kappa=0.2a^{-1}$. (e)-(h): The simulated eigenfunctions of the Airy modes (upper panels) for the structure in (d) at $k_y=0$, and the theoretical envelope functions $\left|\alpha_n\right|^2$ calculated from Eq. \eqref{c22} (lower panels). The mode indices $n$ are: (e) $n=1$; (f) $n=2$; (g) $n=3$; (h) $n=21$.}
\label{figs4}
\end{figure}

Figure \ref{figs4}(a) shows the unit cell of the periodic reference structure, which consists of one circular air hole in a square lattice. Figure \ref{figs4}(b) shows the simulated eigenfunction profile ($\left|H_z\right|^2$ at the center of the silicon slab) at the tip $\mathbf{k}=\Gamma$ of the quadratic band. The eigenfunction is primarily distributed in the corners of the unit cell.

Figure \ref{figs4}(d) shows the dielectric function (upper panel) and the corresponding superpotential (lower panel) of an Airy sample with potential strength $\kappa=0.2a^{-1}$, which is the same structure as in Fig. \ref{figs2}. We then numerically calculate the eigenstate profiles by solving Eq. \eqref{a1} with \textsc{Legume} \cite{Legume}. The eigenfunctions ($\left|H_z\right|^2$ at the center of the slab) of Airy resonances at $k_y=0$ are plotted in the upper panels of Figs. \ref{figs4}(e)(f)(g)(h) with mode index $n=1$, $n=2$, $n=3$ and $n=21$, respectively. The corresponding theoretical envelope functions $\left|\alpha_n\right|^2$ obtained from Eq. \eqref{c22} are plotted in the lower panels. Here, the normalization factor $D_n$ in Eq. \eqref{c22} is omitted, and the $\left|\alpha_n\right|^2$ is normalized to have a maximum value of $1$, instead. We find good quantitative agreement between the simulation and the theory. Moreover, the eigenfunction is indeed the product of the envelope function ($\alpha_n$) in the lower panel times the wavefunction in the reference structure ($\mathbf{H_{k_0}}$) in Fig. \ref{figs4}(b), as predicted in Eq. \eqref{b23}. In all the theory plots, only the real part of the effective mass $\frac{1}{2m}=0.651$ is considered, and the imaginary part of the effective mass is ignored. The effects of the imaginary part of the effective mass on the Airy mode spatial profiles are examined later in Section 6. All the following derivations in this section regarding the eigenfunction profile of Airy resonances consider a real effective mass. From Eq. \eqref{c22}, the eigenfunction of the Airy resonances has a truncation at $x=0$, which means that the $n^{\mathrm{th}}$ Airy mode keeps the first $n$ lobes in the Airy function in its wavefunction. In other words, the $n=1$ Airy mode has one lobe; the $n=2$ state has two lobes, etc. When $n\to\infty$, we approach the continuous limit that the eigenfunction $\alpha_n$ converges to a complete Airy function, and the Airy mode frequency separation $\omega_{n+1}-\omega_n$ tends to zero.

In the experiment, we are not able to directly image the eigenstates of the Airy resonances for the reasons we explain later in Section 6,. However, the main lobe (the brightest lobe) of the wavefunction is clearly visible in the experimental data at all probing frequencies. We then seek a way to quantify the position of the main lobe as a function of the probing frequency.

We define $x_{\mathrm{max}}$ as the position of the main lobe in the $x$ direction:
\begin{equation}
x_{\mathrm{max}}(n)\equiv \mathrm{arg}\ \max_x\ \left|\alpha_n(\mathbf{r})\right|^2
,
\label{c27}
\end{equation}
where $x_{\mathrm{max}}(n)$ is the position of the main lobe for the $n^{\mathrm{th}}$ Airy resonance, such that $\left|\alpha_n\right|^2$ reaches the maximum value at $x=x_{\mathrm{max}}(n)$.

From the property of the Airy function \cite{vallee2010airy}, the Airy function $\mathrm{Ai}(x)$ reaches its maximum value at $x=A_0\approx -1.019$. Hence, according to Eq. \eqref{c22}:
\begin{equation}
x_{\mathrm{max}}(n)
=\left(\frac{1}{2m}\right)^{\frac{1}{3}}\kappa^{-1}\left(\left[\frac{3}{2}\pi\left(n-\frac{1}{4}\right)\right]^{\frac{2}{3}}+A_0\right)
.
\label{c28}
\end{equation}

According to Eq. \eqref{c23}, we get:
\begin{equation}
x_{\mathrm{max}}(n)
=-\frac{2\omega_\mathrm{ref}}{c^2\kappa^3}\omega_{n,k_y}+x_\mathrm{offset}
,
\label{c29}
\end{equation}
where $x_\mathrm{offset}
=\frac{2\omega_\mathrm{ref}^2}{c^2\kappa^3}
-\frac{k_y^2}{2m\cdot c^2\kappa^3}+\left(\frac{1}{2m}\right)^{\frac{1}{3}}\kappa^{-1}A_0$ is an offset in $x$. When $n$ is large, we approach the continuous limit that the frequency separation between states converges to zero. We can then re-write Eq. \eqref{c29} in terms of the excitation frequency:
\begin{equation}
x_{\mathrm{max}}\left(\omega\right)
=-\frac{2\omega_\mathrm{ref}}{c^2\kappa^3}\omega+x_\mathrm{offset}
,
\label{c30}
\end{equation}
where $\omega$ is the excitation frequency. Equation \eqref{c30} is the same as Eq. (7) in the main text.

Figure \ref{figs4}(c) shows the comparison between the simulation and the theory on the position of the main lobe. The theory line is calculated from Eq. \eqref{c30} at $k_y=0$. Each dot in the simulation is obtained from the peak position of $\left|\alpha_n\right|^2$ with the first $n=42$ Airy modes. The simulation matches well with the theory.\\

In addition to the Airy mode frequency (and linewidth) in Eqs. \eqref{c23} and \eqref{c24} and the Airy eigenstate profile in Eq. \eqref{c22}, the third signature of Airy resonances is the bending of light in free space which radiates out from Airy modes.

It is well known that an Airy beam can effectively accelerate in free space \cite{siviloglou2007observation}, which means that the position of the main lobe of an Airy beam $x_{\mathrm{max}}(z)$ (as well as other lobes) bends as a quadratic function of the propagation distance $z$ while propagating in free space. Next, we provide detailed derivations showing that the Airy resonances have a similar property that the trajectory of the out-of-plane radiation also bends as a quadratic function in free space.

After the Airy resonance is excited by an external light field, it effectively becomes a light source and radiates light to free space through out-of-plane radiation. The initial spatial profile of the Airy resonance is a slow-varying Airy function envelope $\alpha_n(\kappa\mathbf{r})$ times a fast-varying Bloch wavefunction $\mathbf{H_{k_0}}$, as described in Eq. \eqref{b23}. We then focus on how the spatial profile of this radiation changes as it propagates in free space.

When propagating in free space, the Maxwell's equation in Eq. \eqref{a1} reduces to a paraxial equation \cite{PhysRevA.11.1365}:
\begin{equation}
\frac{\partial^2 \mathbf{H_0}}{\partial x^2}
+\frac{\partial^2 \mathbf{H_0}}{\partial y^2}
+2i\frac{\omega}{c}\frac{\partial \mathbf{H_0}}{\partial z}=0
,
\label{c31}
\end{equation}
where $\mathbf{H_0}(x,y,z)=\mathbf{H}(x,y,z)e^{-i\frac{\omega}{c}z}$ is the magnetic field envelope (in $z$) which is assumed to vary slowly in the $z$ direction, and $\mathbf{H}(x,y,z)$ is the magnetic field profile in the Maxwell's equations in Eq. \eqref{a1}.

For simplicity, we assume that the excitation light is a plane wave with a conserved $k_y$, and its frequency $\omega$ is exactly the eigen-frequency of the $n^\mathrm{th}$ Airy mode $\omega=\omega_n$. At this point, we ignore the finite Q-factor of any other Airy modes, so only the $n^\mathrm{th}$ Airy mode at $k_y$ is excited. The effect of the finite Q-factor (arising from the complex effective mass) of other modes is examined in detail later in Section 6. The initial condition of $\mathbf{H_0} (x,y,z)$ at $z=0$ is described in Eq. \eqref{b23}:
\begin{equation}
\left.\mathbf{H_0}(x,y,z)\right|_{z=0}=\alpha_{n,k_y}(x)\cdot\left.\mathbf{H_{k_0}}(x,y,z)\right|_{z=0}
,
\label{c32}
\end{equation}
where the $z=0$ surface is defined as the interface between the slab and the air, and $\mathbf{H_{k_0}}$ is the magnetic eigenstate of the Bloch mode at the band tip $\mathbf{k}=\mathbf{k_0}$ in the reference structure.
\begin{figure}[H]
    \centering
    \subfigure[]{\includegraphics[width=0.45\textwidth]{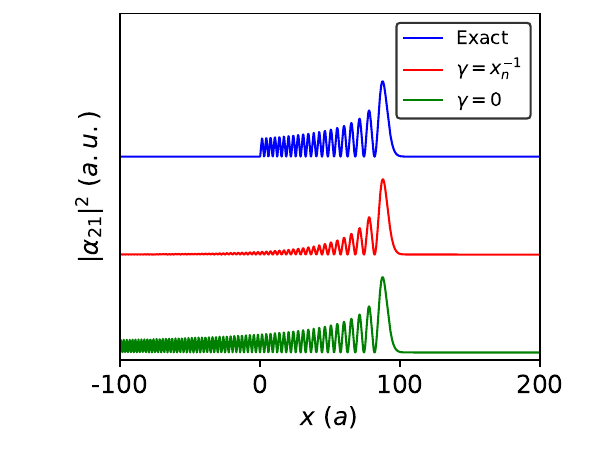}}
    \subfigure[]{\includegraphics[width=0.45\textwidth]{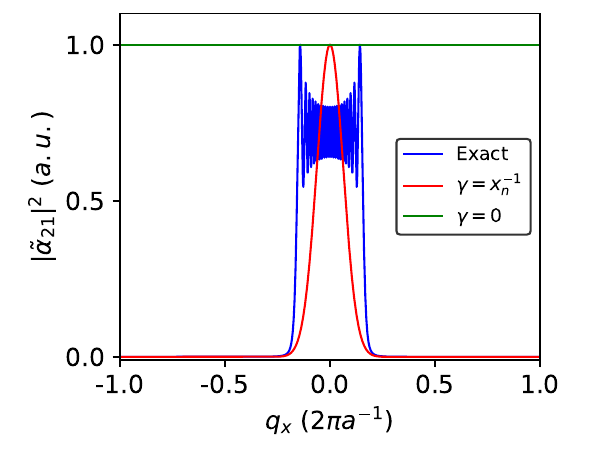}}
    \subfigure[]{\includegraphics[width=0.45\textwidth]{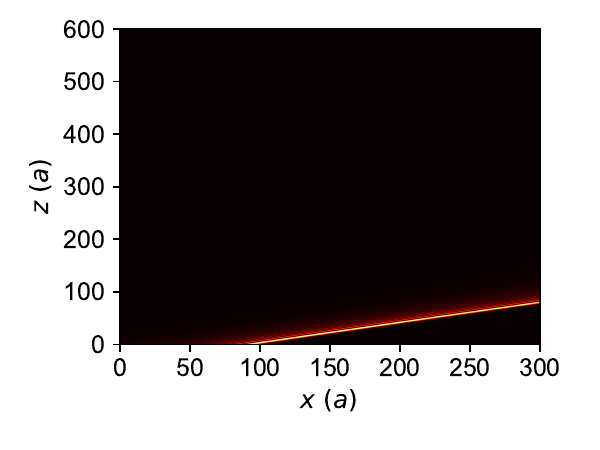}}
    \subfigure[]{\includegraphics[width=0.45\textwidth]{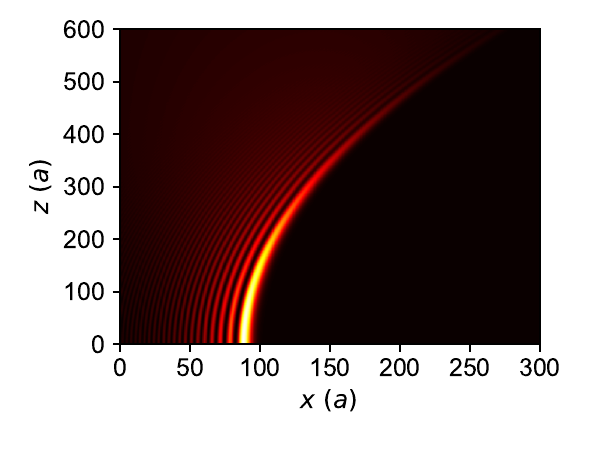}}

\caption{The third signature of Airy resonances: the bending of the Airy mode radiation in free space. (a) The envelope function of the $n=21$ Airy eigenfunction $\left|\alpha_{21}\right|^2$. The blue, red and green lines represent the exact eigenfunction in \eqref{c22} with a sudden truncation at $x=0$, the approximation with an exponential decay in Eq. \eqref{c33} with decay parameter $\gamma=\frac{1}{x_n}$, and the complete Airy function in Eq. \eqref{c22} without the truncation at $x=0$ (which is equivalent to setting $\gamma=0$), respectively. (b) The Fourier component of the envelope function at $q_y=k_y$. Each line represents the Fourier components $\left|\tilde{\alpha}_{21}\right|^2$ calculated from the corresponding eigenfunction in (a). The effective mass is $\frac{1}{2m}=0.651$ and the superpotential strength is $\kappa=0.2a^{-1}$. (c) The intensity of $\left|\mathbf{H_0}(\mathbf{G};x,y,z)\right|^2$ for reciprocal lattice vector $\mathbf{G}=\left(1,0\right)\frac{2\pi}{a}$ from the $n=21$ Airy resonance. The propagation direction has a large angle with respect to the $z$ axis, which indicates the paraxial equation no longer holds. (d) The intensity $\left|\mathbf{H_0}(x,y,z)\right|^2$ of the radiation from $n=21$ Airy resonance as it propagates along the $z$ direction. The parameters of the structure are the same as in Fig. \ref{figs4}(d).}
\label{figs5}
\end{figure}

We notice in Eq. \eqref{c22} that the envelope function $\alpha_{n,k_y}(x)$ has a sudden truncation at $x=0$, which is shown in the blue curve in Fig. \ref{figs5}(a). The sudden truncation makes it difficult to analytically calculate the Fourier components of $\alpha_{n,k_y}(x)$. In the following derivation, to eliminate the sudden truncation at $x=0$, we mathematically introduce an extra exponential decay on top of the Airy function to mitigate this effect:
\begin{equation}
\alpha^{\prime}_{n,k_y}(x,y)
=e^{\gamma \left(x-x_n\right)}\cdot
e^{ik_y y}\cdot
\mathrm{Ai}\left[\left(\frac{1}{2m}\right)^{-\frac{1}{3}}\kappa(x-x_n)\right]
,
\label{c33}
\end{equation}
where $\gamma$ is the factor that characterizes how fast the envelope function decays. Since the sudden truncation occurs at $x=0$ and the main lobe position of the Airy resonance is at $x=x_{\mathrm{max}}(n)$ in Eq. \eqref{c28}, a reasonable choice of the parameter $\gamma$ should be:
\begin{equation}
\gamma=\frac{1}{x_{\mathrm{max}}(n)}\approx
\frac{1}{x_n}
=\left(\frac{1}{2m}\right)^{-\frac{1}{3}}\kappa\left[\frac{3}{2}\pi\left(n-\frac{1}{4}\right)\right]^{-\frac{2}{3}}
.
\label{c33-2}
\end{equation}

Here, we already make the assumption that $n\gg 1$, so the constant $A_0$ in Eq. \eqref{c28} can be ignored. The red curve in Fig. \ref{figs5}(a) shows the approximated eigenfunction in Eq. \eqref{c33} with the decaying parameter from Eq. \eqref{c33-2}. Compared with the complete Airy function as shown in the green curve, the approximation captures the main feature of the truncation at $x=0$ in a continuous way.

Figure \ref{figs5}(b) shows the Fourier component of the envelope function of the $n=21$ Airy resonance. The blue curve is calculated from the exact solution in Eq. \eqref{c22} with a sudden truncation at $x=0$. Compared with the green curve where $\alpha_n$ is a complete Airy function, the truncation at $x=0$ drastically changes the magnitude of the Fourier components. This change is largely captured by our approximation in Eqs. \eqref{c33} and \eqref{c33-2}; the red curve in Fig. \ref{figs5}(b) has approximately the same window size as the exact solution, but with a smoother and more continuous truncation. This approximation allows us to obtain an analytic result.

In this section, the exponential decay approximation in Eq. \eqref{c33} is just a mathematical trick to take the wavefunction truncation at $x=0$ into consideration and make the Fourier components localize in $q_x$. However, in Section 6 we show that this approximation is an experimental reality that arises from the superposition of multiple Airy states induced by the complex effective mass. More details can be found in Section 6.

Next, we use the angular spectrum method \cite{goodman2005introduction} to analytically solve Eq. \eqref{c31} with the initial condition in Eqs. \eqref{c32} and \eqref{c33}. Notice the fact that:
\begin{equation}
\mathrm{Ai}(x)=\frac{1}{2\pi}\int_{-\infty}^{\infty} e^{ikx}e^{i\frac{1}{3}k^3}dk
,
\quad x\in\mathbb{C}
.
\label{c34}
\end{equation}

We get:
\begin{equation}
\begin{aligned}
\tilde{\alpha^{\prime}}_{n,k_y}\left(q_x,q_y\right)
&=\frac{1}{2\pi}\int_{-\infty}^{\infty}\int_{-\infty}^{\infty} dx\ dy\ e^{-i q_x x}e^{-iq_y y}\cdot\alpha^{\prime}_{n,k_y}(x)\\
&=
\left(\frac{1}{2m\kappa^3}\right)^\frac{1}{3}
e^{-i q_x x_n}
e^{i\frac{1}{3}\frac{1}{2m\kappa^3}\left(q_x+i\gamma\right)^3}\delta\left(q_y-k_y\right)
,
\label{c35}
\end{aligned}
\end{equation}
where $\tilde{\alpha^{\prime}}_{n,k_y}\left(q_x,q_y\right)$ is the 2-D Fourier component of $\alpha^{\prime}_{n,k_y}(x,y)$ at $\mathbf{q}=\left(q_x,q_y\right)$.

According to the Bloch theorem, the 2-D Fourier components of $\mathbf{H_{k_0}}(x,y,z)$ are:
\begin{equation}
\begin{aligned}
\left.\tilde{\mathbf{H}}_\mathbf{k_0}\left(q_x,q_y,z\right)\right|_{z=0}
&=\frac{1}{2\pi}\int_{-\infty}^{\infty}\int_{-\infty}^{\infty} dx\ dy\ e^{-iq_x x}e^{-iq_y y}\cdot\left.\mathbf{H_{k_0}}(x,y,z)\right|_{z=0}\\
&=
2\pi\sum_{\mathbf{G}} \mathbf{f(G)}\delta(q_x-G_x)\delta(q_y-G_y)
,
\label{c36}
\end{aligned}
\end{equation}
where $\mathbf{G}$ is the reciprocal lattice vector satisfying $\left(G_x,G_y\right)=(n_x,n_y)\frac{2\pi}{a}$ with $n_x, n_y=0,\pm1,\pm2, \cdots$, and 
\begin{equation}
\mathbf{f(G)}
=\frac{1}{a^2}\iint_{\mathrm{unit\ cell}}dx\ dy\ 
e^{-iG_xx}\cdot e^{-iG_yy} \cdot
\left.\mathbf{H_{k_0}}(x,y,z)\right|_{z=0}
\label{c36-2}
\end{equation}
are the Fourier coefficients. Here, we already use the fact that our quadratic band is centered at $\Gamma$, which means $\mathbf{k_0}=(0,0)$.

Therefore, according to the convolution theorem of Fourier transforms:
\begin{equation}
\begin{aligned}
\left.\tilde{\mathbf{H}}_\mathbf{0}\left(q_x,q_y,z\right)\right|_{z=0}
&=\frac{1}{2\pi}\int_{-\infty}^{\infty}\int_{-\infty}^{\infty} d\tau_x\ d\tau_y\ 
\tilde{\alpha^{\prime}}_{n,k_y}(\tau_x,\tau_y)
\cdot\left.\tilde{\mathbf{H}}_\mathbf{k_0}(q_x-\tau_x,q_y-\tau_y,z)\right|_{z=0}\\
&=\left(\frac{1}{2m\kappa^3}\right)^{\frac{1}{3}}
\sum_{\mathbf{G}} \mathbf{f(G)}
e^{-i\left(q_x-G_x\right)x_n}
e^{i\frac{1}{3}\frac{1}{2m\kappa^3}\left(q_x-G_x+i\gamma\right)^3}
\delta\left(q_y-G_y-k_y\right)
.
\label{c37}
\end{aligned}
\end{equation}

We can see from Eq. \eqref{c37} that the initial condition of the magnetic field $\left.\tilde{\mathbf{H}}_\mathbf{0}\left(q_x,q_y,z\right)\right|_{z=0}$ contains a summation over all the reciprocal lattice vectors $\mathbf{G}$. The magnitude of the contribution from each component $\mathbf{G}$ is proportional to:
\begin{equation}
\left.\tilde{\mathbf{H}}_\mathbf{0}\left(\mathbf{G};q_x,q_y,z,\right)\right|_{z=0}
\propto
e^{-\frac{\gamma}{2m\kappa^3}\left(q_x-G_x\right)^2}
\cdot\delta\left(q_y-G_y-k_y\right)
.
\label{c37-2}
\end{equation}

From Eq. \eqref{c37-2}, we can see that the Fourier component $\left.\tilde{\mathbf{H}}_\mathbf{0}\left(\mathbf{G};q_x,q_y,z,\right)\right|_{z=0}$ is a delta function in the $q_y$ direction located at $q_y=G_y+k_y$. This is a direct consequence of the preserved periodicity along the $y$ direction. Due to the fact that $\frac{\omega_n}{c}<\frac{2\pi}{a}$ and $k_y$ in our experiment is a small number, only the component with $G_y=0$ can leak out of the sample and radiate in the $z$ direction. The contribution from other $G_y=n_y\frac{2\pi}{a}$ with $n_y\neq 0$ will stay inside the sample and decay exponentially in the $z$ direction. Therefore, when calculating the radiation field $\mathbf{H_0}(x,y,z)$, we should only consider the contribution from $G_y=0$ and filter out all the other components of $G_y$.

As for the Fourier component $\left.\tilde{\mathbf{H}}_\mathbf{0}\left(\mathbf{G};q_x,q_y,z,\right)\right|_{z=0}$ in the $q_x$ direction, it is no longer a delta function because the periodicity along the $x$ direction is broken by the superpotential. Instead, it is proportional to $e^{-\frac{\gamma}{2m\kappa^3}\left(q_x-G_x\right)^2}$, which is still a localized function centered at $q_x=G_x$, as shown by the red curve in Fig. \ref{figs5}(b) for the $G_x=0$ case. By using the same argument as in the $q_y$ direction (except this time with $G_x=0$), only the peak centered at $q_x=G_x=0$ can leak out the photonic crystal slab and the components from other $G_x=n_x\frac{2\pi}{a}$ with $n_x\neq 0$ do not contribute to the radiation.

As a result, only the $\mathbf{G}=(0,0)$ component in Eq. \eqref{c37} has non-zero contribution to the radiation. To further illustrate this point, we still keep all the components of $\mathbf{G}$ in the summation and show later in Eq. \eqref{c40} that only the $\mathbf{G}=(0,0)$ component satisfies the paraxial approximation.

In the Fourier space, the paraxial equation in Eq. \eqref{c31} becomes:
\begin{equation}
-\left(q_x^2+q_y^2\right)
\tilde{\mathbf{H}}_\mathbf{0}\left(q_x,q_y,z\right)
+2i\frac{\omega_n}{c}\frac{\partial \tilde{\mathbf{H}}_\mathbf{0}\left(q_x,q_y,z\right)}{\partial z}
=0
.
\label{c38}
\end{equation}

The solution to Eq. \eqref{c38} is:
\begin{equation}
\begin{aligned}
\tilde{\mathbf{H}}_\mathbf{0}\left(q_x,q_y,z\right)
&=e^{-i\frac{c}{2\omega_n}\left(q_x^2+q_y^2\right)\cdot z}
\cdot
\left.\tilde{\mathbf{H}}_\mathbf{0}\left(q_x,q_y,z\right)\right|_{z=0}\\
&=\left(\frac{1}{2m\kappa^3}\right)^\frac{1}{3}
e^{-i\frac{c}{2\omega_n}\left(q_x^2+q_y^2\right)\cdot z}
\sum_{\mathbf{G}} \mathbf{f(G)}
e^{-i\left(q_x-G_x\right)x_n}
e^{i\frac{1}{3}\frac{1}{2m\kappa^3}\left(q_x-G_x+i\gamma\right)^3}
\delta\left(q_y-G_y-k_y\right)
.
\label{c39}
\end{aligned}
\end{equation}

Hence,
\begin{equation}
\begin{aligned}
\mathbf{H_0}(x,y,z)
&=\frac{1}{2\pi}\int_{-\infty}^{\infty}\int_{-\infty}^{\infty} dq_x\ dq_y\ \tilde{\mathbf{H}}_\mathbf{0}\left(q_x,q_y,z\right)
\\&= 
\sum_{\mathbf{G}} \mathbf{f(G)}\cdot 
e^{-\frac{mc^2\kappa^3}{\omega_n^2}\gamma z^2}
\cdot e^{\gamma \left(x-x_n\right)}
\cdot
\mathrm{Ai}\left[\left(\frac{1}{2m}\right)^{-\frac{1}{3}}\kappa\left(x-x_n-\frac{c}{\omega_n}G_x z+i\frac{c}{\omega_n}\gamma z-\frac{mc^2\kappa^3}{2\omega_n^2}z^2\right)\right]\\
&\quad\times 
e^{ik_yy}
\cdot e^{i\left(G_x x+G_y y\right)}
\cdot e^{-i\frac{c}{2\omega_n}\left[\left(G_x-i\gamma\right)^2+\left(G_y+k_y\right)^2\right]z}
\cdot e^{i\frac{mc\kappa^3}{\omega_n}z\left(x-x_n-\frac{c}{\omega_n}G_x z-\frac{mc^2\kappa^3}{3\omega_n^2}z^2\right)}
.
\label{c40}
\end{aligned}
\end{equation}

Similar to Eq. \eqref{c37-2}, the magnetic field $\mathbf{H_0}(x,y,z)$ contains a summation over all the reciprocal lattice vectors $\mathbf{G}$. In the $y$ direction, $\mathbf{H_0}(\mathbf{G};x,y,z)\propto e^{i\left(k_y+G_y\right)y}$, where $G_y=n_y\frac{2\pi}{a}$. When $n_y\neq 0$, $\left|k_y+G_y\right|>\frac{\omega_n}{c}$, which does not satisfy the paraxial approximation condition $\left|k_y+G_y\right|\ll\frac{\omega_n}{c}$, so they do not contribute to the radiation near $(x,y)=(0,0)$. In the $x$ direction, other $G_x=n_x\frac{2\pi}{a}$ (besides the $G_x=0$ component) with $n_x\neq 0$ components all propagate in a large angle with respect to the $z$ axis. Figure \ref{figs5}(c) shows the light propagation for the $\mathbf{G}=(1,0)\frac{2\pi}{a}$ component. The main lobe follows a line that has a large angle with respect to the $z$ axis, which indicates that the paraxial approximation in Eq. \eqref{c31} no longer holds, so it has a minor influence on the light field distribution when measuring nearly along the $z$ direction.

When we only consider the contribution from the $\mathbf{G}=(0,0)$ term, Eq. \eqref{c40} becomes:
\begin{equation}
\begin{aligned}
\mathbf{H_0}(x,y,z)
&= 
\left.\mathbf{f(G)}\right|_{\mathbf{G}=(0,0)}\cdot 
e^{-\frac{mc^2\kappa^3}{\omega_n^2}\gamma z^2}
\cdot e^{\gamma \left(x-x_n\right)}
\cdot
\mathrm{Ai}\left[\left(\frac{1}{2m}\right)^{-\frac{1}{3}}\kappa\left(x-x_n+i\frac{c}{\omega_n}\gamma z-\frac{mc^2\kappa^3}{2\omega_n^2}z^2\right)\right]\\
&\quad\times 
e^{ik_yy}
\cdot e^{-i\frac{c}{2\omega_n}\left(k_y^2-\gamma^2\right)z}
\cdot e^{i\frac{mc\kappa^3}{\omega_n}z\left(x-x_n-\frac{mc^2\kappa^3}{3\omega_n^2}z^2\right)}
.
\label{c41}
\end{aligned}
\end{equation}

Hence, the intensity distribution of the light satisfies:
\begin{equation}
\left|\mathbf{H_0}(x,y,z)\right|^2
= 
\left|\left.\mathbf{f(G)}\right|_{\mathbf{G}=(0,0)}\right|^2\cdot 
e^{-\frac{2mc^2\kappa^3}{\omega_n^2}\gamma z^2}
\cdot \left|e^{\gamma \left(x-x_n\right)}
\cdot
\mathrm{Ai}\left[\left(\frac{1}{2m}\right)^{-\frac{1}{3}}\kappa\left(x-x_n+i\frac{c}{\omega_n}\gamma z-\frac{mc^2\kappa^3}{2\omega_n^2}z^2\right)\right]\right|^2
.
\label{c42-0}
\end{equation}

When $n$ is large, we approach the continuous limit where the frequency separation between states converges to zero. Additionally, we consider the case of $k_y=0$, which is the case in our experiment. We can then re-write Eq. \eqref{c42-0} in terms of the excitation frequency $\omega$ with the information in Eqs. \eqref{c19} and \eqref{c25}:
\begin{equation}
\scalebox{1.00}{$\displaystyle
\left|\mathbf{H_0}\right|^2
= 
\left|\left.\mathbf{f(G)}\right|_{\mathbf{G}=(0,0)}\right|^2 \cdot
e^{-\frac{2mc^2\kappa^3}{\omega^2}\gamma z^2}
\cdot \left|
e^{\gamma \left(x-x_\omega\right)} \cdot
\mathrm{Ai}\left[\left(\frac{1}{2m}\right)^{-\frac{1}{3}}\kappa\left(x-x_\omega+i\frac{c}{\omega}\gamma z-\frac{mc^2\kappa^3}{2\omega^2}z^2\right)\right]\right|^2
,
\label{c42}
$}
\end{equation}
where 
\begin{equation}
x_\omega=\frac{2\omega_{\mathrm{ref}}\left(\omega_\mathrm{ref}-\omega\right)}{c^2 \kappa^3}.
\label{c42-1}
\end{equation}

Equation \eqref{c42} is equivalent to Eq. (9) in the main text.

We notice the fact that:
\begin{equation}
\mathrm{Ai}\left(x+iy\right)
=\mathrm{Ai}\left(x\right)
+i\cdot\mathrm{Ai^{\prime}}\left(x\right)\cdot y
+O\left(y^2\right)
.
\label{c43}
\end{equation}

Therefore, the positions of the lobes in $\left|\mathrm{Ai}(x+iy)\right|^2$ are the same as the lobes in $\left|\mathrm{Ai}(x)\right|^2$ when $y\ll 1$. Back in the case of Eq. \eqref{c42}, all the lobes of $\left|\mathbf{H_0}(x,y,z)\right|^2$ follow the path:
\begin{equation}
x\left(z\right)=\frac{mc^2\kappa^3}{2\omega^2}z^2
+\left.x(z)\right|_{z=0}
,
\label{c44}
\end{equation}
when 
\begin{equation}
z\ll
\left(\frac{1}{2m}\right)^{\frac{1}{3}}\kappa^{-1}\frac{\omega}{\gamma c}
=\left(\frac{1}{2m}\right)^{\frac{1}{3}}\frac{2\omega_\mathrm{ref}\left(\omega_\mathrm{ref}-\omega\right)\omega}{c^3\kappa^4}
,
\label{c45}
\end{equation}
where $x(z)$ is the position (in the $x$ direction) of the lobe at propagation distance $z$, and $\left.x(z)\right|_{z=0}$ is the position of the corresponding lobe at $z=0$. Equation \eqref{c44} is the same as Eq. (10) in the main text.

Equation \eqref{c44} shows that the lobes of the radiation from an Airy resonance follow a quadratic trajectory. We define the beam bending coefficient $\beta$ as the second derivative of the lobe position as it propagates in $z$:
\begin{equation}
\beta\equiv
\frac{\partial^2}{\partial z^2}x(z)=\frac{mc^2\kappa^3}{\omega^2}
.
\label{c46}
\end{equation}

Figure \ref{figs5}(d) shows the light propagation for the $\mathbf{G}=(0,0)$ component calculated from Eq. \eqref{c42}. We can clearly see that the main lobe (as well as all other lobes) follows a quadratic path as predicted in Eq. \eqref{c44}. The intensity of the main lobe decreases as the radiation of the Airy resonance propagates along the $z$ direction. This is because of the imaginary part of the argument in the Airy function in Eq. \eqref{c42}. The propagation distance is estimated in Eq. \eqref{c45}. Although the Airy resonances created in our photonic crystal slab have a fundamental difference compared to the Airy beams, that the Airy resonances have both a slow-varying Airy function envelope and a fast-oscillating Bloch mode, the acceleration feature in the Airy resonances is the same as Airy beams. This is because only the reciprocal lattice vector at $\mathbf{G}=(0,0)$ can contribute to the radiation out of the slab. It is equivalent to treating the fast-oscillating Bloch mode $\mathbf{H_{k_0}(r)}=\sum_{\mathbf{G}}\mathbf{f(G)}\cdot e^{i\mathbf{G\cdot r}}$ as $\mathbf{H_{k_0}(r)}=\left.\mathbf{f(G)}\cdot e^{i\mathbf{G\cdot r}}\right|_{\mathbf{G}=(0,0)}=\left.\mathbf{f(G)}\right|_{\mathbf{G}=(0,0)}$, which is a constant independent of position $\mathbf{r}$. Hence, the Airy resonances reduce to the conventional Airy beam without the fast oscillation and have the same acceleration feature.

\clearpage

\section*{Section 3: Fabrication method}
\addcontentsline{toc}{section}{Section 3: Fabrication method}

In this section, we introduce the fabrication methods for the photonic crystal slabs.

\begin{figure}[H]
    \centering
    %\subfigure[]
    {\includegraphics[width=0.96\textwidth]{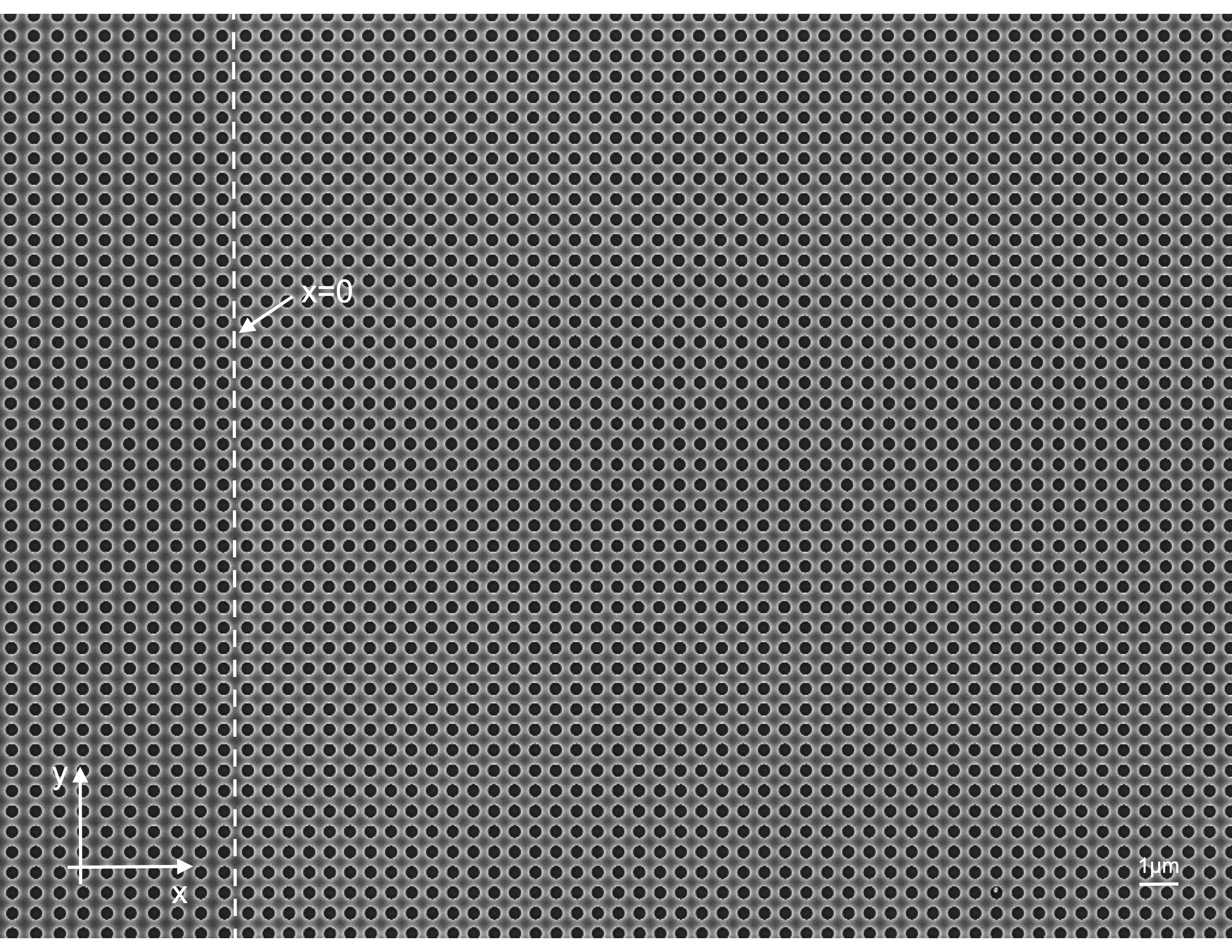}}

\caption{An SEM image of our sample. The strength of the superpotential is $\kappa=0.237a^{-1}$. Only a part of the sample (around $3\%$) is shown in the figure. The total system size of the sample (including the $x<0$ region beyond the potential barrier) is $1mm\times 1mm$ with more than $10^3$  supercells in the $y$ direction. Each supercell contains more than $10^2$ holes in the $x$ direction.}
\label{figs6}
\end{figure}

The photonic crystal samples used in this experiment are fabricated on standard silicon-on-insulator (SOI) wafers from SOITEC. These wafers consist of a $220nm$ thick [110] silicon layer (i.e., the active silicon layer), a buried $2\mu m$ silicon dioxide layer, and a bulk silicon handle wafer. To engrave holes into the silicon layer, we use a standard electron beam lithography (EBL) and inductively coupled plasma etching (RIE) process.

First, a $300nm$ thick layer of ZEON ZEP 520A resist is spin-coated onto the wafer. This resist was chosen for its high resolution, allowing e-beam writing of holes with sub-$10nm$ precision. A thin $20nm$ layer of thermal gold is then deposited on top of the resist using a Lesker Lab-18 thermal evaporator to prevent sample charging during the electron beam lithography step.

The pattern of circular holes is imprinted onto the resist with a Raith EBPG5200 electron beam lithography tool. The gold layer is removed using TFA gold strip, and the sample is submerged in N-amyl acetate to develop the resist. The N-amyl acetate dissolves the areas of the resist exposed by the electron beam during the lithography step.

Next, the holes are etched into the silicon using SF$_6$ and C$_4$F$_8$ plasma gases in a Plasma-Therm Versalock 700 inductively coupled plasma tool. The remaining resist is removed in an O$_2$ plasma. An SEM image of the resulting photonic crystal is shown in Fig. \ref{figs6}. The lattice constant in the $y$ direction is always kept as $a_y=a=629nm$.

\clearpage

\section*{Section 4: Measuring the effective mass}
\addcontentsline{toc}{section}{Section 4: Measuring the effective mass}

In this section, we introduce the experimental setup that characterizes the photonic modes of the sample by measuring its real-space image at various angles and frequencies. Then, we present the method for determining the effective mass of an isolated quadratic band by extracting the resonance frequency and linewidth from the reflection intensity spectrum obtained in the experiment.

\begin{figure}[H]
    \centering
    \subfigure[]{\includegraphics[width=0.32\textwidth]{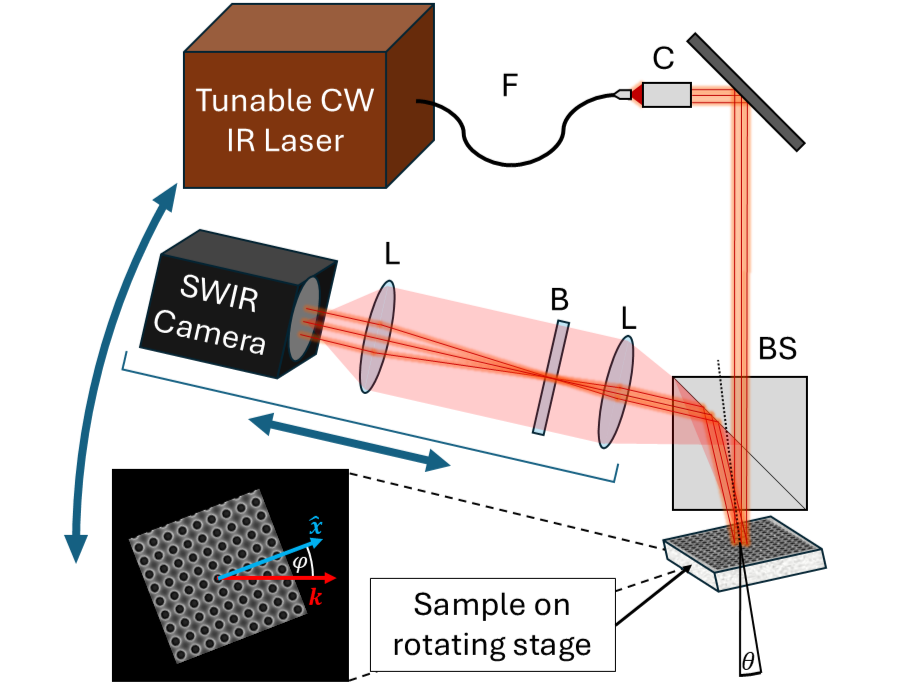}}
    \subfigure[]{\includegraphics[width=0.32\textwidth]{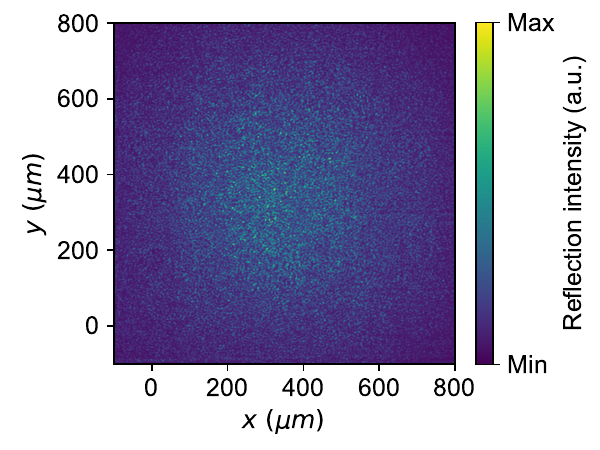}}
    \subfigure[]{\includegraphics[width=0.32\textwidth]{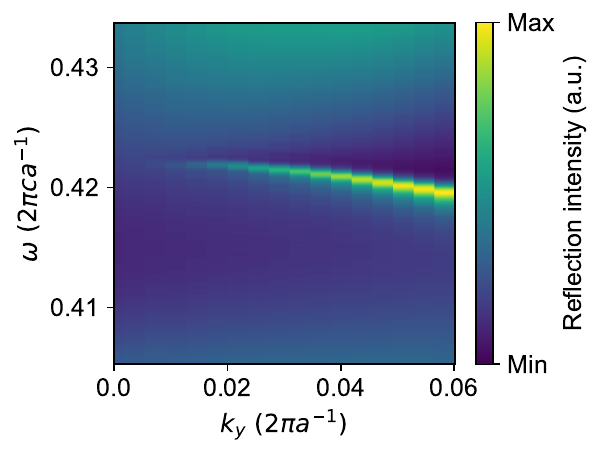}}
    \subfigure[]{\includegraphics[width=0.32\textwidth]{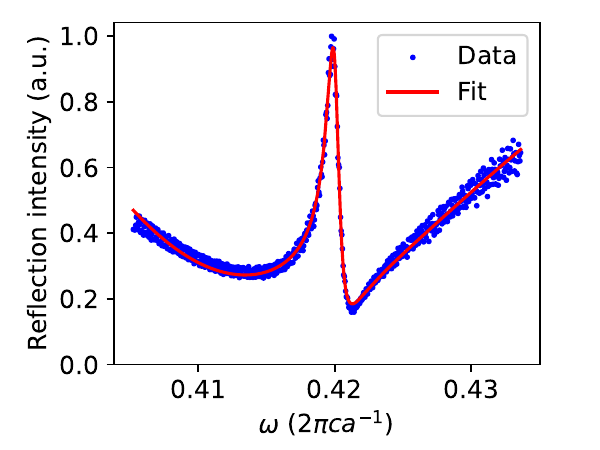}}
    \subfigure[]{\includegraphics[width=0.32\textwidth]{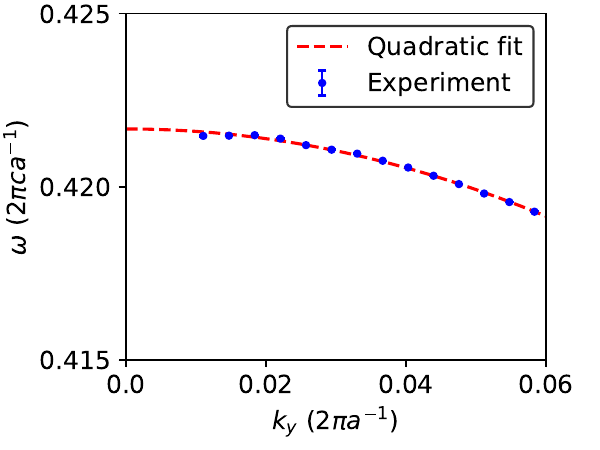}}
    \subfigure[]{\includegraphics[width=0.32\textwidth]{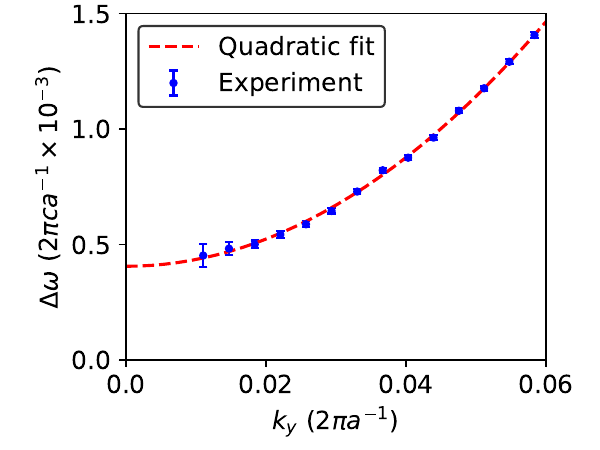}}

\caption{The experimental setup and observables. (a) The experimental setup (variable angle reflection spectroscopy) used to characterize the modes in a photonic crystal slab. The notations in the figure are: F - Fiber optic cable; C - Beam collimator; BS - Beamsplitter; B - momentum filter beam blocker; L - Lenses. On the black inset, the red arrow indicates the direction of the in-plane wavevector. The angle $\varphi$ is defined as the angle between the in-plane wavevector and the $x$-axis of the sample. (b) The experimentally measured real-space image (at $k_x=0$, $k_y=0.0475\ (2\pi a^{-1})$ and $\omega=0.425\ (2\pi c a^{-1})$) obtained from the setup in (a). The sample is periodic with no superpotential. The color shows the reflection intensity. (c) The experimental measured reflection spectrum when sweeping $k_y$ while fixing $k_x=0$. The reflection intensity is obtained from averaging the intensity over all the pixels in (b). (d) The reflection spectrum at $k_x=0$ and $k_y=0.0475\ (2\pi a^{-1})$ measured in (c). The data is fit with a curve containing eight fitting parameters to extract the resonance frequency $\omega_0$ and the linewidth $\Delta \omega$: $\omega_0=\left(0.420\pm5.49\times 10^{-6}\right)\ (2\pi c a^{-1})$, and $\Delta\omega=\left(1.08\times 10^{-3}\pm 1.07\times 10^{-5}\right)\ (2\pi c a^{-1})$. (e) The resonance frequency $\omega_0$ extracted from (c) (blue dots). The resonance frequency is fit with a quadratic function of $k_y$ to obtain the real part of effective mass $\mathrm{Re}(\frac{1}{2m})$ and band tip energy $\mathrm{Re}\left(E_0\right)$. (f) The linewidth $\Delta\omega$ extracted from (c) (blue dots). The linewidth is fit with a quadratic function of $k_y$ to obtain the imaginary part of effective mass $\mathrm{Im}(\frac{1}{2m})$.}
\label{figs7}
\end{figure}

The effective mass of the periodic sample can be determined by measuring the frequency and the linewidth of the isolated quadratic band as a function of the wavevector $\mathbf{k}$. To observe the isolated quadratic band, a ``variable angle spectroscopy" setup shown in Fig. \ref{figs7}(a) was used to measure the real-space intensity distribution of light at the surface of the sample. In this setup, s-polarized light from a tunable CW laser (Keysight Agilent $1450nm-1650nm$) is directed at the sample in a collimated beam so that the incident beam contains light with a narrow range of $k_x$ and $k_y$ ($\Delta k_x\approx\Delta k_y<10^{-3}\ (2\pi a^{-1})$). The beam diameter at the sample plane is approximately equal to the width of our samples ($1mm$). The angle of the incident light relative to the sample ($\theta$ in Fig. \ref{figs7}(a)) is controlled by rotating the sample with a motorized stage, allowing us to control the magnitude of the in-plane wavevector of the light that hits the sample. The direction of the in-plane wavevector can be controlled by rotating the sample with an angle $\varphi$ about its center in the x-y plane. Then the in-plane wavevector of the light is 

\begin{equation}
    \mathbf{k} = \frac{\omega}{c}\sin(\theta)[\cos(\varphi)\mathbf{\hat x} + \sin(\varphi)\mathbf{\hat y}]
    ,
\end{equation}
where $\omega = 2\pi c /\lambda$ is the frequency of the incident beam ($\lambda$ is the wavelength of the laser).

Two lenses focus light from the surface of the sample onto the detector of an SWIR camera, which captures a real-space image of the sample's surface. The lenses are aligned so that at a specific sample stage angle $\theta_0$, light reflected off the sample travels through the center of the lenses. However, the two lenses are large enough to capture light when the stage angle is $\theta_0+\Delta\theta$, so the camera captures the real-space image of the sample for any stage angle $\theta$ in the range $[\theta_0-\Delta\theta, \theta_0 + \Delta\theta]$. In Fig. \ref{figs7}(a), the range of angles that the camera can capture light from is represented by the light red region. The lenses and camera are on a manual rotation stage so that the lens alignment angle $\theta_0$ can be set arbitrarily. 

In measurements described in future sections, a blocker indicated by a ``B" in Fig. \ref{figs7}(a) was placed at the back-focal plane to filter out certain momentum components of the light at the surface of the sample. No blocker was used for the measurements to determine the effective mass of the sample. Future sections also describe measurements where the camera and lenses were moved along the laser beam axis to move the object plane from the sample surface to above the sample surface. The object plane remained at the sample surface in the measurements used to determine the effective mass.  

We emphasize that the real-space imaging setup in Fig. \ref{figs7}(a) does not directly measure the electromagnetic near-field distribution; rather, it only collects and measures the light intensity distribution that radiates away from the sample at the sample surface.

To measure the isolated quadratic band, the laser frequency $\omega$ and stage angle $\theta$ were swept while $\varphi=\pi/2$ was kept constant (effectively sweeping $k_y$ with $k_x=0$), and an image of the sample surface was recorded at each frequency and stage angle. Figure \ref{figs7}(b) shows an example real-space image of our sample taken at $k_y=0.0475 \ (2\pi a^{-1})$ and $\omega =  0.425 \ (2\pi c a^{-1})$. The size of the unit cell ($a=629nm$) is smaller than half of the wavelength of light used to probe the samples ($\lambda/2\approx740nm$), so we cannot directly observe the spatial profiles of the resonant mode on the scale of the unit cell due to the diffraction limit. The average intensity of light over all camera pixels was taken to obtain the total intensity of light radiated from the surface of the sample as a function of $k_y$ and $\omega$. Figure \ref{figs7}(c) displays an image of the isolated quadratic band obtained from this measurement. 

Figure \ref{figs7}(d) shows how the resonance frequency and linewidth are quantitatively extracted from the experimental data. We first fix the value of $k_y$ in the experimental data and obtain a reflection spectrum as shown with blue dots. Next, the spectrum data are fit with a function containing eight parameters \cite{fan2002analysis}:
\begin{equation}
R(\omega)=F_0\left|r_b\left(\omega,c_0,c_1,c_2\right)+Ae^{i\phi}\frac{\frac{\Delta \omega}{2}}{i\left(\omega-\omega_0\right)+\frac{\Delta\omega}{2}}\right|^2,
\label{d1}
\end{equation}
where $R(\omega)$ is the reflection intensity at frequency $\omega$ measured in the experiment, $F_0$ is a factor that arises from the fact that our reflection intensity is measured in arbitrary units, $A$ and $\phi$ are the amplitude and phase of the photonic mode resonance, and $\omega_0$ and $\Delta\omega$ represent the center position and the linewidth (both in the frequency domain) of a photonic mode. $r_b$ is the background reflectivity ($R_b=\left|r_b\right|^2$) with three additional free parameters $c_0$, $c_1$ and $c_2$, which are calculated from the reflectivity of an air ($n_1=1.0$) -- silicon ($n(\omega)=c_0+c_1\omega+c_2\omega^2$) -- silica($n_2=1.5$) system by the transfer matrix method \cite{transfer_matrix} where the silicon layer is $h=0.35a$ thick:
\begin{equation}
r_b=\frac{r_1e^{-i\delta}+r_2e^{i\delta}}{e^{-i\delta}+r_1r_2e^{i\delta}},
\label{d2}
\end{equation}
where $r_1=\frac{\sqrt{\omega^2c^2n_1^2-k_y^2}-\sqrt{\omega^2c^2n^2-k_y^2}}{\sqrt{\omega^2c^2n_1^2-k_y^2}+\sqrt{\omega^2c^2n^2-k_y^2}}$ is the reflectivity of the air -- silicon interface, $r_2=\frac{\sqrt{\omega^2c^2n^2-k_y^2}-\sqrt{\omega^2c^2n_2^2-k_y^2}}{\sqrt{\omega^2c^2n^2-k_y^2}+\sqrt{\omega^2c^2n_2^2-k_y^2}}$ is the reflectivity of the silicon -- silica interface, and $\delta=h\cdot\sqrt{\omega^2c^2n^2-k_y^2}$ is the phase difference generated in the silicon layer. Here, only the reflectivity in the s-polarization is considered because only the s-polarization is used in the experiment.

The reason for a frequency-dependent refraction index $n(\omega)=c_0+c_1\omega+c_2\omega^2$ in the silicon layer is because the slab contains patterns (circular holes), so it cannot be treated as a uniform medium with constant $n$ at all frequencies.

Then, the experimental reflection spectrum $R(\omega)$ is fit with the following eight parameters: $F_0$, $c_0$, $c_1$, $c_2$, $A$, $\phi$, $\omega_0$, and $\Delta\omega$. As shown in Fig. \ref{figs7}(d), our fitting results match the experimental data perfectly. The center frequency and the linewidth of a photonic resonance can be extracted from the fitting results of $\omega_0$ and $\Delta\omega$, and their uncertainties are calculated directly from the uncertainty of the parameters $\omega_0$ and $\Delta\omega$ respectively in the fitting process. For the case in Fig. \ref{figs7}(d), the extracted resonance frequency is $\omega_0=\left(0.420\pm5.49\times 10^{-6}\right)\ (2\pi c a^{-1})$, and the linewidth is $\Delta\omega=\left(1.08\times 10^{-3}\pm 1.07\times 10^{-5}\right)\ (2\pi c a^{-1})$.

From Eqs. \eqref{a1-1}, \eqref{a1-2}, and \eqref{c1}, in the reference structure:
\begin{equation}
\mathrm{Re}\left(\omega_\mathbf{k}\right)
=\omega_{\mathrm{ref}}
-\mathrm{Re}\left(\frac{1}{2m}\right)\cdot \frac{c^2}{2\omega_{\mathrm{ref}}}\left|\mathbf{k}\right|^2
,
\label{d3}
\end{equation}
\begin{equation}
\Delta\omega_\mathbf{k}
=\mathrm{Im}\left(\frac{1}{2m}\right)\cdot
\frac{c^2}{\omega_\mathrm{ref}}\left|\mathbf{k}\right|^2
,
\label{d4}
\end{equation}
where $\omega_{\mathrm{ref}}$ is the tip frequency in the reference structure. Figure \ref{figs7}(e) and Fig. \ref{figs7}(f) show the resonance frequency and linewidth (and their uncertainties) extracted from the experimental data in Fig. \ref{figs7}(c). Some data near $k_y=0$ are missing; due to the BIC at $\Gamma$, the resonance features near $k_y=0$ are too weak to be identified. We can see that both the resonance frequency and the linewidth vary as a function of $O\left(k_y^2\right)$, which is consistent with Eqs. \eqref{d3} and \eqref{d4}. Then, quadratic fits are performed in Fig. \ref{figs7}(e) and Fig. \ref{figs7}(f) to obtain the effective mass and the band tip frequency in the reference structure: $\frac{1}{2m}=0.628+0.124i$, $\omega_\mathrm{ref}=0.422\ (2\pi c a^{-1})$. These band parameters are used to plot the theoretical lines in the main text. We notice that in Fig. \ref{figs7}(f), the linewidth $\Delta\omega$ at $k_y=0$ is not zero but a finite small value. We attribute this non-zero intercept to the absorption loss and fabrication disorder in our sample. 

\clearpage

\section*{Section 5: The first signature of Airy resonances: Airy mode frequency separation and linewidth}
\addcontentsline{toc}{section}{Section 5: The first signature of Airy resonances: Airy mode frequency separation and linewidth}

In this section, we explain the experimental data regarding the first signature of the Airy resonances: the Airy mode frequency separation and linewidth. 

As predicted in Eq. \eqref{c23} and Fig. \ref{figs3}, a single quadratic band in the periodic structure splits into several discrete Airy levels after a linear superpotential is imposed on the system. Each Airy level is quadratically dispersive in $k_y$. 

\begin{figure}[H]
    \centering
    \subfigure[]{\includegraphics[width=0.45\textwidth]{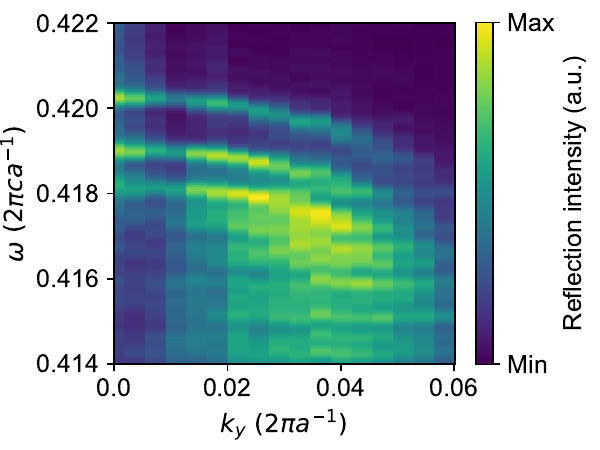}}
    \subfigure[]{\includegraphics[width=0.45\textwidth]{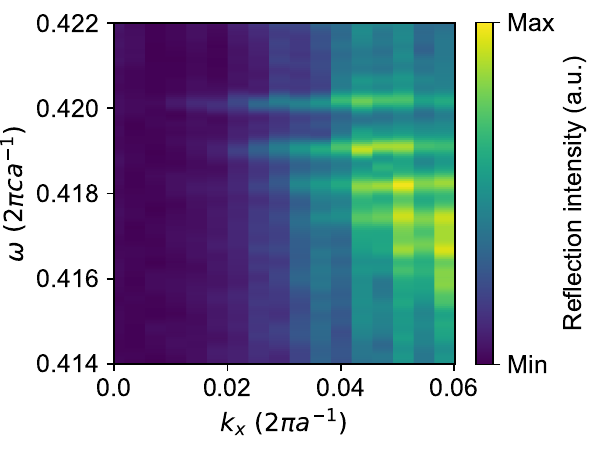}}
    \subfigure[]{\includegraphics[width=0.45\textwidth]{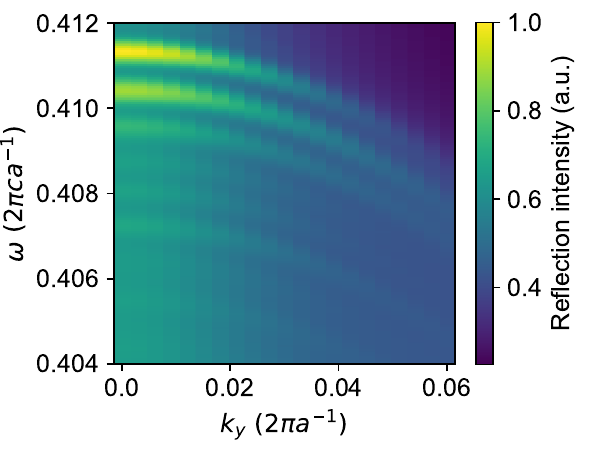}}
    \subfigure[]{\includegraphics[width=0.45\textwidth]{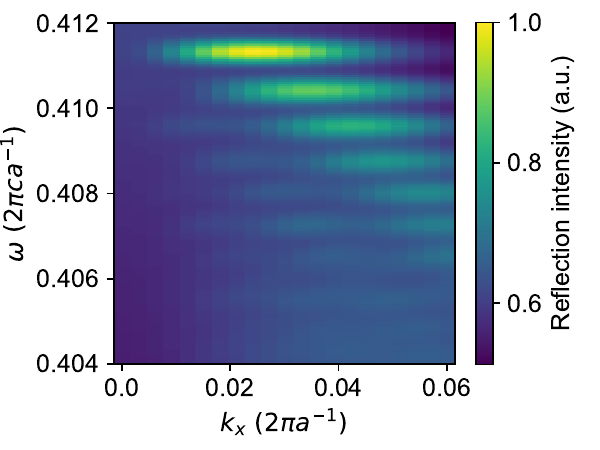}}

\caption{The experimental and simulation results on Airy band structure. The strength of the superpotential is $\kappa=0.152a^{-1}$. (a) The experimental band structure when fixing $k_x=0.03\ (2\pi a^{-1})$ and sweeping $k_y$. (b) The experimental ``band structure" when fixing the wavevector of the incidence beam at $k_y=0$ and sweeping $k_x$. (c) The simulated band structure with the same conditions in (a). (d) The simulated band structure with the same conditions in (b).}
\label{figs8}
\end{figure}

To measure the bands of the Airy samples, the variable angle spectroscopy setup from Fig. \ref{figs7}(a) was used. A momentum space blocker (B in Fig. \ref{figs7}(a)) was used to block the specular reflected light. The light from the Airy resonances emits at a different $k_x$ than the specular reflected light. Real-space images of the sample were taken as the stage angle and the frequency were swept. The average intensity over pixels in the spatial area where the Airy resonance occurs (More details can be found in Section 6 on the expected location of the Airy resonances) was taken to obtain the intensity of the light from the Airy resonance as a function of the wavevector ($k_x$ or $k_y$) and the frequency ($\omega$). The resulting measurements for the sample with $\kappa = 0.152a^{-1}$ are shown in Figs. \ref{figs8}(a) and (b). 

Figure \ref{figs8}(a) shows the experimental measured band structure of Airy levels along $k_y$. The sample is excited with a plane wave and the light intensity reflected from the sample is measured. Here, unlike Fig. \ref{figs7}(c) where the intensities in all the pixels in the real-space image are averaged to obtain the reflection intensity, we only average the intensity of the pixels near the main lobe of the Airy mode to reduce the fluctuation from the background. More details can be found in Fig. \ref{figs11}(a) in Section 6. In order to obtain clear band features at $k_y=0$, we set a non-zero small $k_x$ ($k_x=0.03\ (2\pi a^{-1})$) in the incident light to eliminate the effect of BIC. We can clearly see in the experimental data that there are several discrete Airy levels in the vicinity of $\mathbf{k}=\Gamma$. Each Airy level is quadratically dispersive as predicted in Eq. \eqref{c23}. When $n$ becomes larger ($n>3$), the Airy levels are no longer visible in the band structure because the linewidth of the Airy mode increases as $n$ increases, which is predicted in Eq. \eqref{c24}. When the linewidth of the Airy mode is comparable to the separation between Airy levels, we can no longer distinguish discrete Airy levels. When $k_y$ becomes larger, it is also difficult to identify discrete Airy levels due to the increase of linewidth shown in Eq. \eqref{c24}.

Figure \ref{figs8}(b) shows the experimentally measured ``band structure" of Airy levels along $k_x$. The sample is excited with a plane wave with $k_y=0$ and $k_x\neq 0$. The superpotential breaks the periodicity of the dielectric function along the $x$ direction, so $k_x$ is no longer a conserved quantum number. This ``band structure" along $k_x$ can be understood as the optical response of our Airy sample under plane wave excitation. The intensity of response at different $k_x$ reflects how well the Airy mode can couple to the exciting plane wave. In Fig. \ref{figs8}(b), we observe several flat discrete Airy levels. This is because the resonance frequency only depends on $k_y$ but not on $k_x$. Moreover, we find that different Airy levels (with different $n$) have different brightest $k_x$, indicating that different Airy levels require a different plane wave of $k_x$ to best excite them. At $k_x=0.03\ (2\pi a^{-1})$, which is the value of $k_x$ used in Fig. \ref{figs8}(a), only the first $n=3$ Airy levels can be properly excited, which is consistent with the result in Fig. \ref{figs8}(a).

Figures \ref{figs8}(c)(d) show the corresponding simulation results under the same conditions as Figs. \ref{figs8}(a)(b). The simulation results are obtained from \textsc{Tidy3D}, a simulation software developed by \textsc{Flexcompute} which uses the Finite-Difference Time-Domain (FDTD) method \cite{Minkov:24}. We find good quantitative agreement between the experiment and the simulation.\\

Next, the frequency separation and linewidth of the Airy resonance at $k_y=0$ are extracted from the experimental data in Fig. \ref{figs8}(b).

\begin{figure}[H]
    \centering
    \subfigure[]{\includegraphics[width=0.45\textwidth]{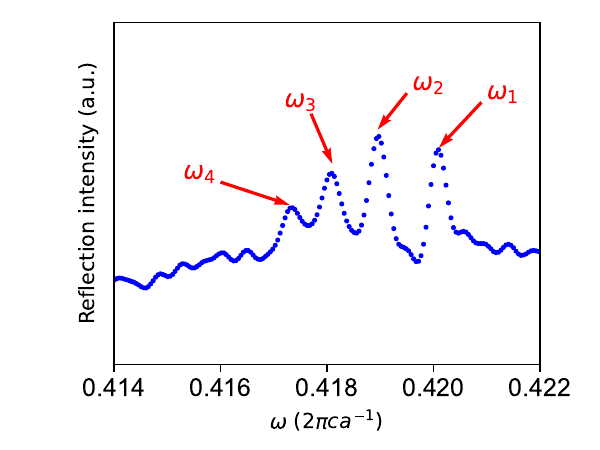}}
    \subfigure[]{\includegraphics[width=0.45\textwidth]{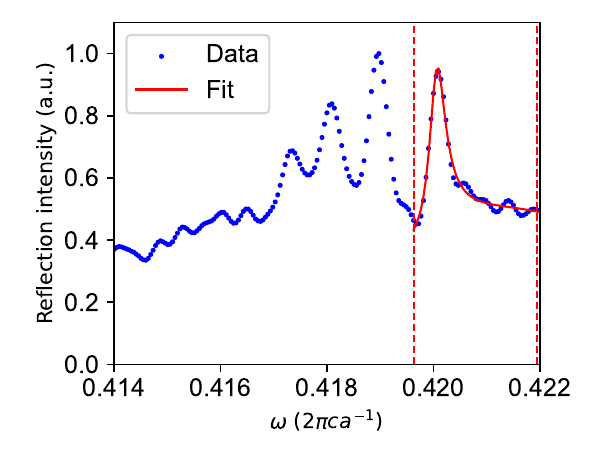}}

\caption{The experimental method to extract the Airy mode frequency separation and linewidth. (a) The reflection intensity spectrum measured at $k_y=0$ and $k_x=0.0398\ (2\pi a^{-1})$. The experimental parameters are the same as Fig. \ref{figs8}. (b) The method to extract the linewidth of the $n=1$ Airy resonance. The fitting method is the same as Fig. \ref{figs7}(d).}
\label{figs9}
\end{figure}

Figure \ref{figs9}(a) shows the reflection intensity spectrum after the data in Fig. \ref{figs8}(b) are fixed at $k_x=0.0398\ (2\pi a^{-1})$. We observe several discrete peaks corresponding to the resonances of Airy levels. The frequency of each Airy level is determined by the frequency of each peak, as marked with red arrows in Fig. \ref{figs9}(a). The Airy level separation is then calculated from the frequency difference between the peaks. The data in Fig. 2(d) in the main text are obtained by the method described above.

Then, we extract the linewidth of the Airy resonances from the reflection spectrum in Fig. \ref{figs9}(a). Equation \eqref{d1} describes the reflection spectrum when there is only one resonance state. However, this is no longer valid in this case because there are several Airy resonances which are very close to each other in frequency. As a result, the reflection spectrum in Eq. \eqref{d1} should be modified as:
\begin{equation}
R(\omega)=F_0\left|r_b\left(\omega,c_0,c_1,c_2\right)+\sum_n A_ne^{i\phi_n}\frac{\frac{\Delta \omega_n}{2}}{i\left(\omega-\omega_n\right)+\frac{\Delta\omega_n}{2}}\right|^2,
\label{g1}
\end{equation}
where $R(\omega)$, $F_0$, $r_b\left(\omega,c_0,c_1,c_2\right)$ are defined as in Eq. \eqref{d1}, and  $A_n$, $\phi_n$, $\omega_n$, $\Delta \omega_n$ are the amplitude, phase, resonance frequency, and linewidth of the $n^{\mathrm{th}}$ Airy resonance. We can see that for each additional state added, four more fitting parameters need to be considered. The effect of overfitting makes it difficult to precisely obtain the fitting parameters. However, the frequency separation between the $n=1$ and $n=2$ Airy resonances is always larger than the linewidths of the $n=1$ and the $n=2$ Airy levels, so the reflection intensity lineshape near the resonance frequency of the $n=1$ state is not significantly influenced by other modes. Hence, we use the same method as described in Eq. \eqref{d1} to fit the experimental spectrum in Fig. \ref{figs9}(a) within the frequency range near the $n=1$ resonance peak to extract the linewidth of the $n=1$ Airy state. Figure \ref{figs9}(b) shows the method described above to extract the linewidth of the $n=1$ Airy resonance. The red dashed lines show the fitting region near the $n=1$ peak. The fit linewidth is $\Delta \omega_0=\left(3.36\times 10^{-4} \pm 1.67\times 10^{-5}\right)\ (2\pi c a^{-1})$.

The data in Fig. 2(e) in the main text are obtained by the method described above. In Fig. 2(e) in the main text, the linewidth of the $n=1$ Airy state is generally larger than the theoretical prediction, which can be explained by material absorption loss and fabrication disorder mentioned in Fig. \ref{figs7}(f). For the linewidths of Airy resonances other than $n=1$, the frequency separation between the modes is comparable to the linewidth, which makes it hard to separate the lineshape in the spectrum or quantitatively extract their linewidths.

\clearpage

\section*{Section 6: The second signature of Airy resonances: Airy spatial mode profile}
\addcontentsline{toc}{section}{Section 6: The second signature of Airy resonances: Airy spatial mode profile}

In this section, we explain the experimental data regarding the second signature of the Airy resonances: the Airy spatial mode profile.

As predicted in Eq. \eqref{c22} and Fig. \ref{figs4}, the eigenstate of an Airy resonance is an Airy function with a truncation at $x=0$. Experimentally, we measured the Airy spatial mode profile with the experimental setup shown in Fig. \ref{figs7}(a).

To measure the spatial profile of the Airy resonances, the variable angle spectroscopy setup shown in Fig. \ref{figs7}(a) was used. The camera and lenses were aligned so that the specular reflected light traveled through the center of the lenses and camera. To increase the resolution of the images, no momentum space blocker was used in these measurements. For the sample with $\kappa = 0.108a^{-1}$, the stage angle ($\theta$) and the sample angle ($\varphi$) were set so that the in-plane wavevector of the light was $\left(k_x,k_y\right)=\left(-0.03,0\right)\ (2\pi a^{-1})$. This wavevector was selected because it coupled to the Airy resonances more efficiently than other wavevectors. The frequency of the laser was swept over the range $\omega \in [0.381, 0.434]\ (2\pi c a^{-1})$, while real-space images at the surface of the sample were captured. These images can probe the long-range envelope function of the resonances ($\alpha_0(\kappa\mathbf{r})$ from Eq. \ref{b23}) but cannot probe the short-range spatial profiles of the modes ($\mathbf{H_{k_0}}(\mathbf{r})$ in Eq. \ref{b23}) due to the diffraction limit. 

\begin{figure}[H]
    \centering
    \subfigure[]{\includegraphics[width=0.45\textwidth]{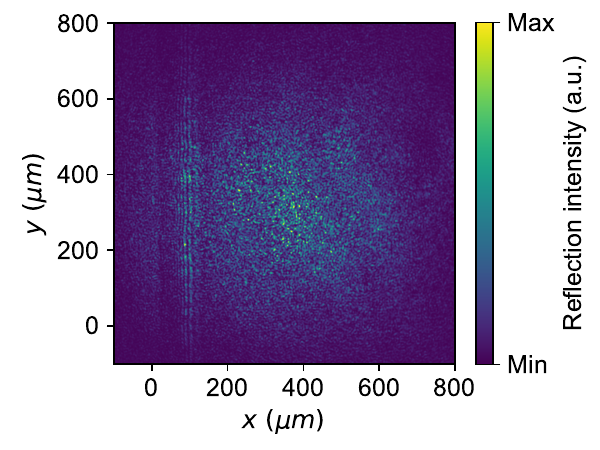}}
    \subfigure[]{\includegraphics[width=0.45\textwidth]{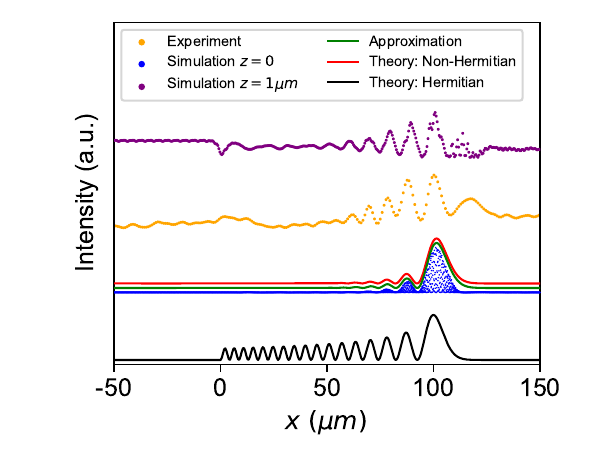}}

\caption{The experimental and simulation results on the wavefunction of Airy resonances. (a) The experimentally measured wavefunction of Airy resonances. (b) Comparison between the experimental data and simulation results.}
\label{figs10}
\end{figure}

Figure \ref{figs10}(a) shows the reflected light intensity at wavevector $\left(k_x,k_y\right)=\left(-0.03,0\right)\ (2\pi a^{-1})$ and frequency $\omega=0.416\ (2\pi c a^{-1})$ in the Airy sample with the potential strength $\kappa=0.108a^{-1}$. As we can see in the figure, apart from a long-range envelope function, there are also fast-varying features indicated by the small bright spots. We attribute these bright spots to specular reflection off of the sub-wavelength structures in our sample forming complex interference patterns. Since the envelope function is uniform in the $y$ direction, and only the Airy-function-shaped envelope functions of the eigenstates are of our interest, the experimental data in the $y$ direction is averaged. The averaged reflection intensity as a function of $x$ is plotted as orange dots in Fig. \ref{figs10}(b).

We then run the simulation under the same conditions as in the experiment with the use of \textsc{Tidy3D} \cite{Minkov:24}. The electromagnetic field intensity distribution at the surface of the sample ($z=0$) is collected and plotted in Fig. \ref{figs10}(b) as blue dots. Compared with the theoretical analysis in Eq. \eqref{c22}, which is plotted as the black line, the intensity of the lobes of both the experimental data and the simulation results decay rapidly to the left, such that only the first few lobes are distinguishable. This effect can be explained by the non-Hermitian effect arising from the imaginary part of the effective mass, $\mathrm{Im}\left(\frac{1}{2m}\right)$.

In the previous derivation in Fig. \ref{figs4} related to the eigenfunction of the Airy resonances, we treat the effective mass $\frac{1}{2m}$ as a real number, so according to Eq. \eqref{c24}, all the Airy resonances have zero linewidth. Hence, a plane wave with a single frequency can excite at most one Airy mode. However, this is not the case in both the experiment and the simulation, where the effective mass is complex. As a result, the Airy resonances have finite linewidths and therefore have non-zero optical response even in the case that the frequency of the incident light is not equal to the eigen-frequency of the mode. To consider this effect, we assume that the optical response has a Gaussian lineshape:
\begin{equation}
r_n\left(\omega\right)\propto
\frac{1}{\Delta\omega_n}e^{-\frac{4\left(\omega-\omega_n\right)^2}{\left(\Delta\omega_n\right)^2}}
,
\label{e1}
\end{equation}
where $r_n$ is the reflectivity response of the Airy resonance with mode index $n$, and $\omega_n$ and $\Delta\omega_n$ are the corresponding eigen-frequency and linewidth. The lineshape in Eq. \eqref{e1} has a width of $\Delta \omega_n$. 

Then, the reflection intensity distribution can be calculated as:
\begin{equation}
I(\omega,x)\propto\left|\sum_{n}r_n(\omega)\alpha_n(x)\right|^2
,
\label{e2}
\end{equation}
where $I\left(\omega,x\right)$ is the intensity distribution.

The theoretical optical response calculated from the non-Hermitian theory in Eq. \eqref{e2} is plotted in Fig. \ref{figs10}(b) as a red curve. In order to clearly see the curve, an additional offset is added in the vertical direction. It agrees very well with the simulation results (blue dots). It is worth mentioning that for simplicity, we assume the normalization factor $D_n$ in Eq. \eqref{c22} to be a positive real number. The theoretical line should become more accurate if the phase of $D_n$ is also considered.

We further notice that the envelope function of the optical response in Fig. \ref{figs10}(b) can be described by the exponential decay approximation in Eq. \eqref{c33}:
\begin{equation}
I\left(\omega,x\right)\approx
\left|e^{\gamma \left(x-x_n\right)}\cdot
\mathrm{Ai}\left(\left(\frac{1}{2m}\right)^{-\frac{1}{3}}\kappa\left(x-x_n\right)\right)\right|^2
,
\label{e3}
\end{equation}
where $\gamma$ is the decay parameter. 

In the continuous limit where $n\gg 1$, we re-write $x_n$ in terms of frequency $\omega$ at $k_y=0$:
\begin{equation}
I\left(\omega,x\right)\approx
\left|e^{\gamma \left(x-x_\omega\right)}\cdot
\mathrm{Ai}\left(\left(\frac{1}{2m}\right)^{-\frac{1}{3}}\kappa\left(x-x_\omega\right)\right)\right|^2
,
\label{e4}
\end{equation}
where $x_\omega$ is the offset of
the Airy function from the potential barrier as defined in Eq. \eqref{c42-1}. Equation \eqref{e4} is the same as Eq. (6) in the main text.

Then, the parameter $\gamma$ is fit by the simulation data and the approximation is plotted in Fig. \ref{figs10}(b) as a green curve. To avoid overlap between the curves and to better show the results, an additional offset is added in the vertical direction. The approximation (green curve) matches very well with the simulation data (blue dots) and the theoretical profile (red curve). The decay parameter $\gamma$ is no longer a mathematical trick in Eq. \eqref{c33} to make the Fourier components localize, but also an experimental reality that can describe the optical response after considering the non-Hermitian effect arising from the complex effective mass.

Moreover, we notice that the experimental data (orange dots) are a little different from the simulated result (blue dots); specifically, there is no fast-oscillating Bloch term in the experimental data. This can be explained by the difference between the near-field electromagnetic field and the radiative real-space imaging. 

For the experimental real-space imaging setup in Fig. \ref{figs7}(a), the camera can only capture the light intensity at the sample surface that radiates in the out-of-plane $z$ direction, which is fundamentally different from the electromagnetic near-field distribution that contains components below the light line. According to Eq. \eqref{c40}, the magnetic field distribution at the surface of the sample $z=0$ is:
\begin{equation}
\begin{aligned}
\mathbf{H_0}(x,y)
&= 
e^{ik_yy} \cdot
e^{\gamma \left(x-x_\omega\right)}
\cdot
\mathrm{Ai}\left[\left(\frac{1}{2m}\right)^{-\frac{1}{3}}\kappa\left(x-x_\omega\right)\right]
\sum_{\mathbf{G}} e^{i\left(G_x x+G_y y\right)}\mathbf{f(G)}
.
\label{e5}
\end{aligned}
\end{equation}

Similar to the reasoning in Eq. \eqref{c37-2}, only the $\mathbf{G}=(0,0)$ component in Eq. \eqref{e5} can leak out of the slab and radiate in the $z$ direction. Other components with $\mathbf{G}\neq(0,0)$ stay inside the slab and decay exponentially in the $z$ direction because they are under the light line. Hence, the radiation component in the magnetic field at the sample surface becomes:
\begin{equation}
\begin{aligned}
\mathbf{H_0}(x,y)
&= 
\left.\mathbf{f(G)}\right|_{\mathbf{G}=(0,0)}
\cdot e^{ik_yy} \cdot
e^{\gamma \left(x-x_\omega\right)}
\cdot
\mathrm{Ai}\left[\left(\frac{1}{2m}\right)^{-\frac{1}{3}}\kappa\left(x-x_\omega\right)\right]
.
\label{e6}
\end{aligned}
\end{equation}

Equation \eqref{e6} no longer contains the fast-oscillating Bloch term, which is consistent with our experimental data.

To verify this point, we simulate the intensity distribution at a small distance above the sample surface $z=1\mu m$, instead of right at the sample surface $z=0$. The near-field components that are below the light line will decay exponentially in $z$ so the intensity at $z=1\mu m$ is dominated by the radiation field. The simulation result is plotted in Fig. \ref{figs10}(b) with a purple line. It matches the experimental data very well.\\

Next, we focus on the position of the main lobe of the Airy eigenfunction as a function of the frequency. As predicted in Eq. \eqref{c30}, the position of the main lobe is a linear function of the probing frequency. Experimentally, we measure the wavefunction of the Airy resonances over a wide range of frequencies to observe this effect.

\begin{figure}[H]
    \centering
    \subfigure[]{\includegraphics[width=0.32\textwidth]{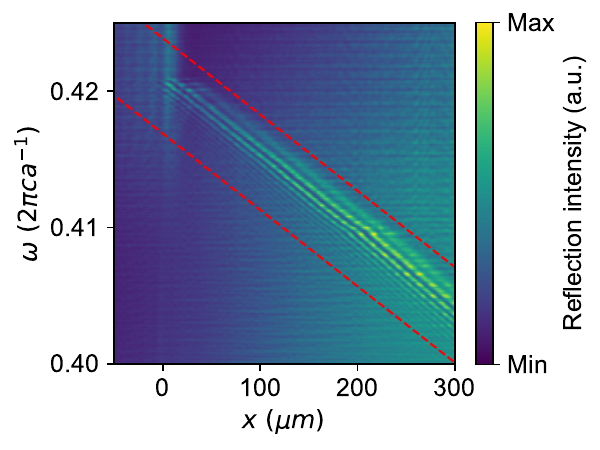}}
    \subfigure[]{\includegraphics[width=0.32\textwidth]{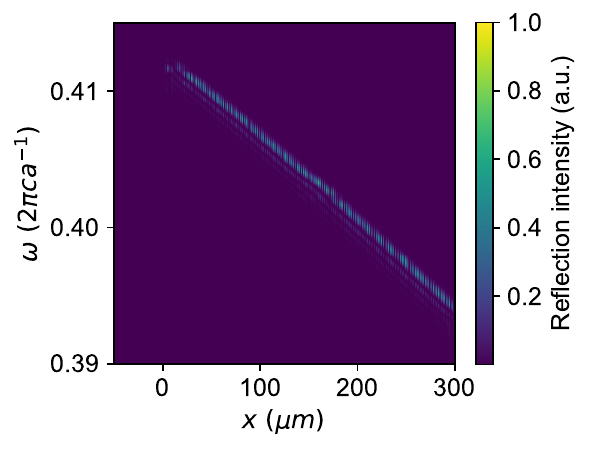}}
    \subfigure[]{\includegraphics[width=0.32\textwidth]{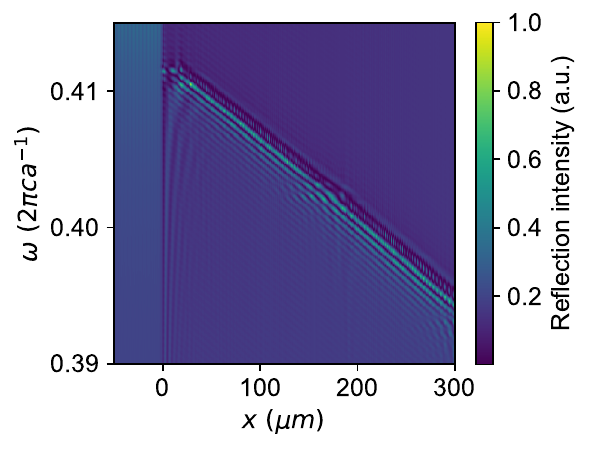}}

\caption{The experimental and simulation results on the wavefunction of the Airy resonances at different frequencies. (a) The experimentally measured wavefunction of Airy resonances at different frequencies. (b) The simulation results under the same conditions as (a). The intensity distribution at the sample surface $z=0$ is plotted. (c) The simulation results under the same condition in (a), but the intensity distribution is collected at $z=1\mu m$ above the sample surface.}
\label{figs11}
\end{figure}

Figure \ref{figs11}(a) shows the experimentally measured Airy mode wavefunction at different frequencies. The sample and experimental conditions are the same as in Fig. \ref{figs10}. Figure \ref{figs11}(b) shows the corresponding simulation results under the same conditions. The reflection intensity is collected and recorded at the sample surface $z=0$. We observe in both experiment and simulation that the main lobe position is indeed a linear function of the probing frequency.

Compared to the experimental data, the simulated near-field distribution in Fig. \ref{figs11}(b) also has an extra fast-varying oscillation feature, as predicted in Eq. \eqref{b23}. Due to the reason we explained in Eqs. \eqref{e5} and \eqref{e6}, we are not able to observe the fast-oscillating Bloch term in our experimental setup, but only the slow-varying Airy envelope function. Moreover, the length scale of the fast-oscillating Bloch term is smaller than the diffraction limit of the light probing our sample, so we cannot experimentally observe any features at this length scale. In Fig. \ref{figs11}(c), we use the same simulation conditions as in Fig. \ref{figs11}(b), but collect and record the reflection intensity at $z=1\mu m$ above the sample surface to eliminate the components below the light line. The simulation result in Fig. \ref{figs11}(c) agrees very well with the experimental data in Fig. \ref{figs11}(a).

We notice that the background reflection intensity in the simulations (both Figs. \ref{figs11}(b) and (c)) is low and uniform. However, the background in the experimental data (Fig. \ref{figs11}(a)) is stronger than the simulation, both in intensity and fluctuation. To minimize the influence of the background, only the intensity of the pixels near the main lobe of the Airy mode, instead of the whole sample, is averaged when obtaining the total reflection intensity in Fig. \eqref{figs8}. The red dashed lines in Fig. \ref{figs11}(a) show the region at each frequency in which the intensity of the pixels are averaged.

\clearpage

\section*{Section 7: The third signature of Airy resonances: Bending of the Airy mode radiation in free space}
\addcontentsline{toc}{section}{Section 7: The third signature of Airy resonances: Bending of the Airy mode radiation in free space}

In this section, we explain the experimental data regarding the third signature of the Airy resonances: the bending of the Airy mode radiation in free space.

We show in Eq. \eqref{e3} that the optical response from the Airy sample can be described by an Airy function with an additional exponential decay. Then, as predicted in Eqs. \eqref{c42} and \eqref{c44} and Fig. \ref{figs5}, the main lobe (as well as all other lobes) of the radiation from Airy resonances follows a quadratic trajectory, which is known as the acceleration or bending in the free space. We experimentally measure this effect by moving the imaging plane away from the sample surface and record the real-space reflection intensity distribution as a function of the propagation distance $z$.

\begin{figure}[H]
    \centering
    \subfigure[]{\includegraphics[width=0.45\textwidth]{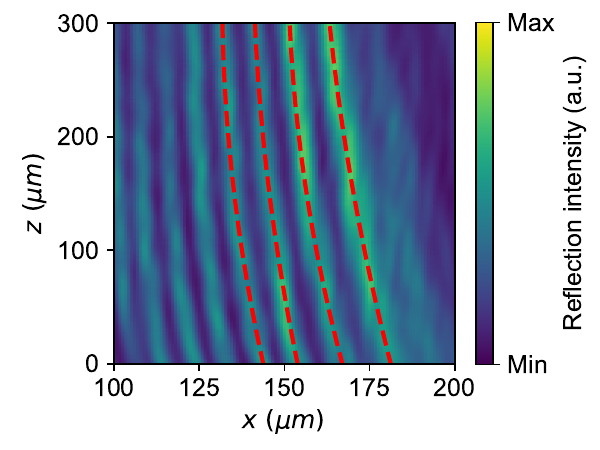}}
    \subfigure[]{\includegraphics[width=0.45\textwidth]{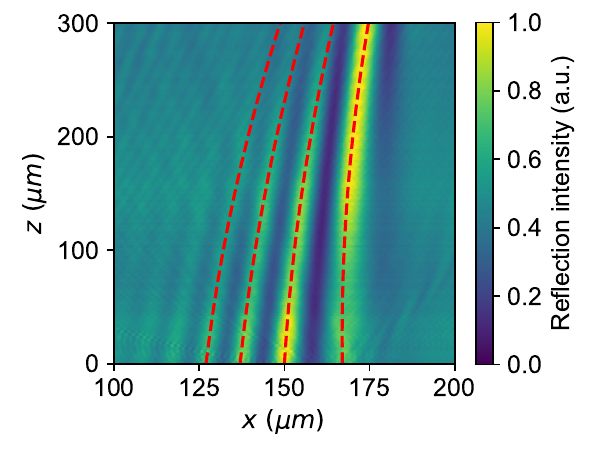}}

\caption{The experimental and simulation results on the bending of the Airy mode radiation in free space. (a) The experimentally measured reflection intensity distribution as a function of the propagation distance $z$. (b) The simulation results under the same conditions as (a).}
\label{figs12}
\end{figure}

To measure the beam evolution in free space, we use the setup in Fig. \ref{figs7}(a). The lens and camera are initially aligned such that the object plane is located at and parallel to the surface of the sample. For all measurements of the beam evolution, $k_y = 0$ and $k_x$ was chosen for each $\kappa$ to maximize the coupling between the plane wave and the Airy resonances. The frequency is fixed at $\omega = 0.411\ (2\pi c a^{-1})$. The lenses and camera are then smoothly moved backward along their center axis with a constant speed using a motor over a distance of $500\mathrm{\mu m}$, smoothly moving the object plane from the surface of the sample $z=0$ to $z=500\mu m$ above the sample while the camera records images. From this setup, we can determine the intensity of the light distribution above the sample as a function of position $I(x, y, z)$. Since the resonances are spatially uniform in the $y$ direction, the intensity is averaged in the $y$ direction to obtain $I(x,z)$, which is shown in Fig. \ref{figs12}(a). 

Figure \ref{figs12}(a) shows the experimentally measured reflection intensity distribution at each propagation distance $z$. We can clearly see that all the lobes are bending, as indicated by red dashed lines. In order to minimize the fluctuation, we use a quadratic fit to determine the path of the first four lobes:
\begin{equation}
x(z)=c_2z^2+c_1z+c_0
,
\label{f1}
\end{equation}
where $c_0$, $c_1$ and $c_2$ are fitting parameters. Compared with Eq. \eqref{c44}, and according to the definition of the bending $\beta$ in Eq. \eqref{c46}, we obtain the bending in the experimental data:
\begin{equation}
\beta
=\frac{\partial^2 x(z)}{\partial z^2}
=2c_2
.
\label{f2}
\end{equation}

Equations \eqref{f1} and \eqref{f2} describe the method used to obtain the experimental data in Fig. 4(b) in the main text. The average value of the quadratic coefficient $c_2$ is then used to plot the quantity in Fig. 4(b). The standard deviation of the bending from different lobes is plotted as the error bar.

Figure \ref{figs12}(b) shows the simulation results carried out by \textsc{Tidy3D} \cite{Minkov:24} under the same conditions as in Fig. \ref{figs12}(a). The intensity profile is normalized at every $z$. We again observe the bending of the first four lobes as marked with red dashed lines. According to Eq. \eqref{c44}, the ``initial velocity" (which is defined as $v=\frac{\partial x(z)}{\partial z}$) should be zero in the theory. Figure \ref{figs12}(b) verifies that at $z=0$ the propagation direction of the Airy mode radiation is nearly perpendicular to the sample surface. However, in the experimental data in Fig. \ref{figs12}(a), the initial velocity is not zero. This is because of the alignment error in our experimental setup: the imaging plane is not moved exactly in the $z$ direction, but along a path slightly tilted from the $z$ axis. As a result, the $x$ coordinates in the experimental data have a linear offset depending on $z$, which is the origin of the initial velocity in the experimental data. We note that this linear offset does not change the value of the bending $\beta$ in the second derivative, so the experimental method described in Eqs. \eqref{f1} and \eqref{f2} is still valid.

We also notice that more lobes can be identified in the experimental data in Fig. \ref{figs12}(a) compared with the simulation results in Fig. \ref{figs12}(b), as well as the previous experimental data in Fig. \ref{figs11}(a), meaning that the decaying parameter $\gamma$ defined in Eq. \eqref{e4} is smaller. This is due to the setting of the camera exposure time when measuring the data in \ref{figs12}(a). For the bending measurement, we try to identify the trajectory of the lobes as many as possible, so the bending can be measured more precisely as the average among all the lobes. Hence, the exposure time of the camera is increased to be able to measure more lobes even if their intensity is relatively low. As a result, some camera pixels with large intensity are saturated. The intensities of the lobes become more similar, so the decay parameter $\gamma$ appears to be smaller.

\end{document}